\documentclass{aa}  

\usepackage{graphicx}       
\usepackage{natbib}
\usepackage{longtable}
\usepackage{supertabular}

\newcommand{\BCDG}{{BCDG}}

\newcommand{\Ha} {H$\alpha$}
\newcommand{\Hb} {H$\beta$}
\newcommand{\HII}{\textsc{H\,ii\ }}

\newcommand{\Lo} {$L_{\odot}$}
\newcommand{\Mo} {$M_{\odot}$}
\newcommand{\Myr}{Myr}          

\newcommand{\NIR} {\emph{NIR}}

\newcommand{\WRBUMP} {WR bump}
\newcommand{\Zo} {$Z_{\odot}$}

\newcommand{\STARBURST}{{\sc Starburst~99}}

\newcommand{\kms}{km\,s$^{-1}$}

\newcommand{\tableline}{\hline}

\newcommand{\HeII}{\ion{He}{ii}}

\newcommand{\WHb}{$W$(H$\beta$)}

\newcommand{\abox}{12+log(O/H)}

\newcommand{\nodata}{...}
\begin{document}

   \title{Massive star formation in Wolf-Rayet galaxies\thanks{Based on observations made with NOT (Nordic Optical Telescope), INT (Isaac Newton 
Telescope) and WHT (William Herschel Telescope) operated on the island of La Palma jointly by Denmark, Finland, Iceland, Norway and Sweden (NOT) or 
the Isaac Newton Group (INT, WHT) in the Spanish Observatorio del Roque de Los Muchachos of the Instituto de Astrof\'\i sica de Canarias. }}

   \subtitle{III. Analysis of the O and WR populations}

   \author{\'Angel R. L\'opez-S\'anchez
          \inst{1,2}
		  \and
		  C\'esar Esteban\inst{2,3}
          }

   \offprints{\'Angel R. L\'opez-S\'anchez, \email{Angel.Lopez-Sanchez@csiro.au}}

\institute{CSIRO Astronomy \& Space Science / Australia Telescope National Facility, PO\,BOX\,76, Epping, NSW\,1710, Australia \and Instituto de  
Astrof{\'\i}sica de Canarias, C/ V\'{\i}a L\'actea S/N, E-38200, La Laguna, Tenerife, Spain \and Departamento de Astrof\'{\i}sica de la Universidad  
de La Laguna, E-38071, La Laguna, Tenerife, Spain}


   \date{Received October 9, 2009; Accepted March 28, 2010}

 
  \abstract
   {}
{We perform a comprehensive multiwavelength analysis of a sample of 20 starburst galaxies that show a substantial
population of Wolf-Rayet (WR) stars. In this paper, the third of the series, we present the analysis of the O and WR star populations within these  
galaxies.}
  {We study the spatial localization of the WR-rich clusters via the detection of the blue \WRBUMP, mainly composed by the broad \ion{He}{ii}  
$\lambda$4686 line and attributed to WN stars, and the red \WRBUMP, composed by the broad \ion{C}{iv} $\lambda$5808 line observed in WC stars. We  
perform a detailed fitting of the nebular and broad emission lines within these broad features and derive the numbers of WN, WC and O stars using (i)  
the standard assumption of constant WR luminosities and (ii) considering metallicity-dependent WR luminosities. We then compare our results with the  
predictions given by evolutionary synthesis models and with previous empirical results.}
 {We report the detection of blue \WRBUMP\ in 20 regions, but the red \WRBUMP\ is only detected in six. Aperture effects and the exact positioning of  
the slit onto the WR-rich bursts play a fundamental role in their detection. The nebular \ion{He}{ii} $\lambda$4686 line is detected in 21 regions;  
its intensity clearly decreases with increasing metallicity. We derive an empirical estimation of the WNL/(WNL+O) ratio using the intensity of the  
broad \ion{He}{ii} line assuming metallicity-dependent WR luminosities. As expected, the total number of WR stars increases with increasing  
metallicity, but objects with \abox$<$8.2 show a rather constant WR/(WR+O) ratio. The computed WCE/WNL ratios are different than those empirically  
found in nearby star-forming galaxies, indicating that the observed galaxies are experiencing a strong and very short burst. Considering  
metallicity-dependent WR luminosities, our data agree with a Salpeter-like IMF in all regimes.}
  {We consider that the contribution of the WCE stars is not negligible at low metallicities, but deeper observations are needed to detect the red  
\WRBUMP\ because of the decreasing of the WR luminosities with decreasing metallicity. Although available models reproduce the WR  
properties at high metallicities fairly well, new evolutionary synthesis models for young starbursts including all involved parameters (age, metallicity,  
star-formation history, IMF and WR stars properties such as metallicity-dependent WR luminosities, stellar rotation and the WR binnary channel) are  
absolutely needed to perform an appropriate comparison with the observational data.}

\titlerunning{Massive star formation in Wolf-Rayet galaxies III: O and WR populations}

\authorrunning{L\'opez-S\'anchez \& Esteban}

   \maketitle
%

\section{Introduction}

Despite their relatively low number and short lifetime in terms of evolutio\-nary timescales, massive stars have a fundamental influence on the  
interstellar medium (ISM) and galaxy evolution: they generate most of the ultraviolet ionizing radiation in galaxies, powering the far-infrared  
luminosities through the heating of dust; they enrich the interstellar medium, not only by returning nuclear processed material during their whole  
lifetime \citep{Maeder81} but also in supernova explosions; finally, they deposit mechanical energy, first via strong stellar winds and later as  
supernovae \citep{Abbott82}, which are a triggering mechanism of star formation \citep{Woodward78}. 
Massive stars are the progenitors of the most energetic phenomena known today, the long gamma-ray bursts (GRB), as they collapse after supernova  
explotions into black holes \citep{WoosleyHeger06}.
The most massive ($M\geq$ \mbox{25 \Mo}\ for \Zo) luminous (10$^5$ to 10$^6$ \Lo) and hot (temperatures $\sim$50~000~K) O stars evolve to the  
Wolf-Rayet phase between 2 and 5 \Myr\ since their birth. A WR star is interpreted as a central He-burning core that has lost the main part of its  
H-rich envelope via strong winds \citep{Conti76,Maeder90,Maeder91}. This stage of the most massive stars lasts only some few hundreds of thousands of  
years ($t_{WR}\leq5\times10^5$~yr) before they explode as Type Ib/Ic supernovae  \citep{MeynetMaeder05}.

The broad emission features that characterize the spectra of WR stars are often observed in extragalactic \HII regions. Actually, these Wolf-Rayet  
galaxies make up a very inhomogeneous class of star-forming objects, all having in common ongoing or recent star formation which has produced stars  
massive enough to evolve to the WR stage \citep{SCP99}.
The blend of the broad \HeII\ $\lambda$4686, \ion{C}{iii}/\ion{C}{iv} $\lambda$4650 and \ion{N}{iii} $\lambda$4640 emission lines constitutes the  
{\bf blue \WRBUMP}; it mainly originates in WN stars with a minor contribution of WC stars.
The original blue \WRBUMP\ detection was made in the blue compact dwarf galaxy He~2-10 by \citet*{Allen76}. Later, \citet{KS81} recognized the  
importance of this \WRBUMP\ to characterize the mode of star formation in starbursts (burst versus continuous mode) and developed the first method to  
empirically derive the WR/O ratio.
The blend of the \ion{C}{iii} $\lambda$5698 and \ion{C}{iv} $\lambda$5808 broad emission lines constitutes the {\bf red \WRBUMP}. \ion{C}{iv} 
$\lambda$5808 is the strongest emission line in WC stars, but it is barely seen in WN stars. The first detections of the red \WRBUMP\ in integrated  
galaxy spectra were reported by \citet{KS86} and \citet{Dinerstein86}, but it seems to be more difficult to detect than the blue \WRBUMP\ and it is  
always weaker than the blue \WRBUMP\ \citep{GIT00,FCCG04}. Actually, the red \WRBUMP\ is rarely observed at low metallicities. However, both WN and  
WC stars have been detected in the lowest metallicity ($Z\sim$1/50~\Zo) galaxy known, I~Zw~18 \citep{Izotov97,Legrand97,DeMello98}.

Although the main feature of WR galaxies is a broad \HeII\ $\lambda$4686 emission line, 
a considerable fraction of objects also show the nebular (narrow) \HeII\ $\lambda$4686 line. 
This emission line is very rarely found in 
Galactic \HII regions \citep{Garnett91,Schaerer97} but common in planetary nebulae, 
and has been sometimes observed in giant \HII regions where no WR stars have been detected.
Sources with sufficient photons of energy $>$54 eV are needed to produce the nebular \HeII\ 
emission line, so its origin is still puzzling \citep{Garnett91,G04}. Some observations concluded that 
this nebular line is intimately linked with the appearance of hot WR stars \citep{Schaerer96,SV98,CH06}
but recent analyses suggest that O stars may also contribute to the \ion{He}{ii} ionizing flux at low metallicities
(Brichmann, Kunth \& Durret 2008, hereafter BKD08). In one way or in another, studies of the nebular \HeII\ emission line give clues about the  
physical processes in \HII regions, the ionizing fluxes of starbursts and their contribution to the ionization of the intergalactic medium (Garnett  
et al. 1991; Schaerer et al. 1998; Stasi\'nska 1998). This is the reason why \citet*{SCP99} also compiled all the extragalactic objects (54) showing  
the nebular \HeII\ $\lambda$4686 emission line in their catalog of WR galaxies. The analysis performed by \citet{TI05} listed 465 Blue Compact  
Dwarf galaxies (\BCDG s) showing nebular \ion{He}{ii} $\lambda$4686 emission.

Wolf-Rayet stars can be individually observed in our Galaxy, the Magellanic Clouds, in some galaxies of the Local Group (M~31, M~33, IC~10, NGC~6822,  
IC~1613) and in the brightest galaxies of the Sculptor group (NGC~300 and NGC~55). 
A recent review of the physical properties of WR stars was presented by \citet{Crowther07}.
Thanks to the analysis of these individual WR stars a quantitative classification in WN, WC and WO stars could be established. This led the  
developing of stellar population synthesis models tailored to the analysis of massive star populations in young starburst galaxies. The best example  
of these models are those presented by Schaerer \& Vacca (1998; hereafter SV98), which provide detailed predictions of many stellar and nebular  
UV/optical features. Famous \STARBURST\ models \citep{L99,VL05} also include predictions for the WR and O stars populations. Except for some very low  
metallicity objects, a good agreement between the observations and the synthesis models of SV98 is usually found. For objects with subsolar  
metallicity (mostly \BCDG s), this comparison indicates fairly short timescales of star formation, with burst ages lower than 2--4 Myr, which is the  
initial mass function (IMF) compatible with a Salpeter slope, but requires the existence of high-mass stars 
(Guseva et al. 2000; De Mello et al. 1998; Mas-Hesse \& Kunth 1999, Fernandes et al. 2004; Buckalew, Kobulnicky \& Dufour 2005). 
However, some authors \citep{Zhang07} suggested a dependence of the slope of the IMF on metallicity. 
The timescales of star formation in high-metallicity enviroments are more extended than those found at low metallicity, with burst durations of  
\mbox{4--10~Myr} \citep{SGIT00,FCCG04}, although observations are also explained assuming a superposition of several bursts.

We performed a detailed photometric and spectroscopic analysis of a sample of 20 WR galaxies. Our main aim is the study of the formation of 
massive stars in starburst galaxies and the role that the interactions with or between dwarf galaxies and/or low surface brightness objects have in 
its triggering mechanism. In Paper~I \citep{LSE08} we explained the motivation of this work, compiled the list of the analyzed 
WR galaxies (Table~1 of Paper~I) and presented the results of optical/\NIR\ broad-band and \Ha\ photometry. In Paper~II \citep{LSE09} we presented  
the results of the analysis of intermediate-resolution long-slit spectroscopy of 16 objects of our sample of WR galaxies --the results for the other  
four objects were published separately. In many cases, two or more slit positions were used 
to analyze the most interesting zones, knots or morphological structures belonging to each galaxy or even surrounding objects. In this paper  
we analyze the localization and the properties of the O and WR populations within those galaxies. Papers~IV and V will present the global  
multiwavelength analysis of our WR galaxy sample considering all available data.

In Sect.~2 we describe the detection of the WR features. The procedure used to fit the broad and nebular emission lines in the blue and red \WRBUMP s is  
explained in Sect.~3. We analyze the detection of the nebular \ion{He}{ii} $\lambda$4686 line in Sect.~4. The quantitative analysis of the WNL and WCE star  
populations in described in Sect.~5. In Sect.~6 we analyze the metallicity dependence of the WR/(WR+O) ratio. The comparison with \citet{SV98} and  
\STARBURST\ \citep{L99,VL05} models is shown in Sect.~7. We analyze the derived WCE/WNL ratio in Sect.~8, while the dependence of the IMF on the WR/(WR+O)  
ratio is discussed in Sect.~9. Finally, we compile our main conclusions in Sect.~10.

\section{Detection of the WR features}

\begin{table*}[t!]
\centering
  \caption{\footnotesize{Detection of WR features and the nebular \ion{He}{ii} emission line for the galaxies analyzed in this work. }}
  \label{wrdata}
  \tiny
  \begin{tabular}{l@{\hspace{4pt}}c  c@{\hspace{8pt}}c c   c@{\hspace{8pt}}cc    c@{\hspace{8pt}}cc  c@{\hspace{8pt}}cc   }
    \noalign{\smallskip}
    \tableline
	\noalign{\smallskip}

 & & \multicolumn{2}{c@{\hspace{4pt}}}{\Hb} &  & 
\multicolumn{2}{c@{\hspace{4pt}}}{\HeII\ $\lambda$4686} & & 
\multicolumn{2}{c@{\hspace{4pt}}}{Blue\ \WRBUMP$^c$} & &
\multicolumn{2}{c@{\hspace{4pt}}}{Red\ \WRBUMP$^d$} & \\

\cline{3-4}
\cline{6-7}
\cline{9-10}
\cline{12-13}            
\noalign{\smallskip}

Galaxy & Zone &  

$I$  &  
$-W$   & 
$\eta$ &

$I$/$I$(\Hb) & 
$-W$ & &

$I$/$I$(\Hb) & 
$-W$ & &

$I$/$I$(\Hb) & 
$-W$ \\

 & & 
(a) & [\AA] & & 
(b) & [\AA] & &
(b) & [\AA] & &
(b) & [\AA] & \\

\noalign{\smallskip}    
  \tableline
\noalign{\smallskip}
 
HCG 31& AC$^e$&46.13$\pm$0.18 &  91.1$\pm$2.1 & 0.25 & 0.9$\pm$0.3 & 0.66$\pm$0.22 &  & 3.2$\pm$1.9 &  2.5$\pm$0.6 & &  2.4$\pm$0.6 & 3.1$\pm$0.8 \\
                & F1    &  3.187$\pm$0.016 &   218$\pm$13  & 0.80 & 2.8$\pm$0.3 & 4.3$\pm$0.4 & & \nodata & \nodata & & \nodata & \nodata  \\
                & F2    &   2.28$\pm$0.19  &   256$\pm$43  & 0.80 & 2.2$\pm$0.5 & 3.9$\pm$1.5 & & 0.95$\pm$0.35 & 3.4$\pm$1.6  &&\nodata&\nodata\\
Mkn 1087        & C$^f$ &   78.1           &  \nodata      & 0.25 & \nodata & \nodata & & 10.6 & \nodata & & \nodata & \nodata  \\
    Haro 15     &  C    &  23.3$\pm$1.1    &  16.4$\pm$1.1 & 0.40 & 0.49: & 0.08:  & & 2.82: & 0.5: & & \nodata &\nodata \\
                &  A    &  23.4$\pm$0.9    &  75.7$\pm$4.2 & 0.50 & 1.7$\pm$0.4 & 1.0$\pm$0.2 & & 1.4$\pm$0.5 & 1.0$\pm$0.3 & & \nodata&\nodata\\
    Mkn 1199    & C$^e$ &  74.2$\pm$3.1    &  21.4$\pm$1.3 & 0.50 & 0.24: & 0.05:  & & 6.3$\pm$1.8 & 1.3$\pm$0.3 & & 9.4$\pm$2.3 & 2.8$\pm$0.7  \\
    Mkn 5       &  A    &  17.7$\pm$0.7    &    75$\pm$5   & 0.40 & 0.9$\pm$0.2 & 1.6$\pm$0.3 & & 1.5$\pm$0.5 & 3.0$\pm$1.2 & & \nodata &\nodata \\
IRAS 08208+2816 &  C    &  12.9$\pm$0.5    &    80$\pm$5   & 0.25 & \nodata & \nodata & & 4.1$\pm$1.3 & 3.3$\pm$1.0 & & 3.5$\pm$1.3 & 3.6$\pm$1.4 \\
IRAS 08339+6517 & \#1   &  36.5$\pm$1.3    &    25$\pm$2   & 0.50 & 0.48: & 0.1: & & 4.1: & 2.2: & & \nodata & \nodata  \\
POX 4 &  C    &  56.0$\pm$1.8  & 200$\pm$9   & 0.60 & 1.20$\pm$0.16 & 1.9$\pm$0.3 & & 1.5$\pm$0.3 & 3.0$\pm$0.9 & & 1.4$\pm$0.2 & 3.9$\pm$0.8  \\
         UM 420 &\nodata&  6.88$\pm$0.27   &   169$\pm$10  & 0.50 & 1.0$\pm$0.2 & 2.0$\pm$0.4 & & 1.5$\pm$0.5 & 2.8$\pm$0.7 & & \nodata&\nodata \\
SBS 0926+606    & A     &  16.2$\pm$0.6    &   125$\pm$6   & 0.60 & 0.7$\pm$0.2 & 1.9$\pm$0.3 & & 0.9:& 3.2: & & \nodata & \nodata \\
SBS 0948+532    &\nodata&  8.44$\pm$0.32   &   213$\pm$11  & 0.60 & 1.3$\pm$0.3 & 1.6$\pm$0.6 & & 1.1$\pm$0.3 & 2.1$\pm$0.7 & & \nodata & \nodata\\
SBS 1054+365    & C     &  14.6$\pm$0.7    &    89$\pm$7   & 0.40 & 0.6$\pm$0.2 & 0.9$\pm$0.3 & & 1.6$\pm$0.5 & 2.4$\pm$0.8 & & \nodata & \nodata\\
SBS 1211+540    &\nodata&  1.84$\pm$0.09   &   135$\pm$10  & 0.50 & \nodata & \nodata & & $<$1.2: & \nodata & & NO OBS.$^g$  & NO OBS.$^g$   \\
SBS 1319+579    & A     &  14.6$\pm$0.5    &   285$\pm$14  & 0.90 & 0.8: & 1.5: & & 0.7: & 1.9: & & \nodata & \nodata  \\
SBS 1415+437    & C     &  18.5$\pm$0.7    &   222$\pm$11  & 0.60 & 2.4$\pm$0.2 & 4.8$\pm$0.3 & & $<$0.9: & \nodata & & \nodata & \nodata \\
III Zw 107      & A     &  22.3$\pm$0.8    &  44.2$\pm$2.3 & 0.50 & \nodata & \nodata & & 3.1$\pm$1.1 & 1.3$\pm$0.4 & & \nodata & \nodata \\
Tol 9           & C     &  23.4$\pm$0.9    &  33.3$\pm$2.0 & 0.50 & 0.33$\pm$0.15 & 0.24$\pm$0.7 & & 7.9$\pm$1.3 & 2.5$\pm$0.4 & & 5.8: & 1.9: \\
Tol 1457-262    & A     &   222$\pm$8      &   101$\pm$6   & 0.50 & 1.9$\pm$0.4 & 1.5$\pm$0.3 & & \nodata & \nodata & & \nodata & \nodata \\
                & B     &  47.7$\pm$2.3    &    83$\pm$7   & 0.40 & 2.9$\pm$0.3 & 2.1$\pm$0.2 & & \nodata & \nodata & & \nodata & \nodata \\
ESO 566-7       &\nodata&  10.2$\pm$0.8    &    13$\pm$2   & 0.25 & \nodata & \nodata & & \nodata & \nodata & & \nodata & \nodata \\				
ESO 566-8       &\nodata& 111$\pm$5  & 95$\pm$7 & 0.50 & 0.74$\pm$0.25 & 0.65$\pm$0.25 & & 6.2$\pm$2.2 & 4.1$\pm$1.5 & & 4.8$\pm$1.2 & 4.2$\pm$1.2 \\
   NGC 5253     & A &  134.5$\pm$0.4  &  234$\pm$5   & 1.00 & 0.10$\pm$0.04 & 0.21$\pm$0.09 & & 0.71$\pm$0.12 & 1.6$\pm$0.3 & & NO OBS.$^g$ & NO OBS.$^g$ \\
                & B &  135.2$\pm$4.2  &  254$\pm$5   & 1.10 & 0.05: & 0.11: & & 0.71$\pm$0.13 & 1.7$\pm$0.3 & & NO OBS.$^g$ & NO OBS.$^g$  \\
                & C$^{e,g}$ &  101.3$\pm$3.3  & 94.0$\pm$1.0 & 0.40 & 0.06: & 0.05: & & 4.13$\pm$0.19 & 3.55$\pm$0.17 & &1.00$\pm$0.14$^h$&NO OBS.$^g$\\
                & D &  25.6$\pm$0.9   & 39.0$\pm$0.5 & 0.30 & 0.40$\pm$0.11 &  0.12$\pm$0.04 & & \nodata & \nodata & & NO OBS.$^g$ & NO OBS.$^g$\\	   
	\noalign{\smallskip}    
  \tableline
  \end{tabular}
    \begin{flushleft}
  $^a$ In units of 10$^{-15}$ erg s$^{-1}$ cm$^{-2}$ and not corrected for extinction.\\
  $^b$ Assuming $I$(\Hb)=100 and reddening-corrected.\\
  $^c$ Broad \ion{He}{ii} $\lambda$4686 emission line only.\\
  $^d$ Broad \ion{C}{iv} $\lambda$5808 emission line only.\\
  $^e$ Broad \ion{N}{iii} $\lambda$4640 also detected. The reddened-corrected $I$/$I$(\Hb) ratios are 0.91, 5.6 and 1.44 for HCG~31~AC, Mkn~1199 and NGC~5253~C, respectively. See text for details.\\
  $^f$ Data from \citet{Va97}, see \citet{LSER04b}. \\
  $^g$ NO OBS indicates that the spectral range where the feature lies was not observed.\\
  $^h$ Broad \ion{C}{iv} $\lambda$4658 detected. The $I$/$I$(\Hb) ratio given here is assuming $L_{\rm WCE}$($\lambda$5808)=$L_{\rm WCE}$($\lambda$4658)/1.71 (SV98). See text for details.
  \end{flushleft}
\end{table*}

\begin{figure*}[t!]
\centering
\begin{tabular}{c@{\hspace{4pt}}c@{\hspace{4pt}}c}
\includegraphics[angle=90,width=0.33\linewidth]{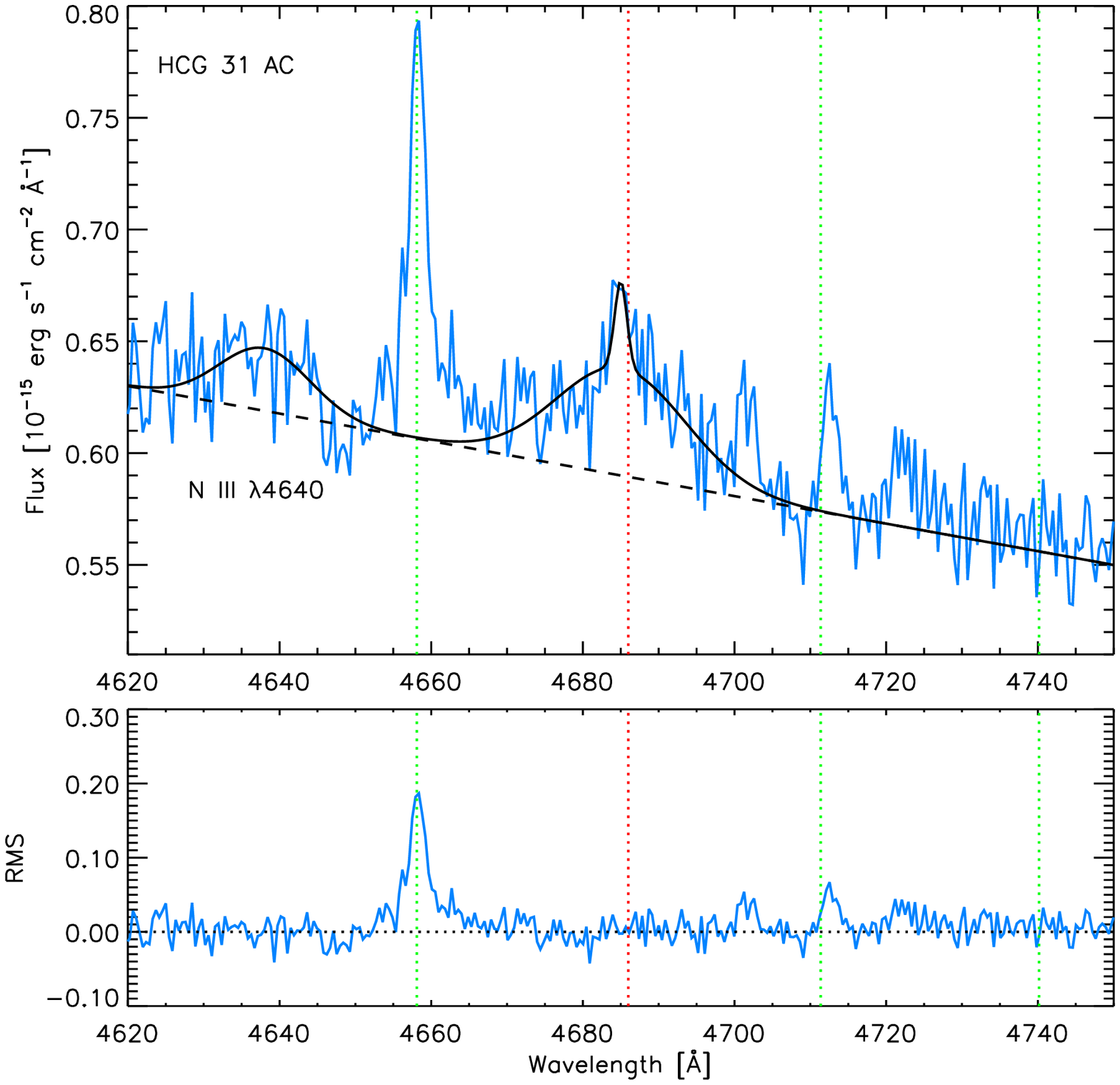} &
\includegraphics[angle=90,width=0.33\linewidth]{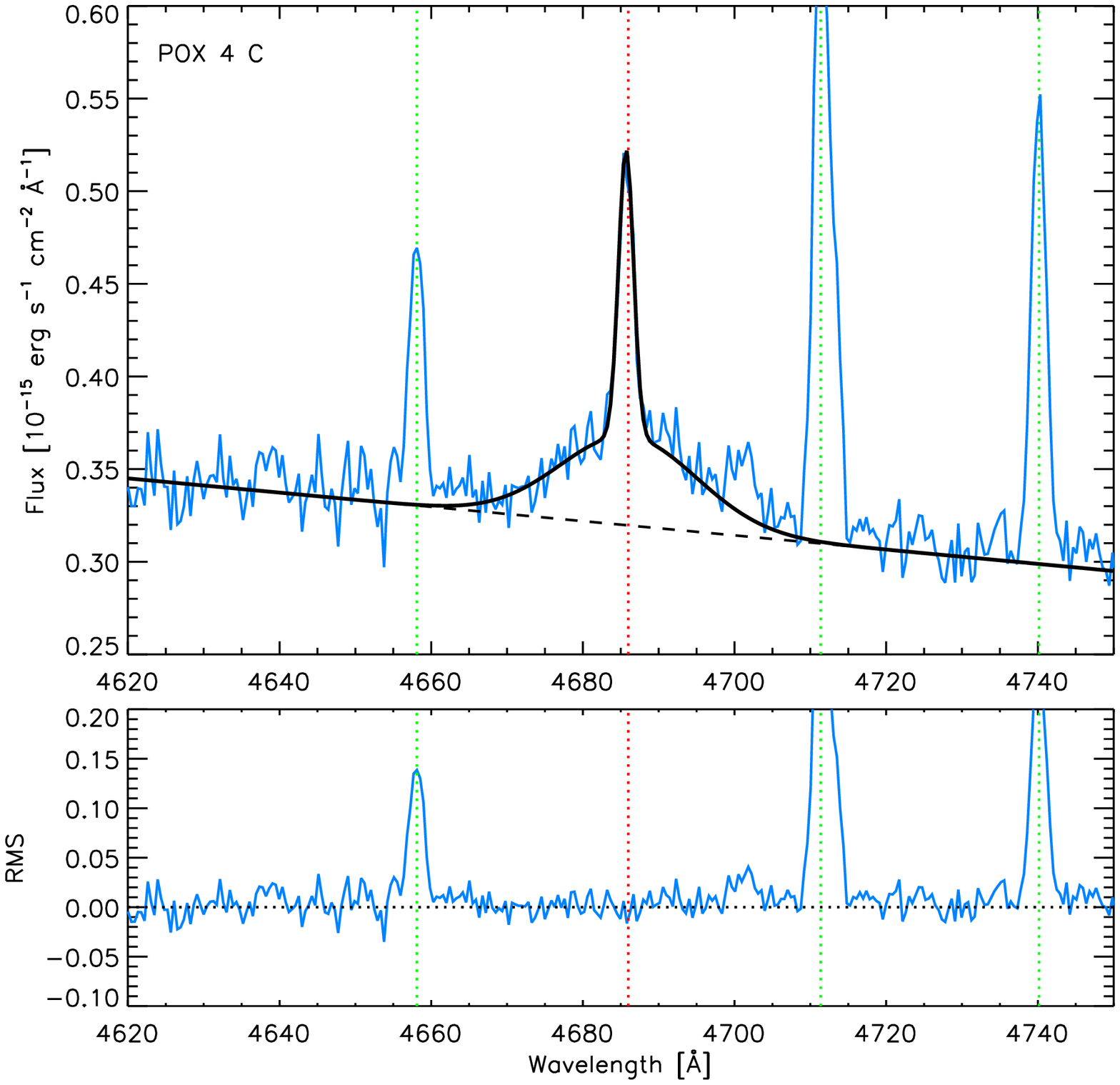} &
\includegraphics[angle=90,width=0.33\linewidth]{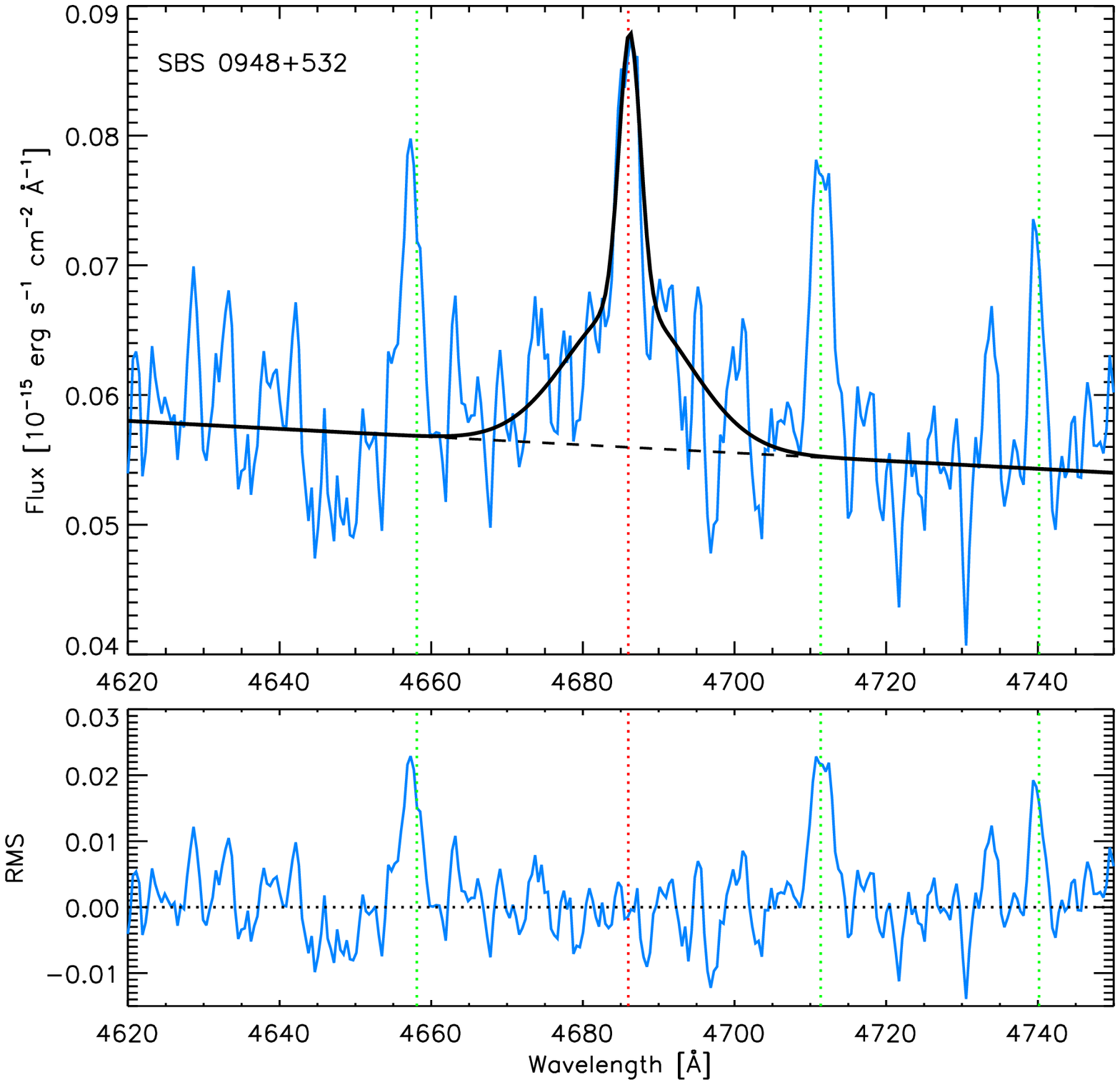} \\
\includegraphics[angle=90,width=0.33\linewidth]{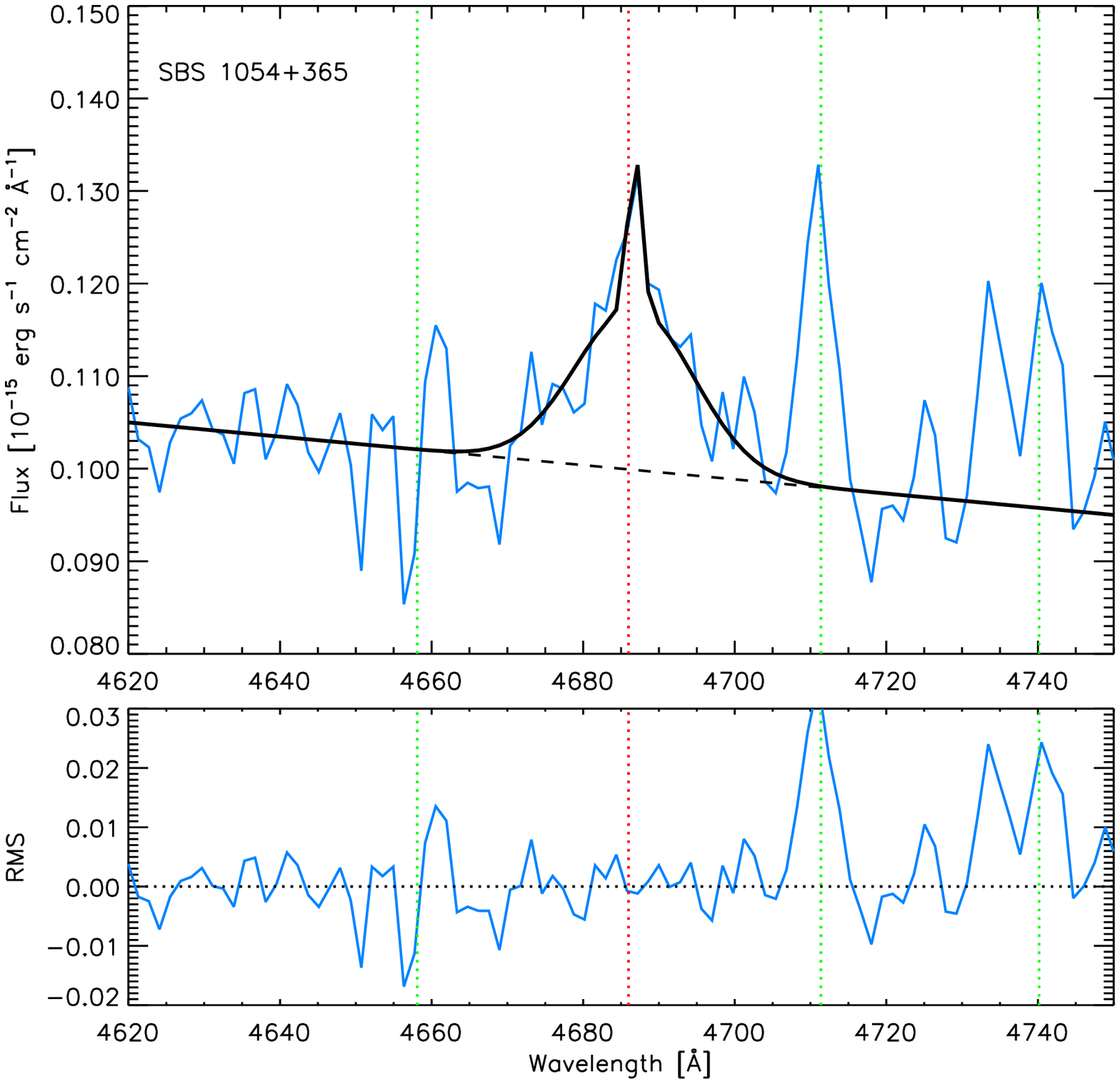} &
\includegraphics[angle=90,width=0.33\linewidth]{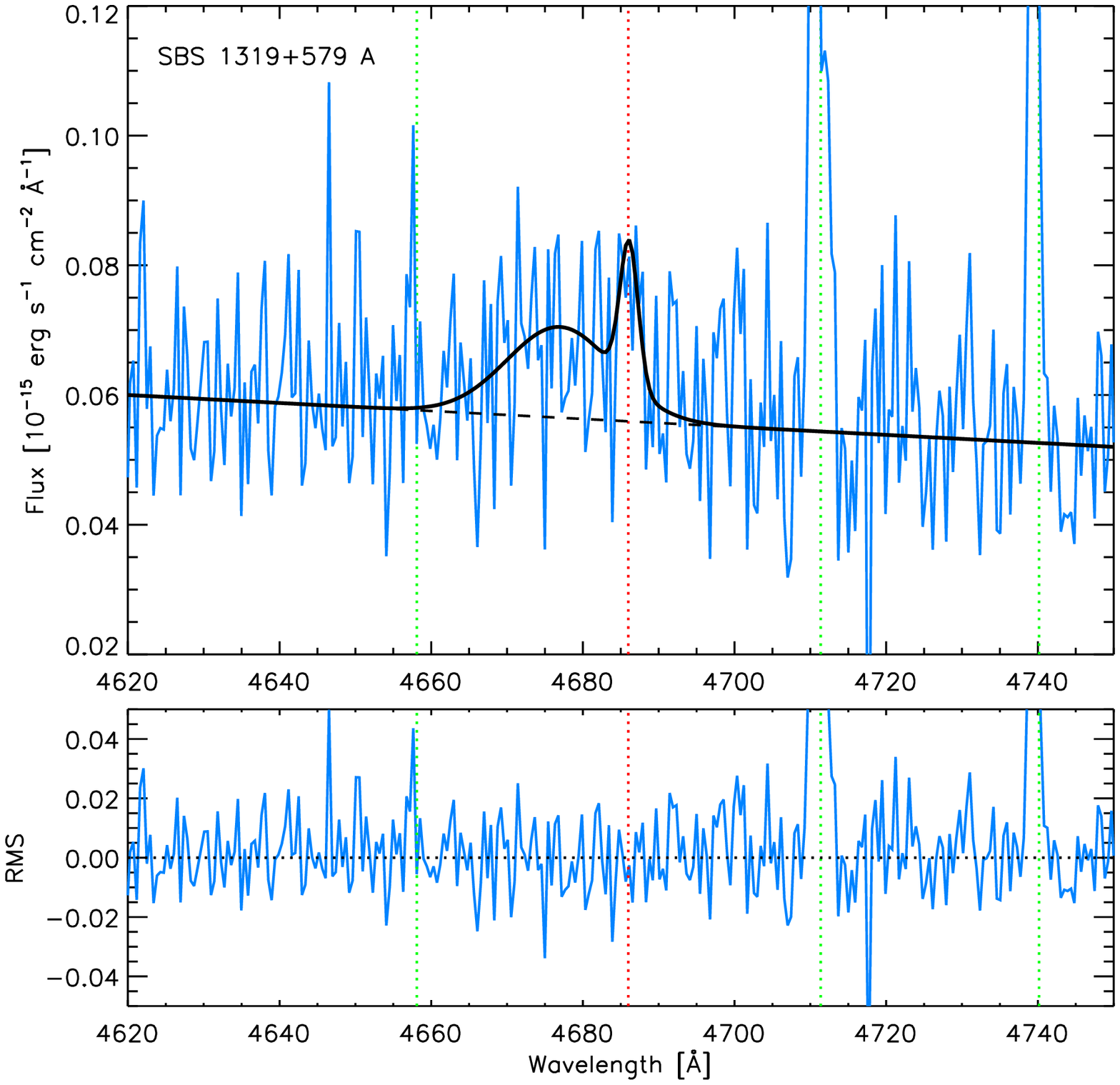} &
\includegraphics[angle=90,width=0.33\linewidth]{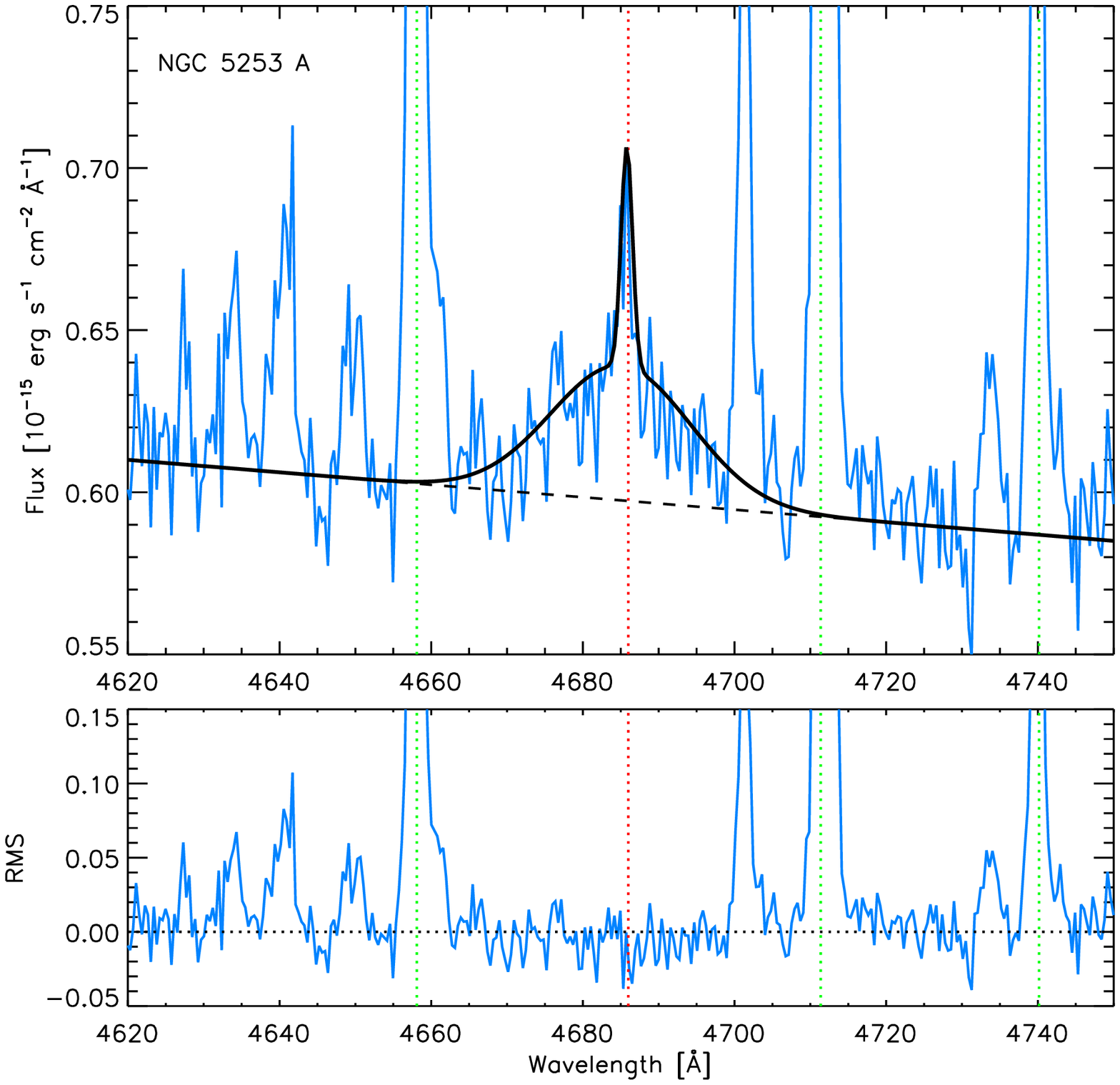} \\
\includegraphics[angle=90,width=0.33\linewidth]{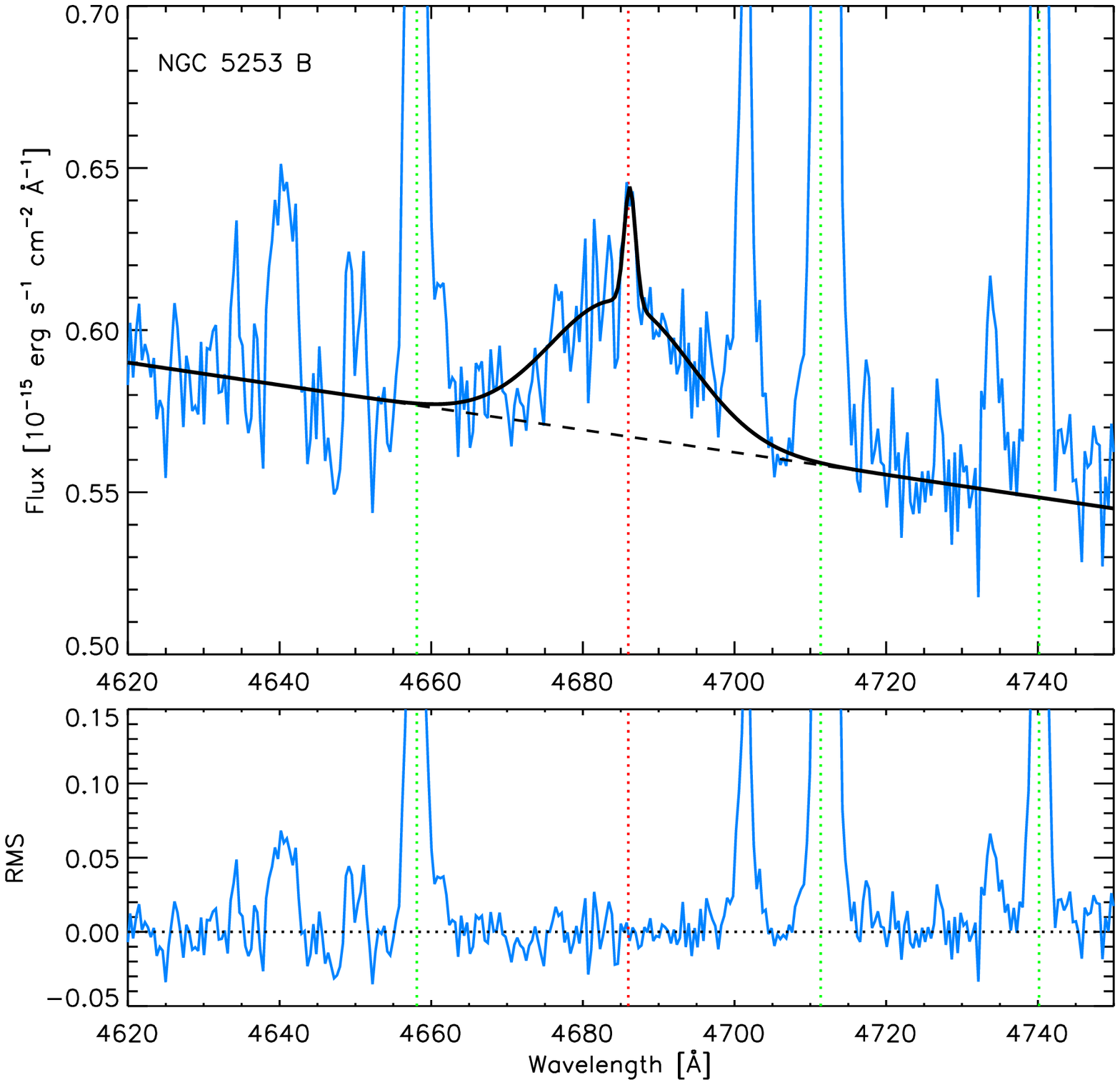} &
\includegraphics[angle=90,width=0.33\linewidth]{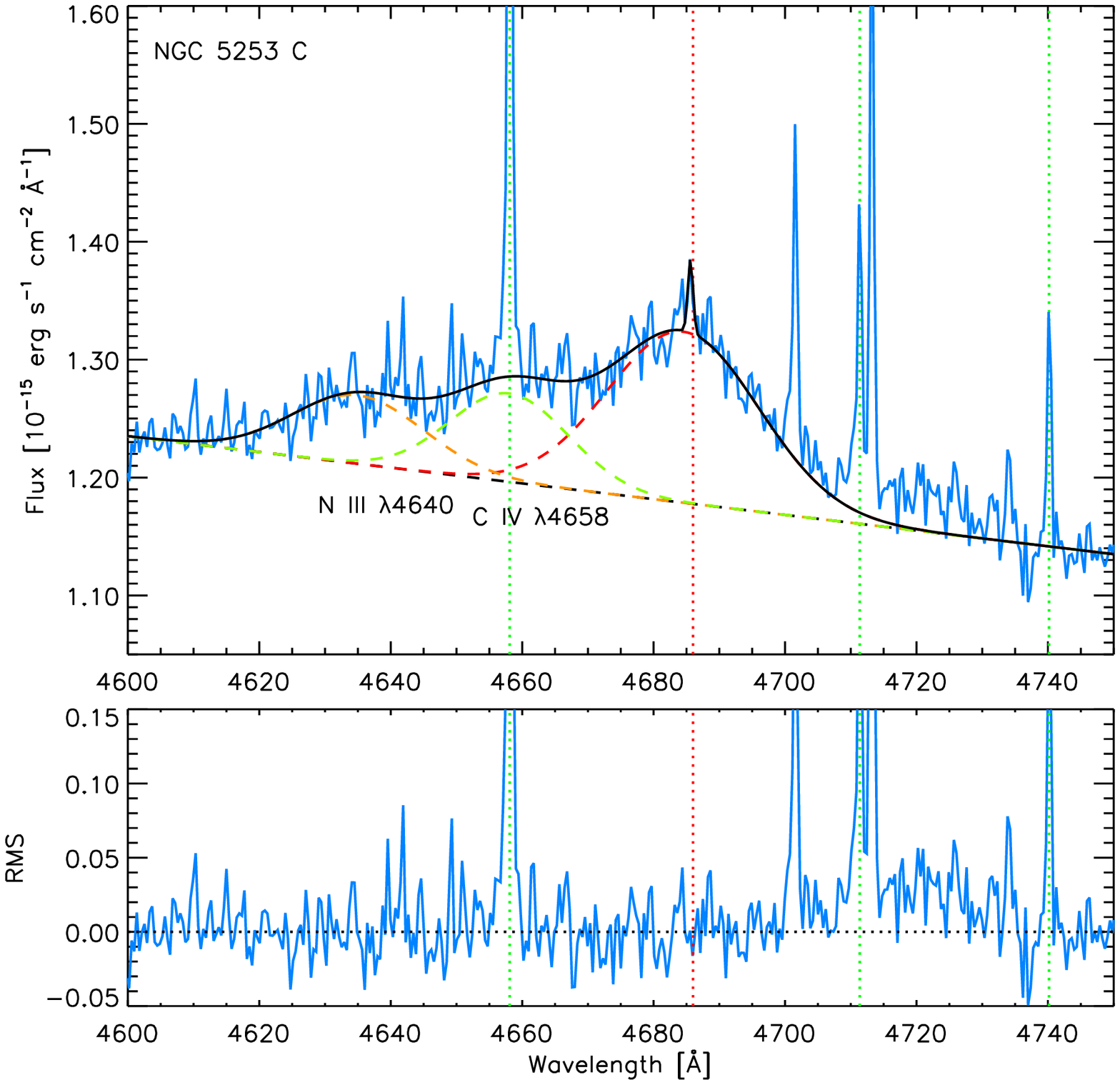} &
\includegraphics[angle=90,width=0.33\linewidth]{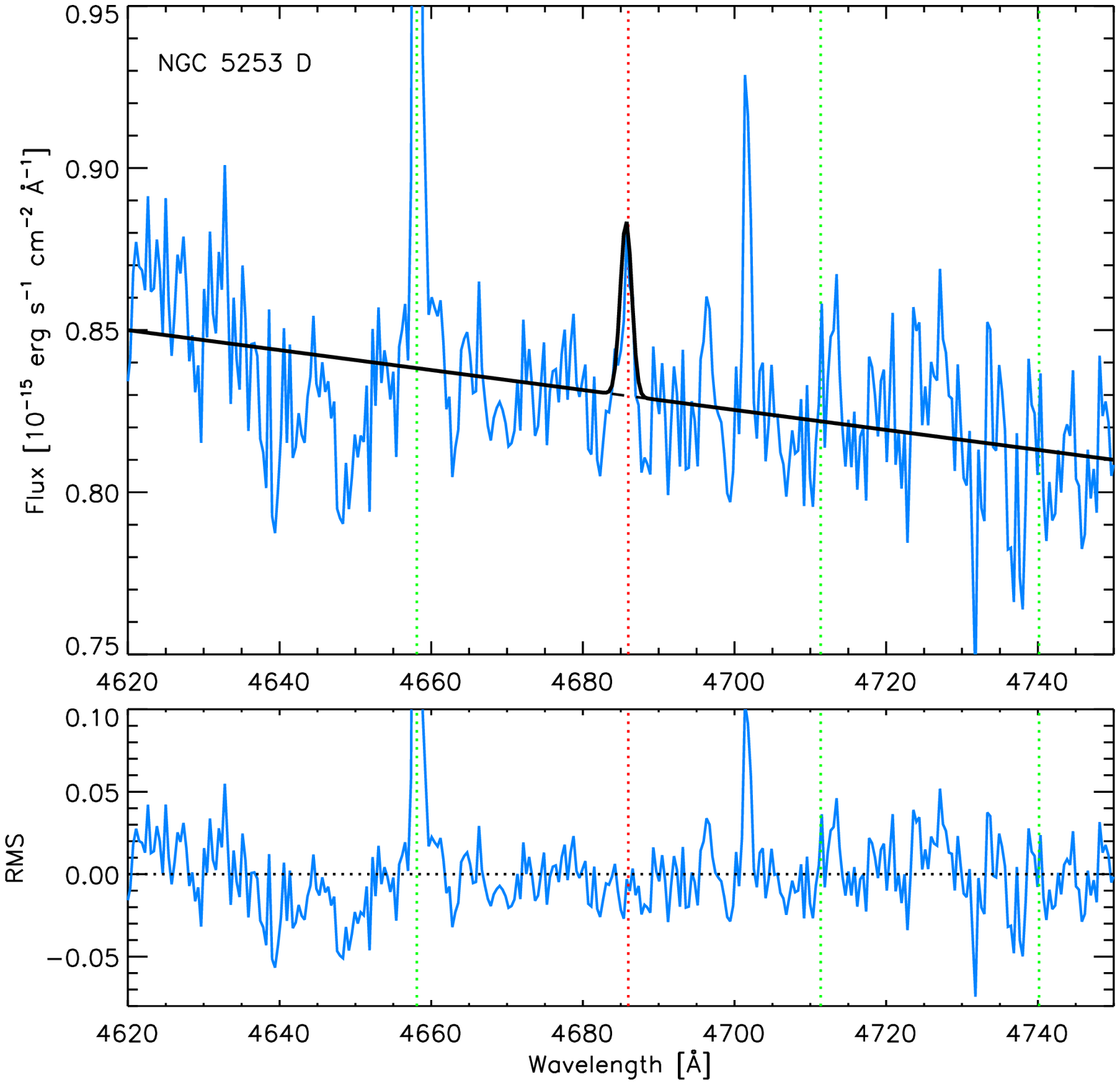} \\
\end{tabular} 
\protect\caption[ ]{\footnotesize{
Examples of fits to the blue \WRBUMP\ for some of the analyzed galaxies in the 4620 -- 4750 \AA\ range.  The red dotted line represents the position  
of the \ion{He}{ii} $\lambda$4686 emission line, the blue dotted lines indicate the position of [\ion{Fe}{iii}] $\lambda$4658 and [\ion{Ar}{iv}]  
$\lambda\lambda$4711,4740 emission lines. The black dashed line represents the continuum fit. The best fit to the observed spectrum (blue continuous  
line) is shown with a black continuous line, and it is usually composed by the superposition of a broad and a narrow \ion{He}{ii} $\lambda$4686  
lines. Region D in NGC~5253 only shows a narrow nebular \ion{He}{ii} component. The broad \ion{N}{iii} $\lambda$4640 emission feature is detected and  
fitted in HCG~31~AC and NGC~5253~C. This last region also shows a broad \ion{C}{iv} $\lambda$4658 emission line, which was included in the fit.  
The NGC~5253~C also shows the three dashed broad Gaussians that composed the blue \WRBUMP. All diagrams include the residual spectrum after  
subtracting the best-fit model to the observed spectrum.}}
\label{wrfig1}
\end{figure*}

As our galaxy sample was extracted from the catalog of Wolf-Rayet galaxies compiled by \citet{SCP99}, we first analyze the presence or absence of  
the WR features in the star-forming bursts within our starburst galaxies and the characteristics of their WR populations. Table~\ref{wrdata} compiles  
the objects or regions where any \WRBUMP\ and/or the nebular \ion{He}{ii} $\lambda$4686 emission line have been detected. 
Figures~36 and 37 in Paper~II show the details of the optical spectrum in the 4600--4750 \AA\ and the 5550--6000 \AA\ ranges, respectively, of all  
important objects (faint regions with very low S/N have been excluded). 

The nebular \ion{He}{ii} $\lambda$4686 emission line is detected in 21 regions within 16 systems (Fig.~36 in Paper~II). It is unambiguously clear  
in 14 objects (HCG~31~AC and F1, Haro~15~A, Mkn~5, POX~4, UM~420, SBS~0926+606~A, SBS~0948+532, SBS~1415+437~C, Tol~1457-262~A and B, ESO~566-8 and  
NGC~5253~A and D), and particularly strong in POX~4. 

The blue \WRBUMP\ is clearly detected in many objects (Fig.~36 in Paper~II), and unambiguously observed in 12 regions (HCG~31 AC, Mkn~1199, Mkn~5,  
IRAS~08208+2816, POX~4, SBS~0948+532, SBS~1054+365, III~Zw~107, Tol~9 and NGC~5253 A,B and C). We indicate the spatial localization of the WR stars  
in each galaxy in Col.~2 of Table~\ref{wrdata}, sometimes they are detected in different regions within the same system. Our observations of  
the galaxy IRAS~08339+6517, no cataloged so far as WR, suggest the detection of this kind of massive stars in its central burst \citep*{LSEGR06}.  
Only in three objects previously listed as WR galaxies (Mkn~1087, SBS~1211+540 and ESO~566-7) do we not detect any feature that can be attributed to this kind of massive stars. We consider that aperture effects and the exact positioning of the slit onto the WR-rich bursts play a  
fundamental role in the detection of the WR features. Furthermore, a good S/N ratio is needed to observe this feature in faint regions. A good  
spectral resolution is also needed to get a proper deblending of the nebular and broad \ion{He}{ii} $\lambda$4686 lines. 

The dominant contribution to the blue \WRBUMP\ is the broad \ion{He}{ii} $\lambda$4686 line, arising in late type WN stars (WNL). Three regions  
(HCG~31 AC, Mkn~1199~C and knot C in NGC~5253) show a clear detection of the broad \ion{N}{iii} $\lambda$4640 line, also mainly originated in WNL  
stars. The absence of the faint \ion{N}{v} $\lambda$4604 indicates the predominance of WNL stars, although some contribution of early-type WN stars  
(WNE) might be still present in the blue \WRBUMP\ (SV98).

\begin{figure*}[t!]
\centering
\begin{tabular}{c@{\hspace{4pt}}c@{\hspace{4pt}}c}
\includegraphics[angle=90,width=0.33\linewidth]{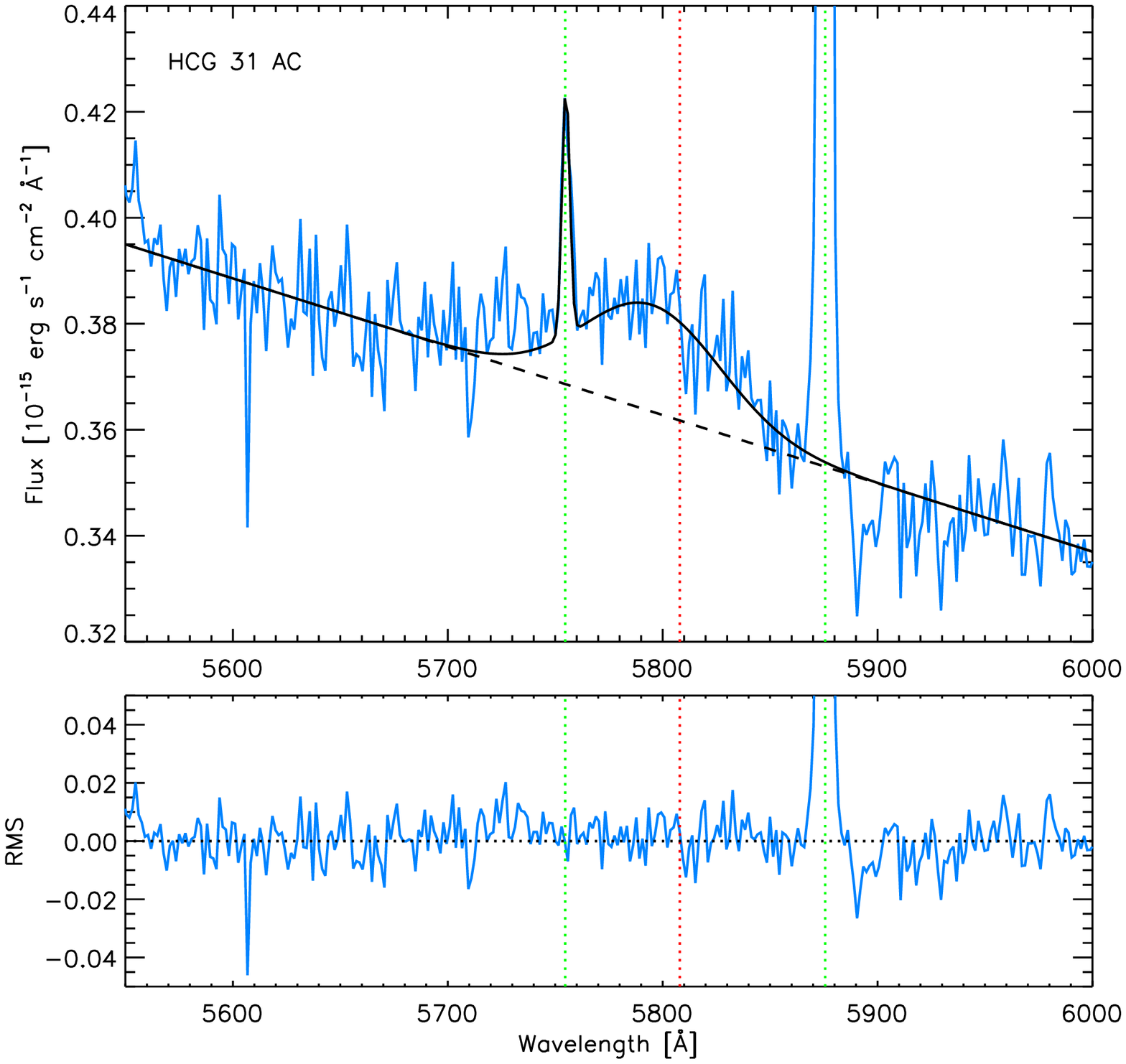} &
\includegraphics[angle=90,width=0.33\linewidth]{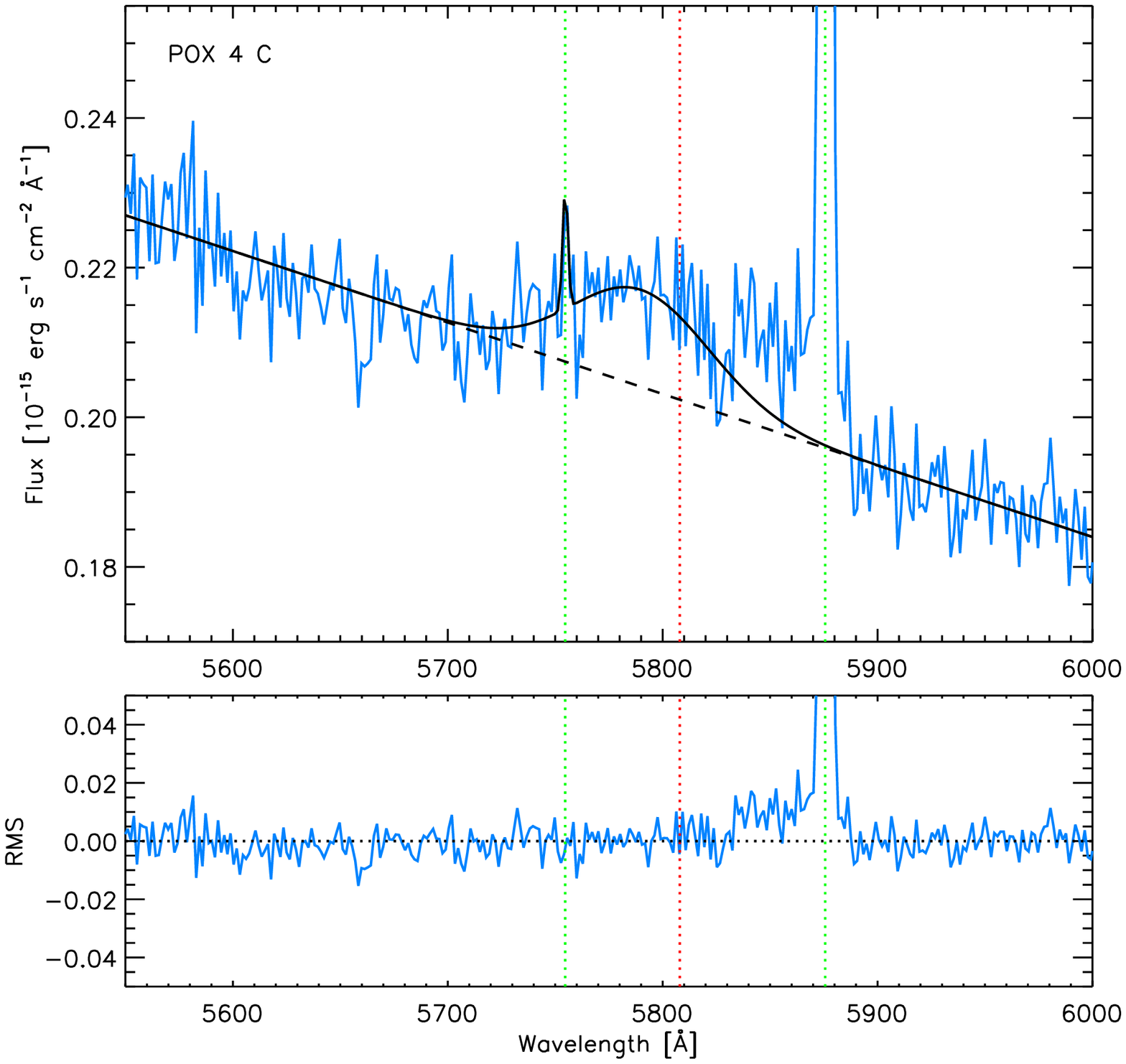} &
\includegraphics[angle=90,width=0.33\linewidth]{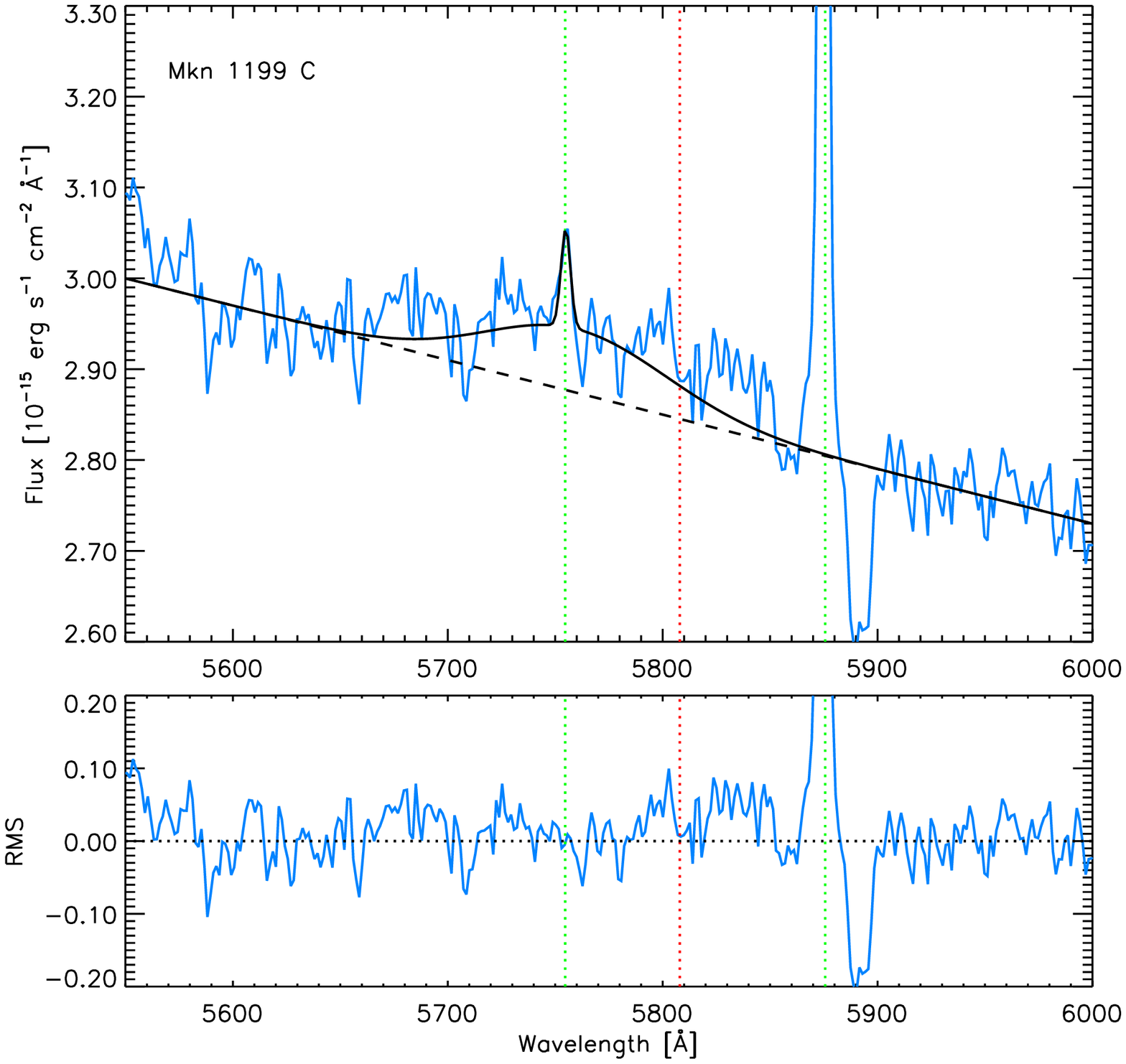} \\
\includegraphics[angle=90,width=0.33\linewidth]{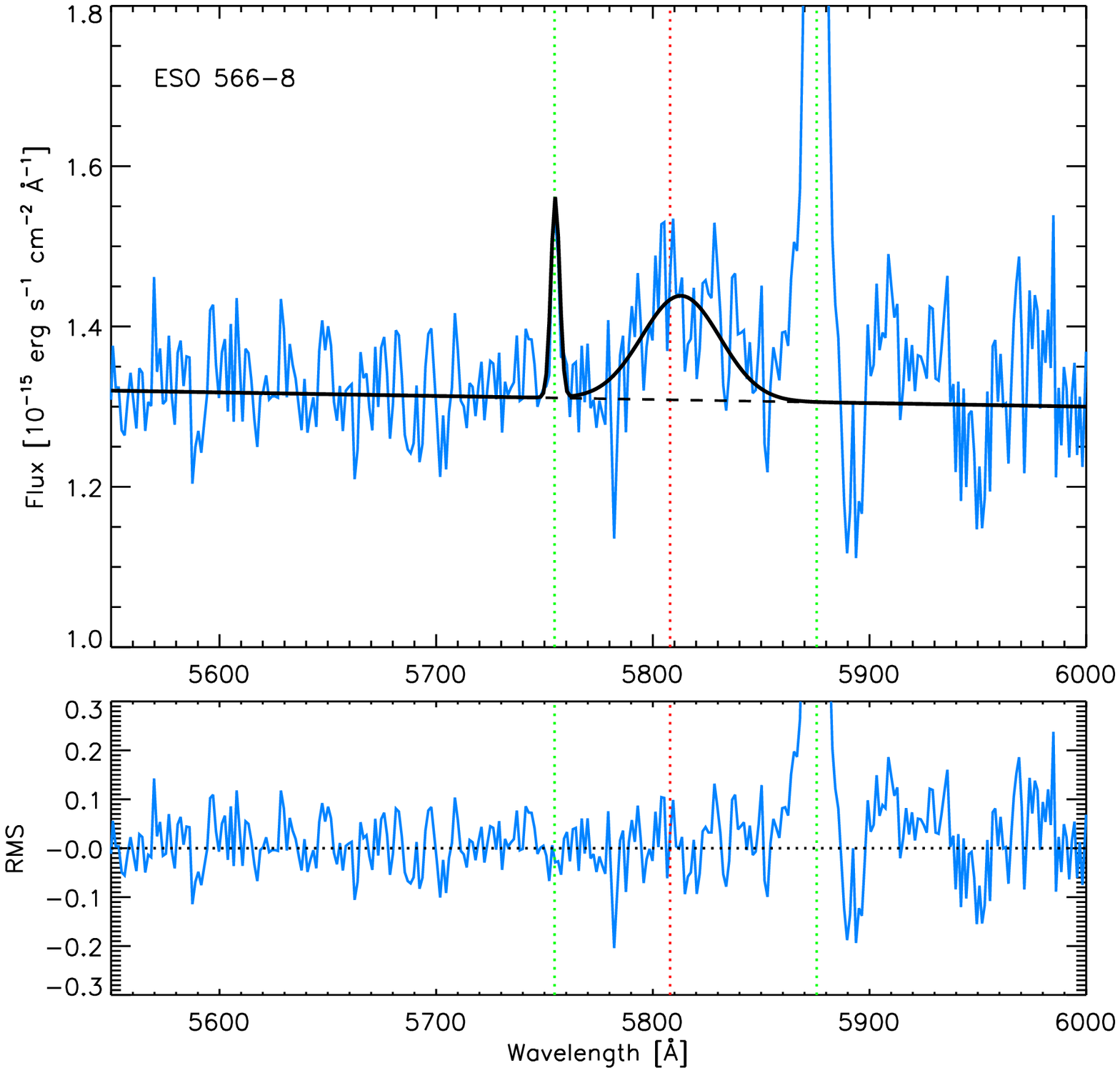} &
\includegraphics[angle=90,width=0.33\linewidth]{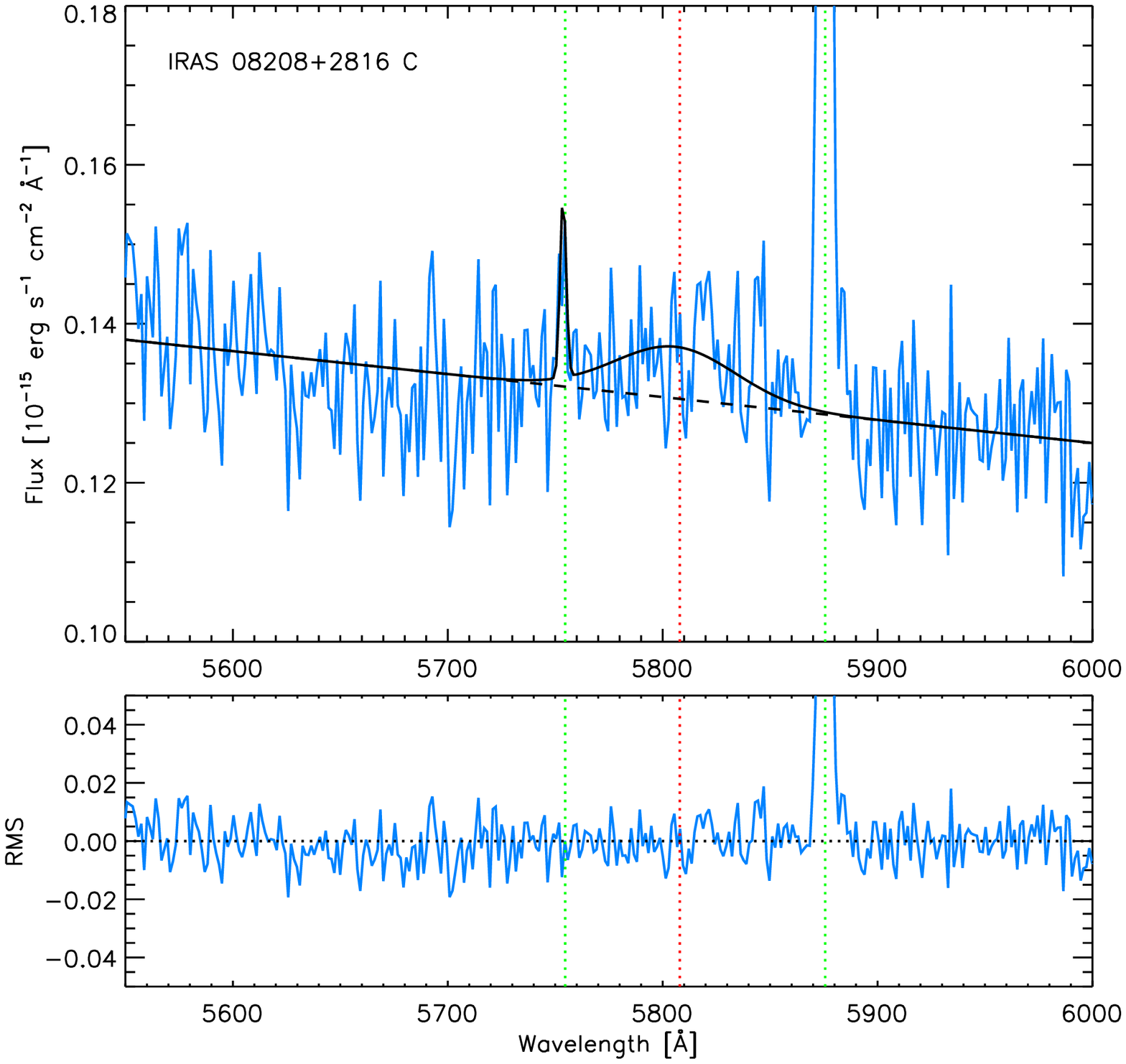} &
\includegraphics[angle=90,width=0.33\linewidth]{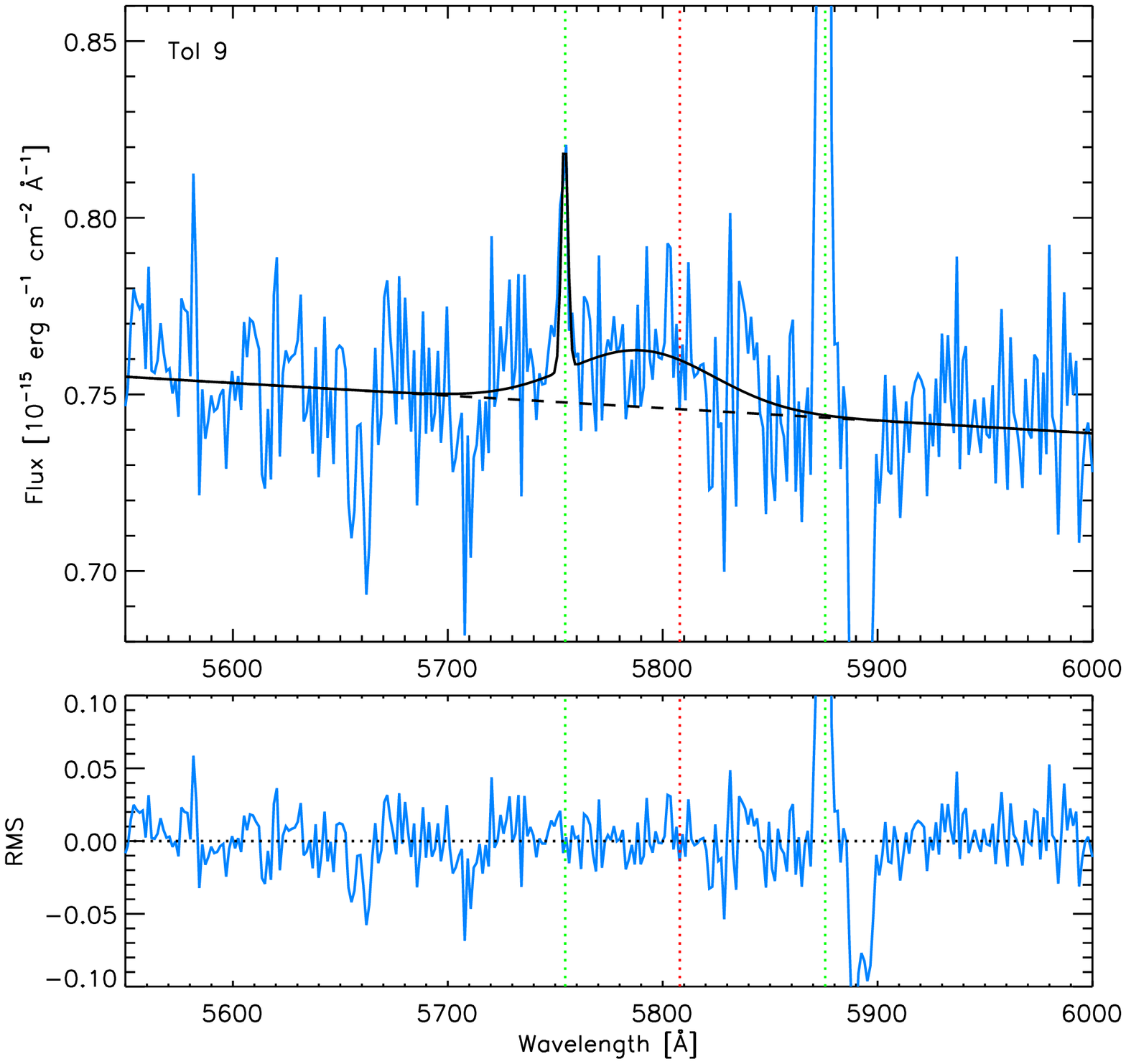} \\
\end{tabular} 
\protect\caption[ ]{\footnotesize{
Examples of fits to the red \WRBUMP\ for all regions for which we have a detection of the broad \ion{C}{iv} $\lambda$5808 line in the 5550 -- 6000  
\AA\ range. The red dotted line represents the position of the \ion{C}{iv} $\lambda$5808 emission line, the blue dotted lines indicate the position  
of [\ion{N}{ii}] $\lambda$5755 and \ion{He}{i} $\lambda$5875 emission lines. The black dashed line represents the continuum fit. The best fit to the  
observed spectrum (blue continuous line) is shown with a black continuous line, and it is composed by a broad \ion{C}{iv} $\lambda$5808 and a nebular  
[\ion{N}{ii}] $\lambda$5755 line. The broad feature is very evident in HCG~31~AC and in POX~4, detected in Mkn~1199 and ESO~566-8, and probably  
observed (but with a very low S/N ratio) in IRAS~08208+2816 and Tol~9. Except in POX~4 and IRAS~08208+2816, the spectra also show a clear \ion{He}{i}  
$\lambda$5875 absorption feature. All diagrams include the residual spectrum after subtracting the best-fit model to the observed spectrum.}}
\label{wrfig2}
\end{figure*}

As was explained in the introduction, the \ion{He}{ii} $\lambda$4686 emission line and the blue \WRBUMP\ are mainly linked to WN  
stars, while the \ion{C}{iv} $\lambda$5808 emission line (the red \WRBUMP) essentially originates in WC stars. As previous studies indicate, the  
red \WRBUMP\ is much more difficult to observe than the blue \WRBUMP. Figure~37 in Paper~II shows the spectral range where the red \WRBUMP\ is  
located for all important objects; only two galaxies (SBS~1211+540 and NGC 5253) are lacking data in this spectral range. There is always a clear  
detection of the \ion{He}{i} $\lambda$5876 line in all spectra, and the faint auroral [\ion{N}{ii}] $\lambda$5755 emission line in also seen in many  
cases. We clearly identify the broad \ion{C}{iv} $\lambda$5808 line in 2 galaxies  (HCG~31~AC and POX~4), detect it in two galaxies (Mkn~1199 and  
ESO~566-8) and it also seems to be observed in other two galaxies (IRAS~08208+2816 and Tol~9). However, we do not see this feature in galaxies for  
which \citet{GIT00} reported its detection (HGC~31~G, Mkn~5, UM~420, SBS~0926+606, SBS~0948+532, SBS~1054+365 and SBS~1319+579A). The 
non-detection of the red \WRBUMP\ in these objects may be a consequence of aperture effects or the position of the slit, i.e., \citet{GIT00} observed  
slightly different regions than the areas we analyze here. However, our spectra usually have higher spatial and spectral resolution (see Paper~II),  
and even higher signal-to-noise ratio (the average SNR value achieved in this part of the spectra for our galaxy sample is 10 -- 12), particularly those   
obtained using the ISIS spectrograph at the 4.2m WHT, than the spectra compiled by \citet{GIT00}, which were obtained using 2m and 4m class  
telescopes (see their Fig.~2). In the past, the non-detection of the red \WRBUMP\ may have been a detection-threshold effect (the slope in  
starbursts rises towards the blue and hence on average there are half as many counts in the red \WRBUMP\ than in the blue \WRBUMP), but we consider  
this not to be the situation here because of both the quality of our spectra and their relatively good SNR.


The non-detection of the red \WRBUMP\ could be explained because WC stars are very rarely formed in low-metallicity environments, as evolutionary  
synthesis models predict and observations suggest \citep{Crowther07}, indeed our galaxy sample is mainly composed by low-metallicity objects.  
However, in some objects the non-detection of the red \WRBUMP\ may be a consequence of a too low S/N ratio. In particular, they should
definitively be observed in NGC~5253 given the quality and depth of our spectra. Indeed, the very broad blue \WRBUMP\ in region C of NGC~5253 strongly  
suggests the broad \ion{C}{iv} $\lambda$4658 emission line, with the nebular [\ion{Fe}{iii}] $\lambda$4658 line on top of it (see  
Fig.~\ref{wrfig1}). We will use this line to estimate the number of WCE stars in this particular zone of NGC~5253. 

The line \ion{C}{iii} $\lambda$5696, much fainter than the \ion{C}{iv} $\lambda$5808 line, is not detected in any case. 
Hence the emission is likely due  to early-type WC stars (WCE).

\section{Fitting of the WR features}

\begin{figure*}[t!]
\includegraphics[angle=270,width=0.5\linewidth]{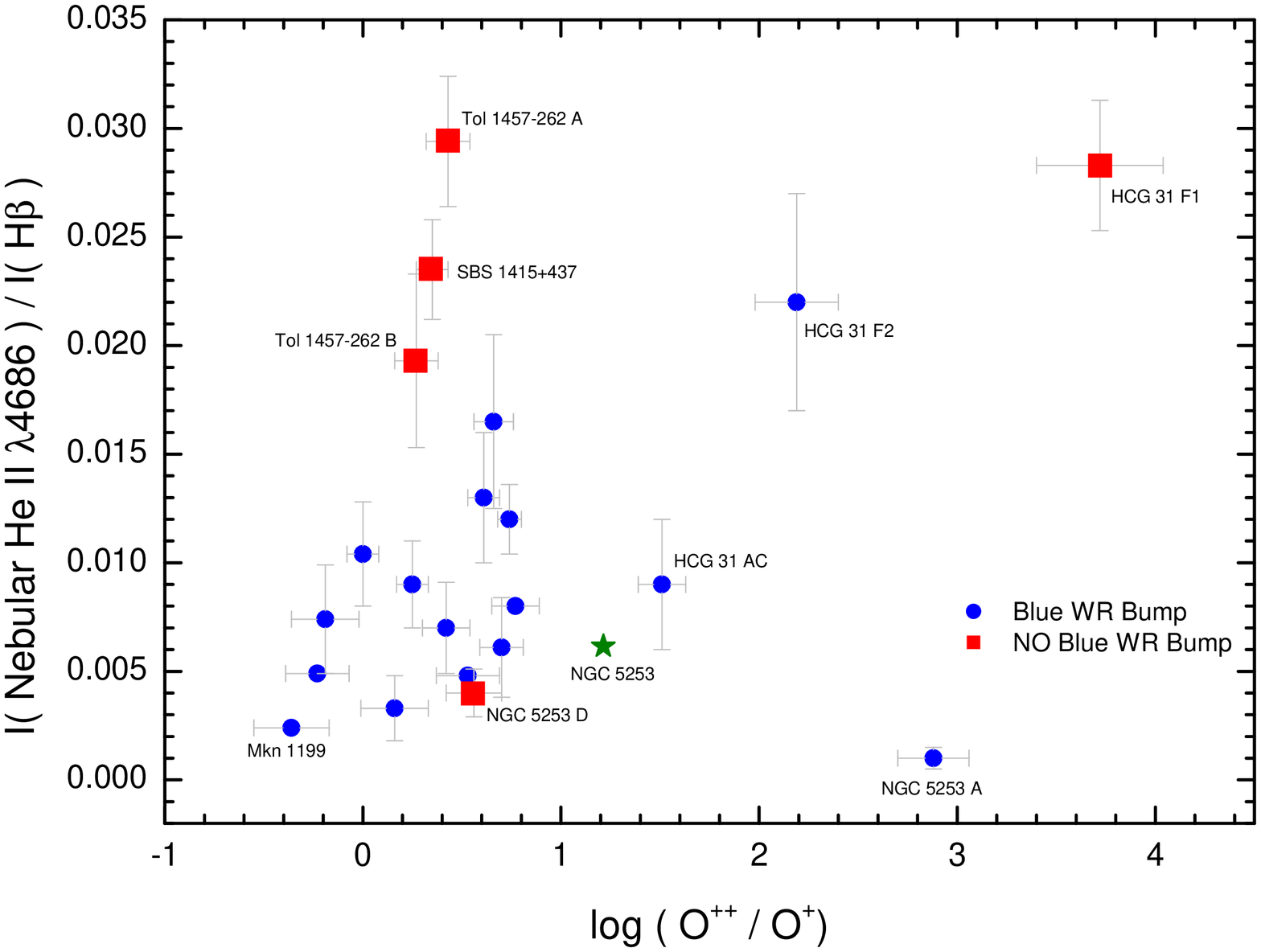}
\includegraphics[angle=270,width=0.5\linewidth]{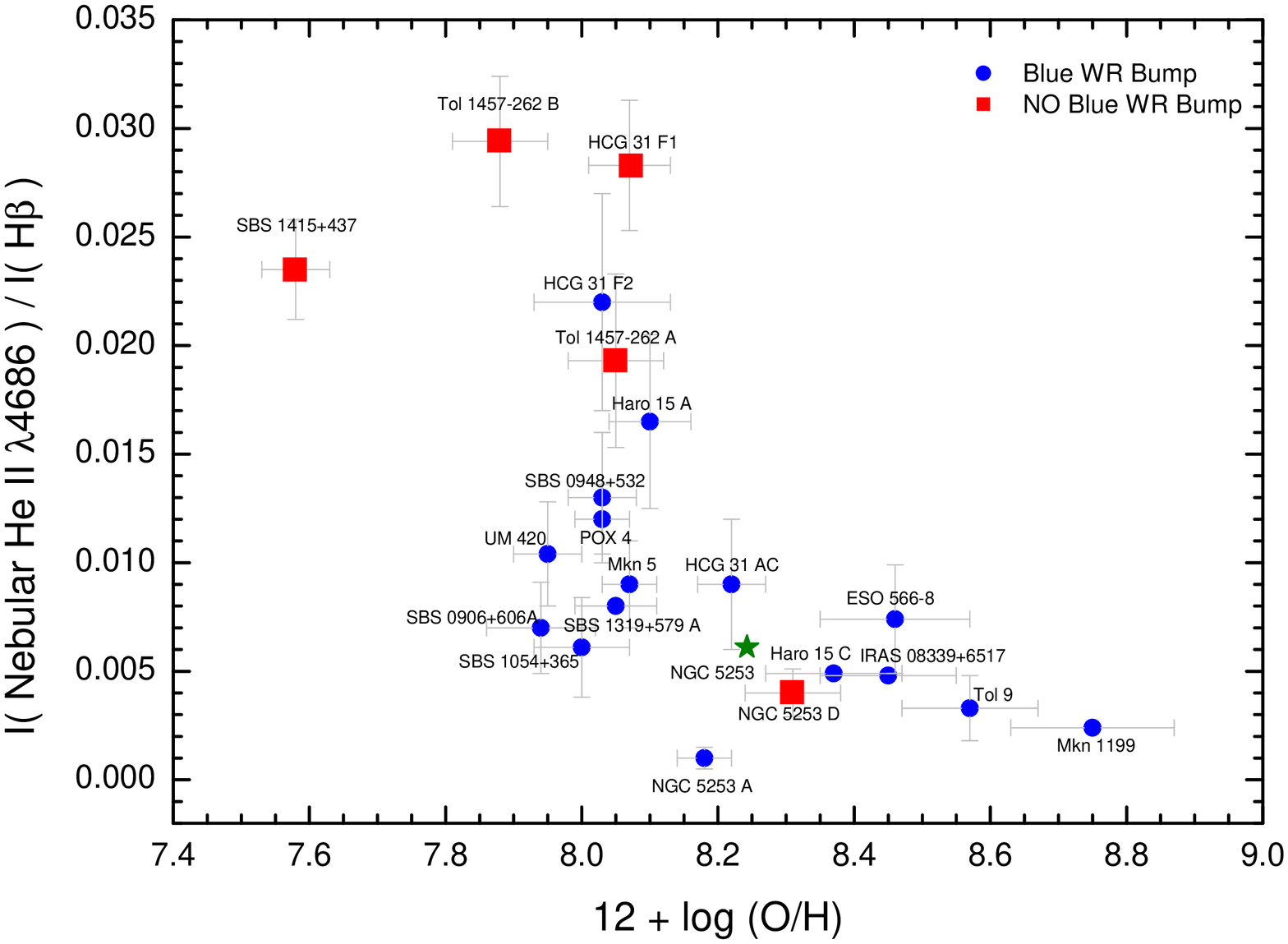}
\protect\caption[ ]{\footnotesize{
Intensity of the nebular \ion{He}{ii} $\lambda$4686 line vs. the O$^{++}$/O$^+$ ratio (\emph{left}) and the oxygen abundance (\emph{right}). In both  
figures, red squares indicate objects in which the nebular \ion{He}{ii} line but not blue \WRBUMP\ are detected. Some objects have been labeled. The  
green star indicates the position of NGC~5253, taking an average value of all analyzed regions.}}
\label{nheii}
\end{figure*}

As BKD08 pointed out, the fitting of the WR features, specially in the blue bump, should be done carefully and not only considering one single Gaussian  
fit. Furthermore, the nebular emission lines within the bump should be properly removed and not included in the flux of the broad stellar lines.  
Following the procedure described in BKD08, we performed a detailed analysis of each spectrum, checking even those spectra without a clear WR  
detection. We fitted a broad and a narrow Gaussian for the stellar and nebular \ion{He}{ii} $\lambda$4686 lines in the blue bump and a narrow  
Gaussian for the nebular [\ion{N}{ii}] $\lambda$5755 emission line ontop a broad Gaussian for the stellar \ion{C}{iv} $\lambda$5808 line in the red  
bump. We considered the typical FWHM of the nebular lines to constrain the width of the nebular \ion{He}{ii} $\lambda$4686 line and a maximum of  
FWHM$\sim$47 \AA\ ($\sigma$=20 \AA) in the blue and FWHM$\sim$59 \AA\ ($\sigma$=25 \AA) in the red for the width of the broad stellar lines. This  
corresponds to FWHM$\sim$3000 \kms\ ($\sigma\sim$1280 \kms), which is an adequate upper limit to the width of individual WR features (BKD08). For the  
broad \ion{He}{ii} $\lambda$4686 line, all the fits yield $\sigma\leq$10 \AA\ except Tol~9 ($\sigma$=13 \AA). The fits to the broad \ion{C}{iv}  
$\lambda$5808 line were set to the maximum value for Mkn~1199 and Tol~9. 
 
Some examples of this fitting are shown in Figs.~\ref{wrfig1} (blue \WRBUMP) and \ref{wrfig2} (red \WRBUMP). The blue line traces our spectra, the  
continuum level is indicated by a dashed black line, and the best fit (usually composed by a broad and a nebular \ion{He}{ii} $\lambda$4686 line in  
the blue bump) is shown with a black continuous line. These figures also show the residual spectrum after subtracting the best-fit model to our  
observed spectrum. As can be seen, the residual spectra are consistent with Gaussian noise plus the nebular emission lines. We did not consider any  
fits to the nebular [\ion{Fe}{iii}] $\lambda$4658 and [\ion{Ar}{iv}] $\lambda\lambda$4711,4740 lines in the blue range and the nebular \ion{He}{i}  
$\lambda$5875 line in the red range because they are clearly seen in the residual spectrum and, considering our spectral resolution, their  
contribution (if any) to the flux of the \WRBUMP s should be negligible. We want to remark that this careful fitting of the broad and narrow lines  
was only performed before by BKD08, who used SDSS data; the majority of our spectra have both a higher spectral resolution and S/N ratio than  
those presented by these authors (compare their Fig.~2 with our Fig.~\ref{wrfig1}). Although BKD08 have a more extensive galaxy sample, the  
additional strength of our study is the detailed analysis we performed for each region and galaxy (see Papers I and II), which yields a better  
determination of their properties (chemical abundances, stellar population, star formation rate, colors, masses, ...). In particular, we used the  
bright doublet [\ion{O}{ii}] $\lambda\lambda$3726,3729 to estimate the total oxygen abundance, and not the much fainter [\ion{O}{ii}]  
$\lambda\lambda$7319,7330 lines, which usually are the only [\ion{O}{ii}] lines available in SDSS data. Furthermore, as we will explain in Sect.~7,  
our galaxy sample is more homogeneous than that used by BKD08, allowing us to perform additional assumptions about their star-formation history.

As seen in Figs.~\ref{wrfig1} and \ref{wrfig2}, some spectra reveal very faint WR features.
This is the case for region A in SBS~1319+579, for which a very faint blue \WRBUMP\ and a nebular \ion{He}{ii} emission line were apparently detected,  
besides their very low S/N ratio. For the central regions of galaxies IRAS~08208+2818 and Tol~9, the fit of a faint broad \ion{C}{iv} $\lambda$5808  
line also suggests  WCE stars. We will consider the values derived from these tentative fits as upper limits to the actual ones.

The broad \ion{N}{iii} $\lambda$4640 line is observed in three objects (HCG~31~AC, Mkn~1199 and NGC~5253~C). We also included this line in the  
fitting of the data, as Fig.~\ref{wrfig1} shows. This broad feature is clearly evident in region C of NGC~5253. Incidentally, this region also shows a  
detection of the broad \ion{C}{iv} $\lambda$4656 emission line, which was also included in the fit of the blue \WRBUMP. 

Table~\ref{wrdata} compiles the results of our best fits for the nebular and broad \ion{He}{ii} $\lambda$4686 and the broad \ion{C}{iv} $\lambda$5808  
lines in those regions where these features have been detected. We indicate both the equivalent width and the reddening-corrected $I$/$I$(\Hb)  
ratios of these lines, including an error estimation for all values. We remind that these features are usually weak, which makes their associated  
uncertainties considerably high in some cases. 

Although we did not observe any WR feature in the two galaxies with the lowest metallicity of our sample (SBS~1211+540 and SBS~1415+437), we  
estimated an upper limit to the broad \ion{He}{ii} $\lambda$4686 flux. These results are also shown in Table~\ref{wrdata}.

\begin{table*}[t!]
\centering
  \caption{\footnotesize{O and WR stellar populations for the galaxies analyzed here in which WR features are detected.}}
  \label{wrpop}
  \tiny
  \begin{tabular}{l@{\hspace{3pt}}c  c@{\hspace{3pt}}ccc   c@{\hspace{3pt}}cc    cccccc}
    \noalign{\smallskip}
    \tableline
	\noalign{\smallskip}

 & & 

\multicolumn{3}{c@{\hspace{4pt}}}{SV98$^a$} & &
\multicolumn{8}{c@{\hspace{4pt}}}{CH06$^b$} \\  
 
\cline{3-5}
\cline{7-14}
\noalign{\smallskip}

Galaxy & Zone &  
WNL  &  O$^c$ &  ${\rm \frac{WNL}{WNL+O}}$ & &
WNL  &  O$^d$ &  WCE$^e$ & O$^f$ & ${\rm \frac{WNL}{WNL+O}}$  
 & ${\rm \frac{WCE}{WCE+O}}$ &     ${\rm \frac{WCE}{WNL}}$  &   ${\rm \frac{WR}{WR+O}}$     &${\rm \frac{WR}{WR+O}}$  \\
\noalign{\smallskip} 
 
& &  
[$\star$] & [$\star$] & (g) & &
[$\star$] & [$\star$] & 
[$\star$] & [$\star$] & (g,h) & (e)  &  (e) &  (e)    & (i)\\

	\noalign{\smallskip}    
  \tableline
\noalign{\smallskip}
 
HCG 31          & AC    & 312 & 10798 & 0.028 & &  427 &  9646 &   203 &  7075 & 0.042 & 0.028 & 0.476 & 0.082 &  0.051$^j$ \\ 
                & F2    & 4.6 &   200 & 0.022 & &  7.1 &   192 &\nodata&   192 & 0.036 &\nodata&\nodata&\nodata&  0.015    \\
Mkn 1087        & C     &7181 & 24680 & 0.225 & & 7994 & 16509 &\nodata& 16509 & 0.326 &\nodata&\nodata&\nodata&  0.115    \\
    Haro 15     &  C    & 347 &  8806 & 0.038 & &  432 &  8271 &\nodata&  8271 & 0.050 &\nodata&\nodata&\nodata&  0.037    \\
                &  A    & 173 &  7963 & 0.021 & &  257 &  7542 &\nodata&  7542 & 0.033 &\nodata&\nodata&\nodata&  0.021    \\
    Mkn 1199    &  C    & 960 &  6063 & 0.137 & &  975 &  5985 &   932 &    94 & 0.140 & 0.909 & 0.956 & 0.953 &  0.127$^j$ \\ 
    Mkn 5       &  A    & 2.7 &   143 & 0.018 & & 4.1  &   134 &\nodata&   134 & 0.029 &\nodata&\nodata&\nodata&  0.022    \\
IRAS 08208+2816 &  C    &1344 & 33338 & 0.039 & & 1715 & 29611 &  932: & 17828:& 0.055 & 0.050:& 0.543:& 0.129:&  0.051    \\
IRAS 08339+6517 & \#1   & 658 &  8444 & 0.072 & &  782 &  7819 &\nodata&  7819 & 0.091 &\nodata&\nodata&\nodata&  0.050    \\
POX 4           &  C    & 122 &  4346 & 0.027 & &  191 &  4058 &   113 &  3464 & 0.045 & 0.032 & 0.590 & 0.081 &  0.022    \\
         UM 420 &\nodata& 398 & 17453 & 0.022 & &  660 & 16134 &\nodata& 16134 & 0.039 &\nodata&\nodata&\nodata&  0.021    \\
SBS 0926+606    & A     &  32 &  1987 & 0.016 & &   54 &  1897 &\nodata&  1897 & 0.028 &\nodata&\nodata&\nodata&  0.014    \\
SBS 0948+532    &\nodata& 230 & 11460 & 0.020 & &  358 & 10921 &\nodata& 10921 & 0.032 &\nodata&\nodata&\nodata&  0.017    \\
SBS 1054+365    & C     &  1.1&    52 & 0.020 & &  1.7 &  48.1 &\nodata&   1.7 & 0.034 &\nodata&\nodata&\nodata&  0.023    \\
SBS 1211+540    &\nodata&$<$0.27&$>$14.5&$<$0.018&&$<$0.58&$>$13&\nodata&$>$13 &$<$0.043&\nodata&\nodata&\nodata& \nodata   \\
SBS 1319+579    & A     &  6.0&   321 & 0.018 & &  9.2 &   312 &\nodata& 312   & 0.029 &\nodata&\nodata&\nodata&  0.011    \\
SBS 1415+437    & C     &$<$1.0& $>$62.8&$<$0.016&&$<$2.4&$>$57&\nodata&$>$57  &$<$0.040&\nodata&\nodata&\nodata& \nodata   \\
III Zw 107      & A     & 309 &  5567 & 0.053 & &  420 &  5010 &\nodata&  5010 & 0.077 &\nodata&\nodata&\nodata&  0.041    \\
Tol 9           & C     & 244 &   981 & 0.199 & &  272 &   842 &  127: &    37:& 0.244 & 0.776:& 0.469:& 0.916:&  0.090    \\			
ESO 566-8       &\nodata&8097 & 53392 & 0.132 & & 9574 & 45975 & 4768  & 15839 & 0.172 & 0.231 & 0.498 & 0.475 &  0.073    \\
   NGC 5253     & A     & 0.73&  35.0 & 0.020 & & 1.03 &  34.2 &\nodata&   34.2& 0.029 &\nodata&\nodata&\nodata&  0.012    \\
                & B     & 0.74&  32.0 & 0.022 & & 1.03 &  31.3 &\nodata&   31.3& 0.032 &\nodata&\nodata&\nodata&  0.012    \\
                & C$^k$ & 3.2 &  49.2 & 0.061 & & 4.2  &  42.8 &  0.65 &  37.7 & 0.090 & 0.017 & 0.154 & 0.114 &  0.084$^j$  \\ 
	\noalign{\smallskip}    
  \tableline
  \end{tabular}
    \begin{flushleft}
  $^a$ We consider  the solar value for the luminosity of a single WNL star to the broad \ion{He}{ii} $\lambda$4686 emission line (SV98 method).\\
  $^b$ We consider a metallicity-dependent value for the luminosity of a single WNL star to  the broad \ion{He}{ii} $\lambda$4686 emission line  (CH06 method).\\
  
  $^c$ Total number of O stars after correcting by the ionization of the WNL stars (Eq.~\ref{GIT00WR}) using the SV98 method.\\
  $^d$ Total number of O stars after correcting by the ionization of the WNL stars (Eq.~\ref{GIT00WR}) using the CH06 method.\\
  $^e$ The number of WCE stars and the WCE/(WCE+O), WCE/WNL and WR/(WR+O) ratios were computed assuming the metallicity-dependent (CH06) method. \\ 
  $^f$ Total number of O stars after correcting by the ionization of the WNL and WCE stars (Eq.~\ref{GIT00WR}) using the CH06 method.\\
  $^g$ This coincides with the WR/(WR+O) ratio when no emission from WCE stars is detected. \\
  $^h$ This considers the total number of O stars after correcting by the ionization of the WNL and WCE stars derived using the CH06 method.\\ 
  $^i$ Using the empirical calibration given by \citet{SV98} (Eq.~\ref{ewr1}).\\
  $^j$ Considering not only the flux of the broad \ion{He}{ii} $\lambda$4686 flux but also the \ion{N}{iii} $\lambda$4640 and \ion{C}{iv}  
$\lambda$4658 fluxes.\\
  $^k$ Number of WCE stars computed using the \ion{C}{iv} $\lambda$4658 line. See text for details.
  \end{flushleft}
\end{table*} 
 
\section{The nebular \ion{He}{ii} line}

As was briefly mentioned in the introduction, there is still some discussion about the origin of the nebular \ion{He}{ii} $\lambda$4686 line. It  
is generally assumed that the hard radiation field of WR stars produces this nebular line in extragalactic objects \citep{Schaerer96L,Schaerer98} but  
some authors are skeptic about this \citep{Garnett91,G04}. The nebular \ion{He}{ii} $\lambda$4686 line may be also produced by hard X-ray radiation  
from either massive binaries or by supernova remnants from previous generation of stars in the range 10-50~Myr  
\citep{PakullAngebault86,VanBeverVanbeveren00,Cerv02}
as some observational studies suggest \citep{Bresolin99,GIT00,StaIzo03,TI05},
especially at the late stages of the star-formation burst. \citet{G04} even suggests that as we do not yet understand the evolutionary status of WR  
stars, it is premature to predict them from stellar evolution models. 

Recently, BKD08 re-examined this issue using the SDSS data. They found that the nebular \ion{He}{ii} $\lambda$4686 line was always detected in  
systems with $\log EW\sim2.5$ as long as the S/N of the spectrum permitted. In the same way, not \emph{every} galaxy where this nebular line was  
detected shows WR features, which discrepancy is more pronounced at lower metallicities. As \citet{GIT00} already said, in those galaxies (usually  
very low-metallicity objects) the broad WR feature was very probably too weak to detect. If low-density stellar winds at low-metallicity lead to a  
higher flux of \ion{He}{ii} ionizing photons 
\citep{SNC02,HadfieldCrowther07}, then normal O stars and not only WR stars would be expect to contribute to the \ion{He}{ii} ionizing flux. BKD08  
finally concluded that although at metallicities higher than 20\% solar the ionization of \ion{He}{ii} is largely produced by WN stars, at lower  
metallicities it is most likely dominated by O stars, although there should be a probable contribution of WN stars which is undetectable because  
of the faintness of their optical features.

We compared the intensity of the nebular \ion{He}{ii} $\lambda$4686 emission line with the oxygen abundance and the $\log$\,O$^{++}$/O$^+$  
ratio. Notice that for NGC~5253 we analyzed four independent regions that are very close one to the other (see L\'opez-S\'anchez et al. 2007),  
and hence we also indicate their average values in the diagrams (green stars). In Fig.~\ref{nheii} (\emph{left}) we see that objects with higher  
excitation tend to show a higher nebular \ion{He}{ii} $\lambda$4686 flux.    
Regions for which the nebular line is observed but the \WRBUMP\ is not (HCG~31~F1, SBS~1415+437, Tol~1457-262~A and B and NGC~5253~D) are dispersed  
throughout the diagram, but all have $\log$\,O$^{++}$/O$^+>$0. On the other hand, the intensity of the nebular \ion{He}{ii} line clearly decreases  
with increasing metallicity (Fig.~\ref{nheii}, right). Objects in the low-metallicity regime [\abox$<$8.2] have the highest values of the  
$I$(\ion{He}{ii})/$I$(\Hb) ratio, and broad WR features are sometimes not detected in their optical spectra, in agreement with the result found by  
BKD08. Hence the nebular \ion{He}{ii} $\lambda$4686 emission line \emph{may} arise from ionizing O stars and not only from WN stars at low  
metallicities. An analysis similar to that shown here involving more galaxies with \abox$<$8.2 and new models of winds of high-mass stars at low  
metallicities are needed to confirm this assumption. Below we will only consider the broad \ion{He}{ii} emission line as a clear indication of WR stars within these starburst galaxies. 

\citet{GIT00} did not find any nebular \ion{He}{ii} $\lambda$4686 line for objects with \abox$>$8.13, concluding that high-metallicity stellar models  
overpredict the number of photons with $\lambda<$228 \AA\ responsible for the ionization of the He$^+$. Our careful fitting of the blue \WRBUMP\  
showed that although faint, this nebular line is present in bursts with higher metallicities.

\section{Computing of the WR ratios}

We estimate the number of WR stars from the luminosity of the broad \ion{He}{ii} $\lambda$4686 line and the broad \ion{C}{iv} $\lambda$5808 line.  
Assuming that only WNL stars contribute to the luminosity of the broad \ion{He}{ii} $\lambda$4686 line, we can derive the number of this subtype WR  
population applying
\begin{eqnarray}
{\rm WNL} = N_{WNL}= \frac{L_{obs}({\rm He\,II}\, \lambda 4686)}{L_{WNL}({\rm He\,II}\, \lambda 4686)}.
\end{eqnarray}
On the other hand, the number of WCE stars can be computed from the luminosity of the broad \ion{C}{iv} $\lambda$5808 line:
\begin{eqnarray}
{\rm WCE} = N_{WCE}= \frac{L_{obs}({\rm C\,IV}\, \lambda 5808)}{L_{WCE}({\rm C\,IV}\, \lambda 5808)}.
\end{eqnarray}
For solar metallicities it is well established that one single WNL star contributes with $L_{WNL}$(\ion{He}{ii} $\lambda$4686)$\sim$1.7$\times  
10^{36}$~erg\,s$^{-1}$, while a single WCE star contributes with $L_{WCE}$(\ion{C}{iv} $\lambda$5808)$\sim$3.3$\times 10^{36}$~erg\,s$^{-1}$  
\citep{VC92}. These are the quantity values commonly used \citep{GIT00,FCCG04,Buckalew05} and the assumptions considered in the theoretical models  
(SV98, SB99). However, there is increasing evidence that the WR line luminosities are reduced at lower metallicities  
\citep{CH06,HadfieldCrowther06,Crowther07}. Indeed, Crowther \& Hadfield (2006; hereafter CH06) showed how reduced line luminosities at 
low-metallicity follow naturally if WR winds are metallicity-dependent. Hence we also considered the metallicity of the star-forming regions to get a  
more appropriate estimation of the number of WR stars. We assumed the broad-line WNL and WCE luminosities given by CH06 for solar and $Z_{\odot}$/50  
metallicites, 
{\small \begin{eqnarray}
L_{\rm WNL}({\rm He\,II}\, \lambda 4686,\, Z_{\odot})=1.6 \times 10^{36}\,  {\rm erg\, s^{-1}},\\
L_{\rm WNL}({\rm He\,II}\, \lambda 4686,\, Z_{\odot}/50)=2.2 \times 10^{35}\,  {\rm erg\, s^{-1}},\\
L_{\rm WCE}({\rm C\,IV}\, \lambda 5808,\, Z_{\odot})=2.5 \times 10^{36}\,  {\rm erg\, s^{-1}},\\
L_{\rm WCE}({\rm C\,IV}\, \lambda 5808,\, Z_{\odot}/50)=4.0 \times 10^{35}\,  {\rm erg\, s^{-1}}, 
\end{eqnarray} }
(see their Tables~3 and 4) and performed a linear fit to these values considering \abox=8.66 and 6.96 for $Z_{\odot}$ \citep*{ASP05} and  
$Z_{\odot}$/50, respectively. We then adopted for each burst the WR luminosities computed from these relations: 
{\small \begin{eqnarray}
\label{formula1} L_{\rm WNL}({\rm He\,II}\, \lambda 4686)= (-5.430 + 0.812x) \times 10^{36}\,  {\rm erg\, s^{-1}}, \\
\label{formula2} L_{\rm WCE}({\rm C\,IV}\, \lambda 5808)= (-8.198 + 1.235x) \times 10^{36}\,  {\rm erg\, s^{-1}}, 
\end{eqnarray} }
with $x$=\abox. Table~\ref{wrpop} compiles the number of WNL stars derived for each burst following both the traditional and the  
metallicity-depended methods, and the number of WCE stars in objects where the red \WRBUMP\ was detected following the CH06 assumption. 
For region~C in NGC~5253, we assumed that 
{\small \begin{eqnarray}
L_{\rm WCE}({\rm C\,IV}\, \lambda 4658)= 1.71 \times L_{\rm WCE}({\rm C\,IV}\, \lambda 5808), 
\end{eqnarray}}
(SV98) to compute the number of WCE within this knot.

\begin{figure}[t!]
\includegraphics[angle=270,width=\linewidth]{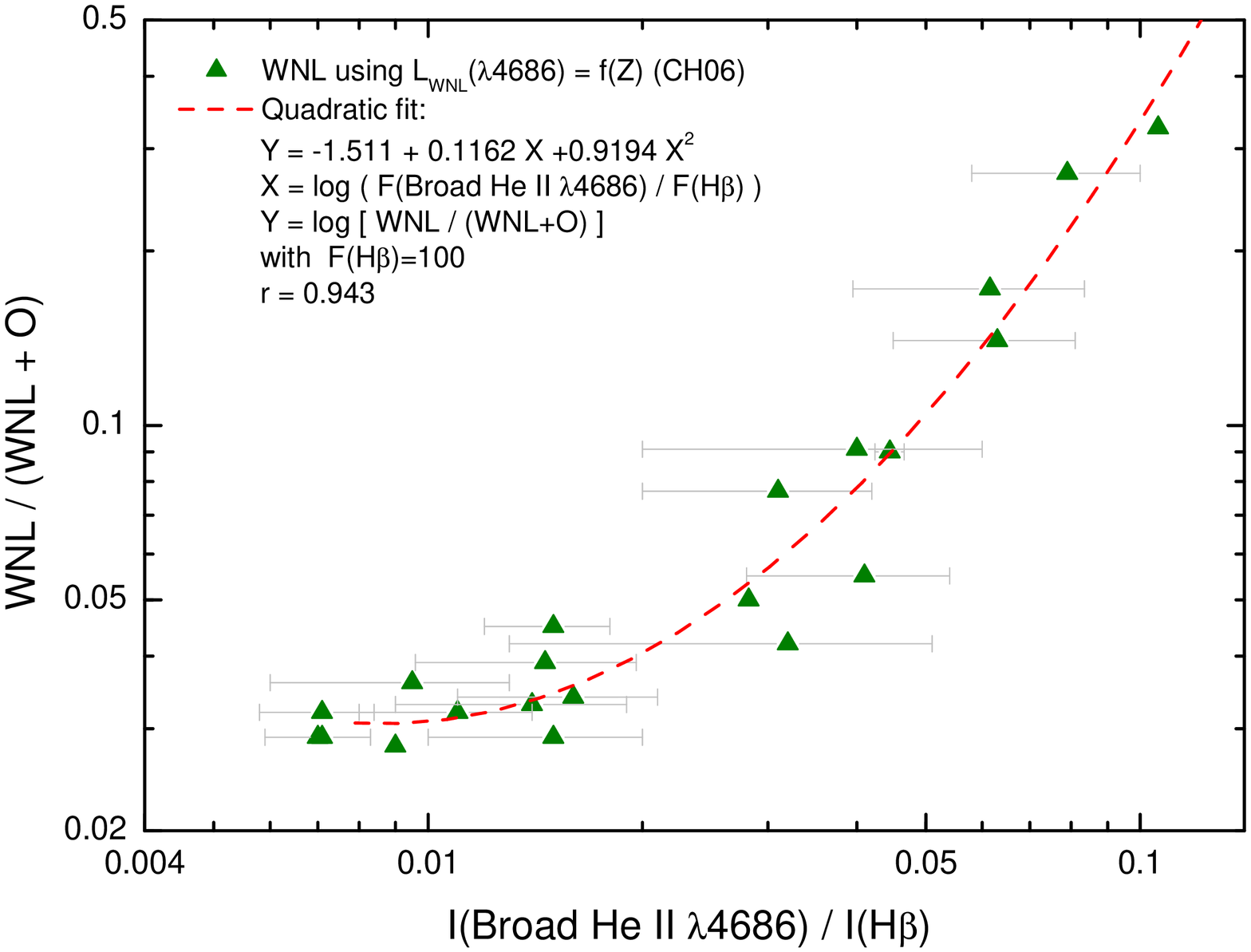}
\protect\caption[ ]{\footnotesize{Number of WNL stars (derived assuming a metallicity-dependence of the WNL luminosities) vs. the reddening-corrected  
flux of the broad \ion{He}{ii} $\lambda$4686 line. A quadratic fit to the data is shown with a dashed red line.}}
\label{wnlheii}
\end{figure}

The total number of O stars is derived via the \Hb\ luminosity, assuming $L_{\rm O7V}$(\Hb)=4.76$\times 10^{36}$ erg s$^{-1}$ for a O7V star,  
$N_{O'}$ = $L$(\Hb)/$L$(O7V \Hb). However, the contribution of the WR stars and other O stars subtypes to the ionizing flux must be considered. This  
is done via the the $\eta(t)\equiv$O7V/O parameter \citep{VC92,V94,SV98}, which depends on the age of the burst. We used SV98 models (see their  
Fig.~21) and the estimated age derived for each burst to get an estimation of $\eta$ (listed in Col.~5 of Table~\ref{wrdata}), and then applied
{\small \begin{eqnarray}
\label{GIT00WR} {\rm O}=N_{\rm O}= \frac{Q_0^{Total}-N_{\rm WNL}Q_0^{\rm WNL}-N_{\rm WCE}Q_0^{\rm WCE} }{\eta_0(t) Q_0^{\rm O7V}},    
\end{eqnarray}}
with  $Q_0^{Total}$, $Q_0^{\rm WNL}$, $Q_0^{\rm WCE}$ and $Q_0^{\rm O7V}$ the total, the WNL, the WCE and the O7V numbers of ionizing photons,  
respectively. Following CH06, we assumed $\log Q_0^{\rm WNL}$=49.4 and $\log Q_0^{\rm WCE}$=49.5, while we adopted $\log Q_0^{\rm O7V}$=49.0 (SV98).  
$Q_0^{Total}$ was derived from $Q_0^{Total}$ = $N_{O'} Q_0^{\rm O7V}$. The total number of O stars estimated for each burst are compiled in  
Table~\ref{wrpop}. Notice that this table includes several values for the total number of O stars depending on the assumed number of WR stars (only  
WNL stars following the standard method, only WNL stars considering the metallicity-dependent method, or WNL and WCE stars assuming the  
metallicity-dependent method). Table~\ref{wrpop} also lists the WNL/(WNL+O) ratios derived for the two methods assumed here, and the WCE/(WCE+O),  
WCE/WNL and WR/(WR+O) ratios computed using the metallicity-dependent method for regions in which WCE stars are detected. The number of WR stars was  
computed assuming WR=WN+WC=WNL+WCE. For galaxies in which WCE stars are not detected, the total number of WR stars should be the number of WNL stars.  
However, we must keep in mind that because of the faintness of the broad \ion{C}{iv} $\lambda$5808 lines (specially at low metallicities), some WCE  
stars contribution may be also expected, and hence the total number of WR stars assuming WR=WNL is just a lower limit to the actual value.   

The last column in Table~\ref{wrpop} shows the WR/(WR+O) ratio obtained when applying the empirical relation given by \citet{SV98} between the intensity  
of the blue \WRBUMP\ and the WR/(WR+O) ratio,
{\small \begin{eqnarray}
\label{ewr1} 
\log \frac{\rm WR}{\rm WR+O} = 0.85 \times \log\frac{I({\rm Blue\, WR\, Bump})}{I({\rm H\beta})}-0.11.
\end{eqnarray}}
We assumed that the luminosity of the blue \WRBUMP\ is the luminosity of the broad \ion{He}{ii} $\lambda$4686 line, and added the contribution of the  
\ion{N}{iii} $\lambda$4640 and the \ion{C}{iv} $\lambda$4658 to the objects for which these broad lines are detected. As we see, the results provided  
by this empirical relation are systematically lower than those derived using the direct method (Col.~5 in Table~\ref{wrpop}) except for the three  
regions in which other broad emission lines are detected in the blue \WRBUMP. Figure~\ref{wnlheii} plots the derived WNL/(WNL+O) ratio using CH06  
data vs. the intensity of the broad \ion{He}{ii} $\lambda$4686 line. As it can be seen, the best fit to the data shown in Fig.~\ref{wnlheii} is a  
quadratic fit,
{\small \begin{eqnarray}
y = -1.511 + 0.1162 x +0.9194 x^2,
\end{eqnarray}} 
where $x=\log [I$(\ion{He}{ii} $\lambda$4686)/$I$(\Hb)] assuming $I$(\Hb)=100 and $y=\log$[ WNL / (WNL+O)]. The correlation coefficient of this fit  
is $r$=0.943. A similar relation between the broad \ion{C}{iv} $\lambda$5808 intensity and the WCE/(WCE+O) ratio seems to be found. However, we did
not attempt to fit the data due to the scarcity of observational points. 

\section{The number of WR stars-metallicity relation}

\begin{figure}[t!]
\includegraphics[angle=270,width=\linewidth]{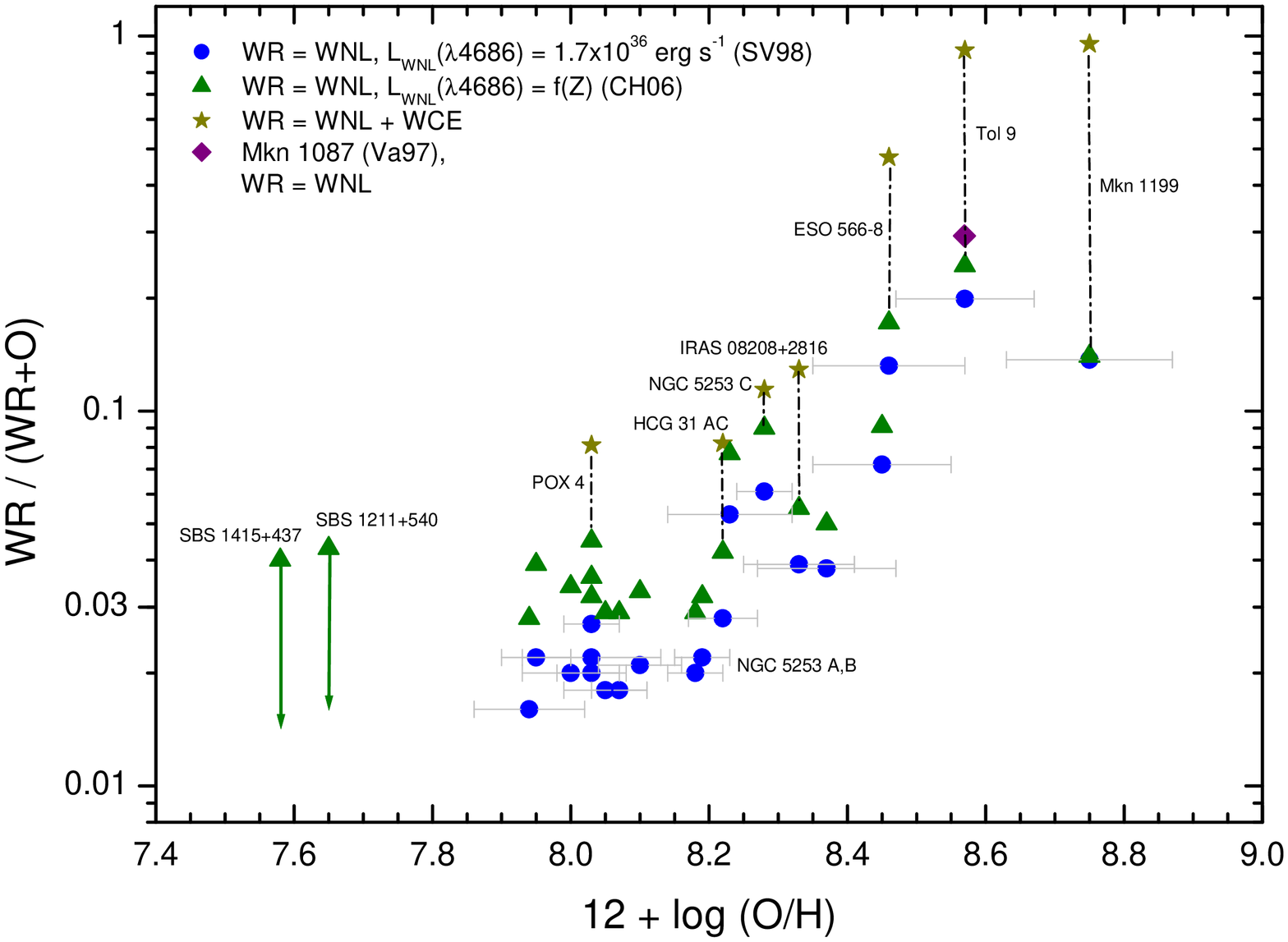}
\protect\caption[ ]{\footnotesize{WR/(WR+O) vs. oxygen abundance for our galaxy sample. Blue points represent WR=WNL assuming a constant luminosity  
for the broad \ion{He}{ii} $\lambda$4686 emission, green triangles indicate WR=WNL considering a metallicity-dependent broad \ion{He}{ii}  
$\lambda$4686. Dark yellow stars plot WR=WNL+WCE assuming the metallicity-dependent WNL and WCE luminosities. 
We include the upper limit to the WR/(WR+O) ratio estimated for the very low-metallicity galaxies SBS~1211+540 and SBS~1415+437. Dashed lines connect  
the WR=WNL and the WR=WNL+WCE values for the galaxies for which we have WCE data. These galaxies have been labelled.}}
\label{wrabox}
\end{figure}

Following evolutionary models of massive stars and population synthesis models (Mas-Hesse \& Kunth 1991; SV98; Maeder \& Meynet 2005), we should  
expect an increase of the number of WR stars with increasing metallicity because of the decrease of the minimum mass that a massive star needs to  
reach the WR phase due to the strong metallicity-dependence of the stellar winds \citep{VinkdeKoter05}. \citet{KS86} were the first that noticed that  
the fraction of WR stars relative to other massive stars increases with increasing metallicity. The analysis performed by \citet{GIT00} also showed a  
tendency between the WR/(WR+O) ratio and the metallicity, and \citet{PSGD02} reported this dependence for high-metallicity objects. Recent studies  
indicate that this correlation is certainly satisfied (Zhang et al. 2007; Crowther 2007; BKD08).
Besides the uncertainties and considerable high scatter of their data, BKD08 indicated that models of stellar evolution including massive binaries by  
\citet*{Eldridge08} appear to well reproduce the relation between the metallicity and the number of WR stars even at low metallicity.

Figure~\ref{wrabox} plots the evolution of the relative numbers of massive stars, WR/(WR+O), with respect to the oxygen abundance for our galaxy sample.  
Blue points represent WR=WNL assuming a constant luminosity for the broad \ion{He}{ii} $\lambda$4686, green triangles indicate WR=WNL considering a  
metallicity-dependent broad \ion{He}{ii} $\lambda$4686. Dark yellow stars plot WR=WNL+WCE assuming the metallicity-dependent of both WNL and WCE  
luminosities. 
From this figure is evident that the number of WR stars increases with increasing metallicity, specially in the high-metallicity regime. 

We should expect some scatter in Fig.~\ref{wrabox} because of the strong dependence on time of the WR star relative numbers during the short WR  
stage of the starburst. That is what may be happening in NGC~5253: regions A and B show a low WNL/(WNL+O) ratio, region C has a very intense blue  
\WRBUMP\ and a high WNR/(WR+O) ratio, but no broad WR feature is found in region D even though all zones are very close. Actually, as it was  
explained by \citet{GIT00}, a galaxy undergoing a star-formation burst would describe a loop in this diagram, quickly reach the WR/(WR+O) maximum  
and after that drop down to slightly higher metallicities because of the chemical enrichment of the surrounding interstellar medium. We consider  
that the scatter in  Fig.~\ref{wrabox} is not so evident because all analyzed objects are undergoing an intense and very probably short  
star-formation event with an age between 3.5 and 5.5~\Myr\ (see Paper~I) and have a very similar nature, with the majority of them \BCDG s.

However, Fig.~\ref{wrabox} may suggest a possible change of the slope of the relation between the number of WNL stars and the metallicity when
considering the metallicity-dependent broad \ion{He}{ii} $\lambda$4686 (green triangles).
For galaxies with  \abox$>$8.2, the WNL/(WNL+O) ratio clearly increases with metallicity, but in the range  \abox=7.9--8.2 it seems to become
constant, with a value of WNL/(WNL+O)$\sim$0.03--0.04 (11 points). 
We cannot confirm that tendency, however, because of the lack of objects in the range \abox=7.6--7.9, and hence more data in the low metallicity
regime are needed to understand the metallicity dependence of the WNL/(WNL+O) ratio.
This behavior was previously noticed by BKD08 and suggested by \citet{GIT00}.  
One of the main problems to acquire these data is that both the WR/(WR+O) ratio and the duration of the WR phase decrease with decreasing metallicity, so  
the number of galaxies with extremely low metallicity --\abox$\leq$7.8-- containing a WR stellar population is expected to be small. Despite the good  
spectral resolution and S/N ratio of our WHT spectra, we did not detect a hint of WR stars in any of the regions within our two lowest metallicity  
galaxies (SBS~1211+540 and SBS~1415+437; their upper limits are plotted in Fig.~\ref{wrabox}), only the nebular \ion{He}{ii} $\lambda$4686 line in  
the brightest region of SBS~1415+437.  
However, WR stars have been detected in the lowest metallicity galaxies known so far: I Zw 18, with $Z\sim$\Zo/50 
\citep{Izotov97,Legrand97,Brown02}
and SBS 0335-052E, with $Z\sim$\Zo/40 \citep{IT99,Izotov99,Papaderos06}.

Ideally, we would also like to compare the WCE and WR=WNL+WCE corrected values between high- and low-metallicity objects, but this is difficult  
because WCE stars are undetected most of the time, in particular at low metallicities. Considering our data plotted in Fig.~\ref{wrabox}, the  
WCE/(WCE+O) ratio also increases with metallicity in galaxies with \abox$\geq$8.2. However, the two lowest metallicity objects in which we detect  
both WNL and WCE stars, POX~4 and HCG~31~AC, have very similar WNL/(WNL+O), WCE/(WCE+O) and WR/(WR+O) ratios (see Table~\ref{wrpop}) despite their  
relatively different oxygen abundance, \abox=8.03 and 8.22, respectively. Again, we state that more data of low-metallicity galaxies with clear  
detections of both the blue and red \WRBUMP s are needed to understand this behavior.

Another interesting information from Fig.~\ref{wrabox} is that for all galaxies with a detection of WCE stars
the derived WR/(WR+O) ratio is considerably higher than when only WNL stars are considered. 
The contribution of these WCE stars is smaller at lower metallicities, but even for POX~4 (the lowest metallicity galaxy where both WNL and WCE stars  
are detected) the WCE stars contribute to the 37\% to the total WR number. This as well as the analysis of the WCE/WNL ratio that we will  
explain below strongly suggests that WCE stars are not only present at low metallicities but also make an important contribution to the total number  
of WR stars.

\begin{figure*}[]
\includegraphics[angle=270,width=0.47\linewidth]{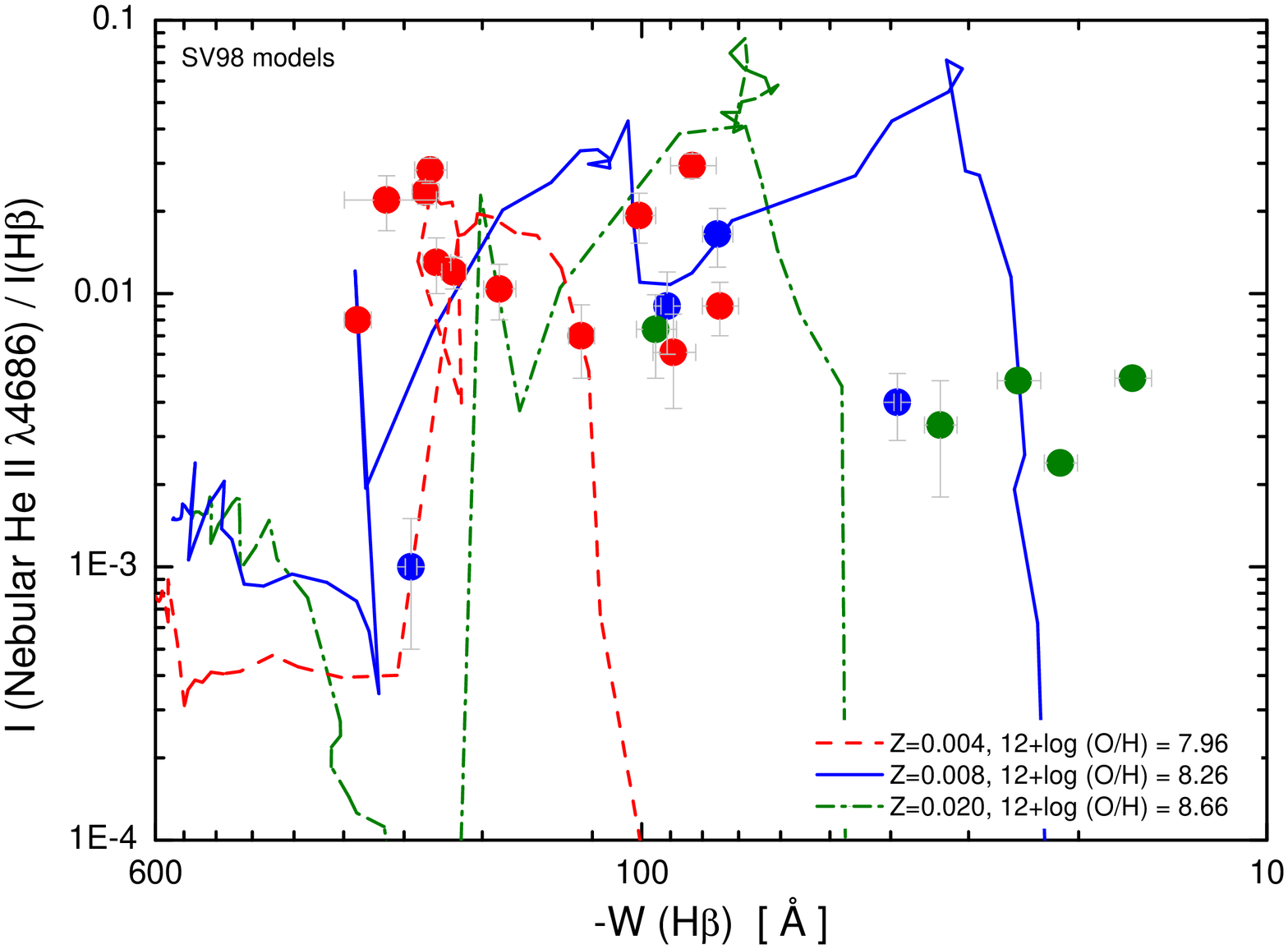}
\includegraphics[angle=270,width=0.47\linewidth]{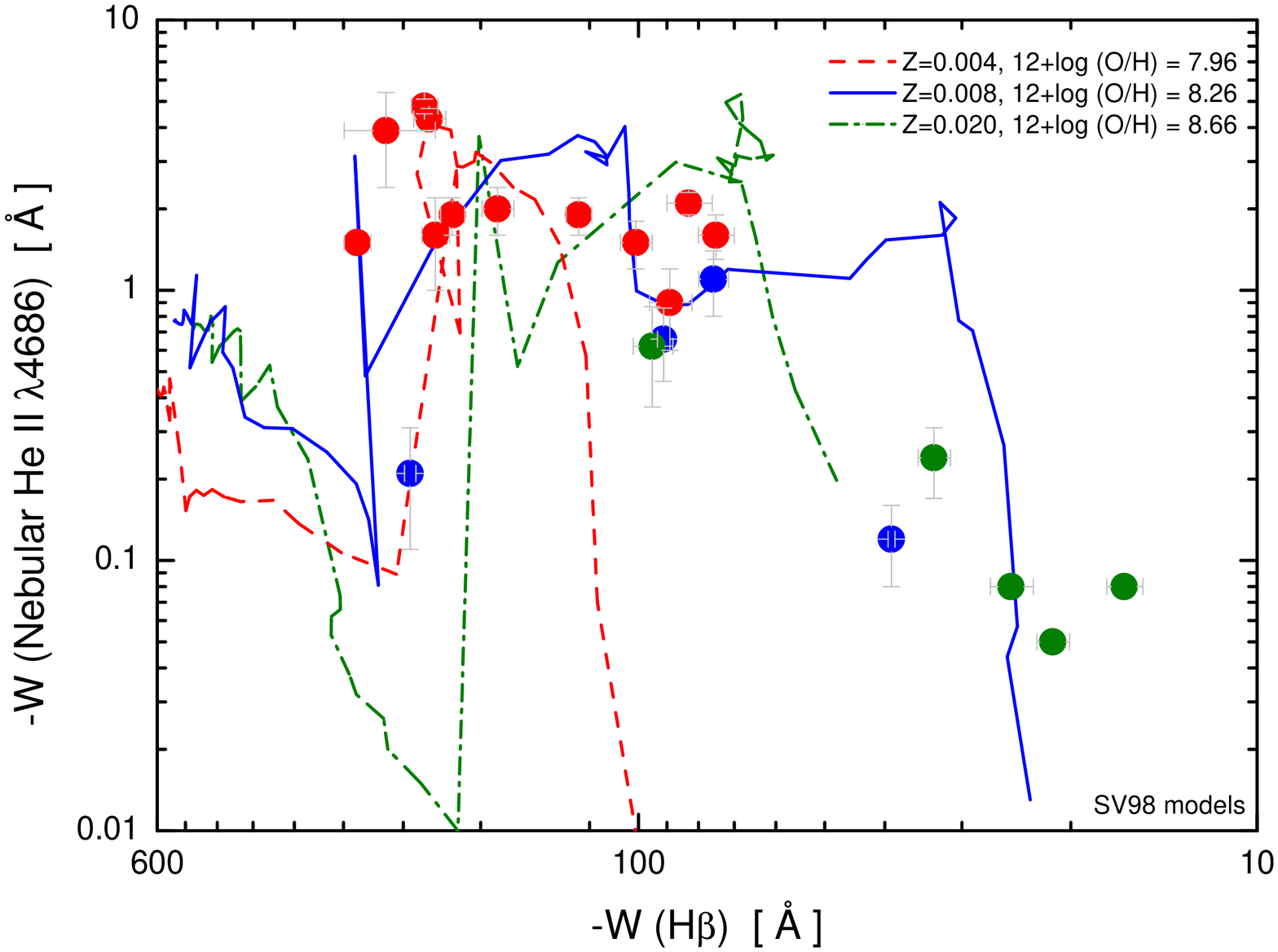}
\protect\caption[ ]{\footnotesize{Intensity (\emph{left}) and equivalent width (\emph{right}) of the nebular \ion{He}{ii} $\lambda$4686 line vs.  
$W$(\Hb) and their comparison with the predictions given by SV98 models. Three different metallicities, $Z$=0.004, 0.008 and 0.020 (\Zo) are plotted  
with dashed red, continuous blue, and dashed-dotted green lines, respectively. Data colors (red, blue and green) indicate the model that best matches  
the metallicity of the object.}}
\label{heiimodel}
\end{figure*}

\section{Comparison with the theoretical models}

The first models that tried to quantify the population of WR stars in starbursts were those presented by \citet*{Arnault89}, who found that the 
star-formation event only occurs during a short period of time compared with the typical lifetime of the massive stars and that the intensity of the blue  
\WRBUMP\ with respect \Hb\ decreases with decreasing metallicity. The models presented by  \citet{MK91}, updated by \citet{Cer94}, were the first  
models that determined the number of WR stars. However, their method had several flaws because they did not separate the stellar emission  
lines within the blue \WRBUMP, which as we see are also blended with the nearby Fe, He and Ar lines from the emission of the nebular gas. With the aim  
of solving this problem, \citet{Kru92} separately synthesized the broad \ion{He}{ii} $\lambda$4686 and \ion{C}{iii/iv} $\lambda$4650.   
\citet{Meynet95} studied the effect of changing the star-formation rate, the IMF, the age and the metallicity in the massive star populations.  
Although the detailed models provided by \citet{LH95} included a good study of the O and WR populations, they failed to reproduce some observational  
features. The evolutionary models presented by \citet*{GVBD95} had the same problem.

Schaerer \& Vacca (1998) (hereafter SV98) constructed evolutionary synthesis models for young starbursts using stellar evolution models, theoretical  
stellar spectra and a compilation of observed line-strengths from WR stars. They explicitly distinguish between several WR stars (WN, WC, WO) whose  
relative frequency strongly depends on the metallicity, and also consider O and Of stars independently. The SV98 models compute the number of O and WR  
stars produced during the starbursts and predict the intensities of the UV and optical emission lines of both the main nebular lines (H and He) and  
the broad stellar lines (He, N and C) as a function of several parameters related to the star-formation episode. These models provide the most reliable  
determinations to date. The basic framework of the SV98 models was recently included in the STARBURST~99 stellar population synthesis model \citep{L99}  
with refined spectra from \citet*{SNC02}. The last release of the SB99 code \citep{VL05} is available  
on-line\footnote{http://www.stsci.edu/science/starburst99/}.   

We considered both the SV98 and SB99 models to compare their predictions with our observational data. In both cases we assumed an instantaneous  
burst with a Salpeter IMF slope $\alpha$=2.35 and $M_{up}$=120 \Mo\  with three different metallicities, $Z$=0.004, 0.008 and 0.020 (\Zo), which  
correspond to an oxygen abundance --in units of \abox-- of 7.96, 8.26 and 8.66, respectively. The SV98 models consider $M_{low}$=0.8 \Mo\ and SB99 models  
assume $M_{low}$=0.1 \Mo. Although this does not affect to the predictions for the massive stars, the SB99 models were created 
assuming Padova tracks in which asymptotic giant branch (AGB) phases with the same metallicity of the gas are included \citep{VL05}. 
The total stellar mass created in the starburst (a normalization factor) was set to 1 \Mo\ and $10^6$ \Mo\ for SV98 and SB99 models, respectively.

Besides that the SV98 models are the most reliable WR galaxy models to date, we remind the reader that nowadays there is a better understanding of the properties of the WR stars than when the SV98 models were released. The main refinement is the metallicity-dependence of the \ion{He}{ii} $\lambda$4686 luminosity (CH06) and  
the influence of the rotation \citep{MeynetMaeder05} and wind loss and binary evolution \citep*{VanBeverVanbeveren03,Vanbeveren+07,Eldridge08} in the  
stellar tracks followed by massive stars. Stellar rotation is expected to predict a longer WR stage, lower WR star masses and thus larger WR  
populations. BKD08 considered some variations in the WR lines fluxes with the metallicity to update the SV98 models, but these models were not  
published anywhere. These authors also remarked the importance of the star-formation histories of the galaxies hosting the WR-rich starclusters,  
combining their models with the predictions given by both the SB99 and the \citet{BC03} 
codes to generate a wide range of models that depend on many parameters. New models 
of evolving starbursts showing WR features that include all these aspects are absolutely needed. 

\begin{figure*}[t!]
\centering
\begin{tabular}{c@{\hspace{4pt}}c}
\includegraphics[angle=270,width=0.47\linewidth]{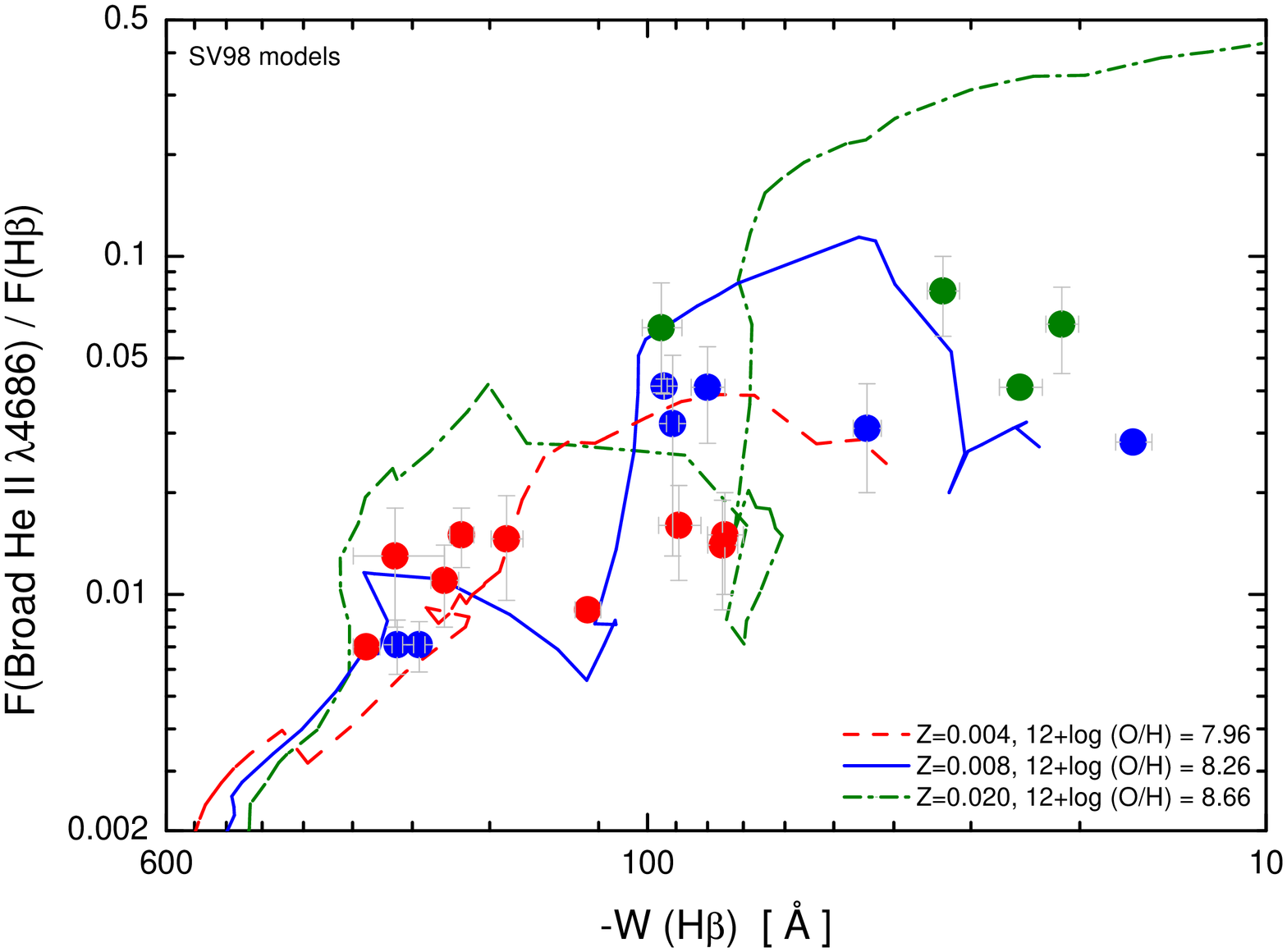} &
\includegraphics[angle=270,width=0.47\linewidth]{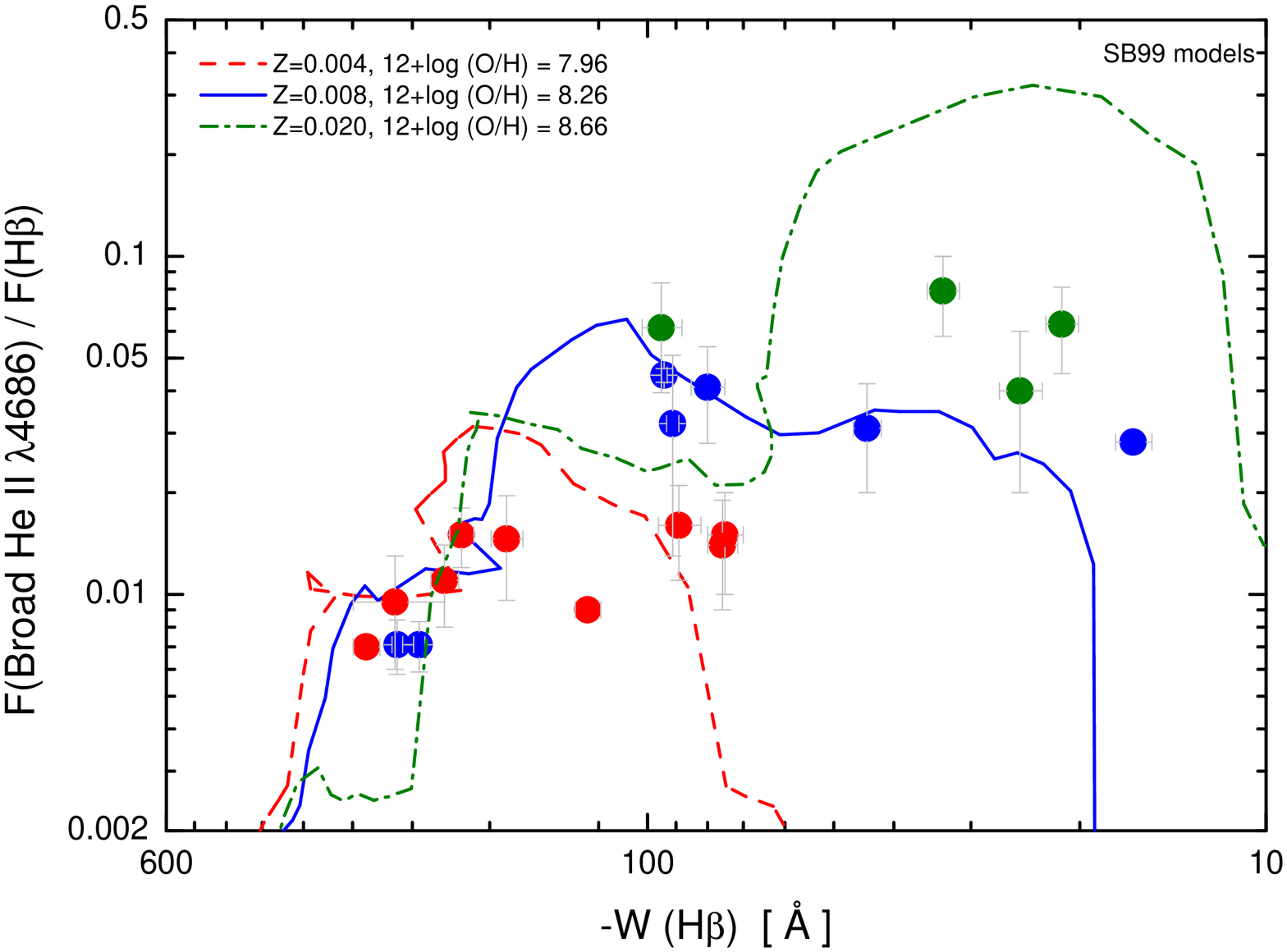} \\
\includegraphics[angle=270,width=0.47\linewidth]{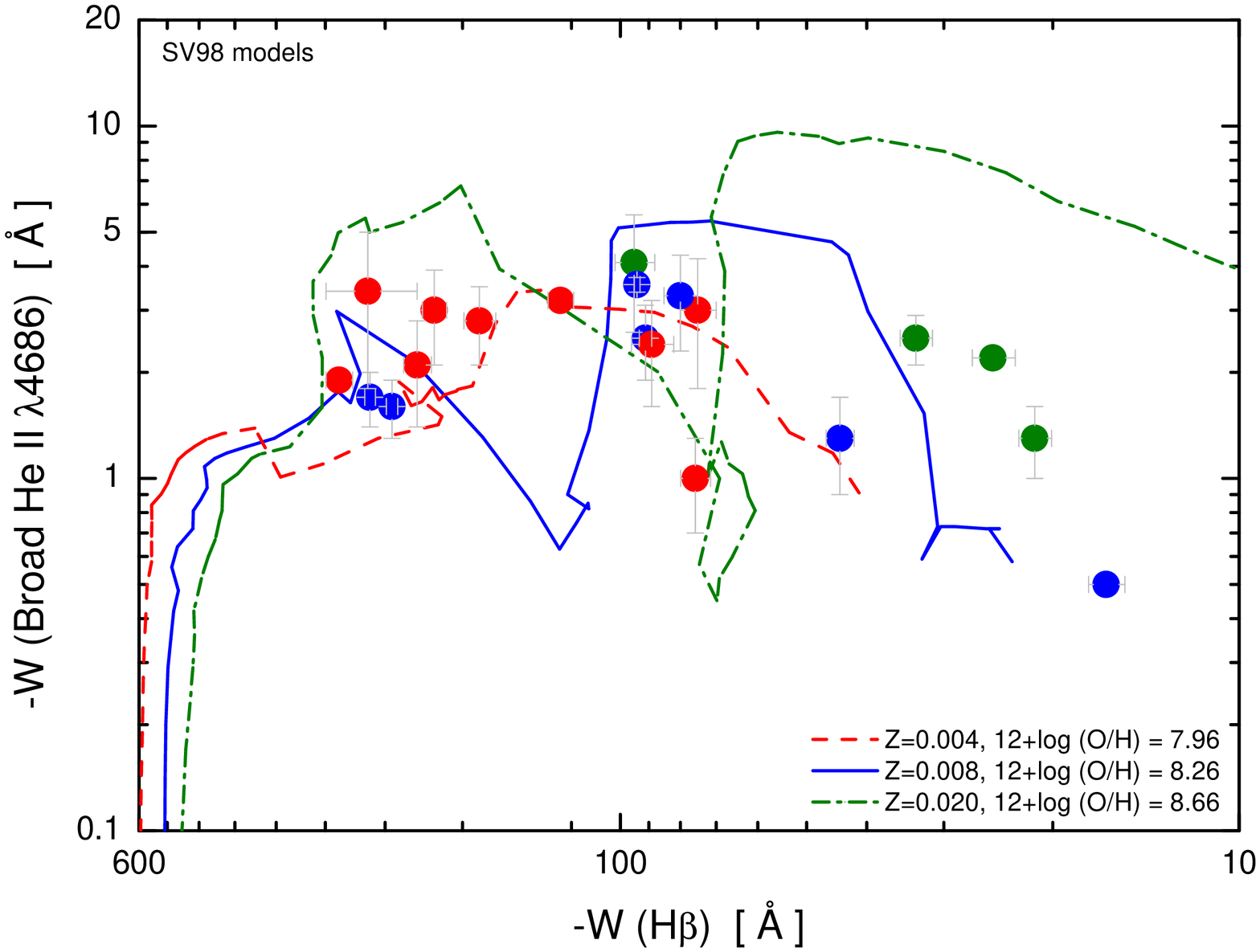} &
\includegraphics[angle=270,width=0.47\linewidth]{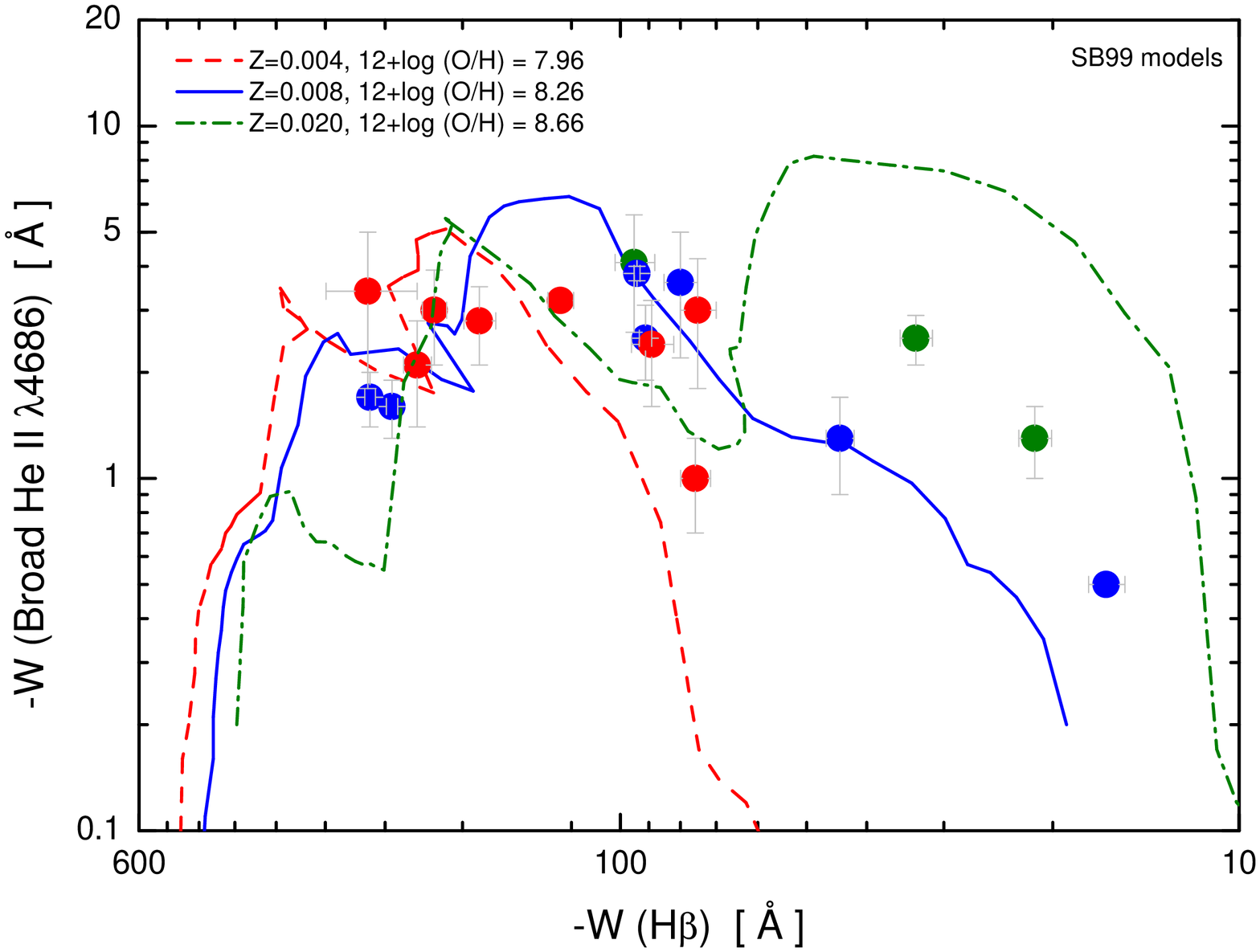} \\
\end{tabular} 
\protect\caption[ ]{\footnotesize{Comparison of the observed intensities (upper row) and equivalent width (lower row) of the broad \ion{He}{ii}  
$\lambda$4686 emission line with the predictions given by SV98 (left) and SB99 (right) models. Symbols are the same as in Fig.~\ref{heiimodel}. 
}}
\label{wrmodels1}
\end{figure*}

\begin{figure*}[t!]
\centering
\begin{tabular}{c@{\hspace{4pt}}c}
\includegraphics[angle=270,width=0.47\linewidth]{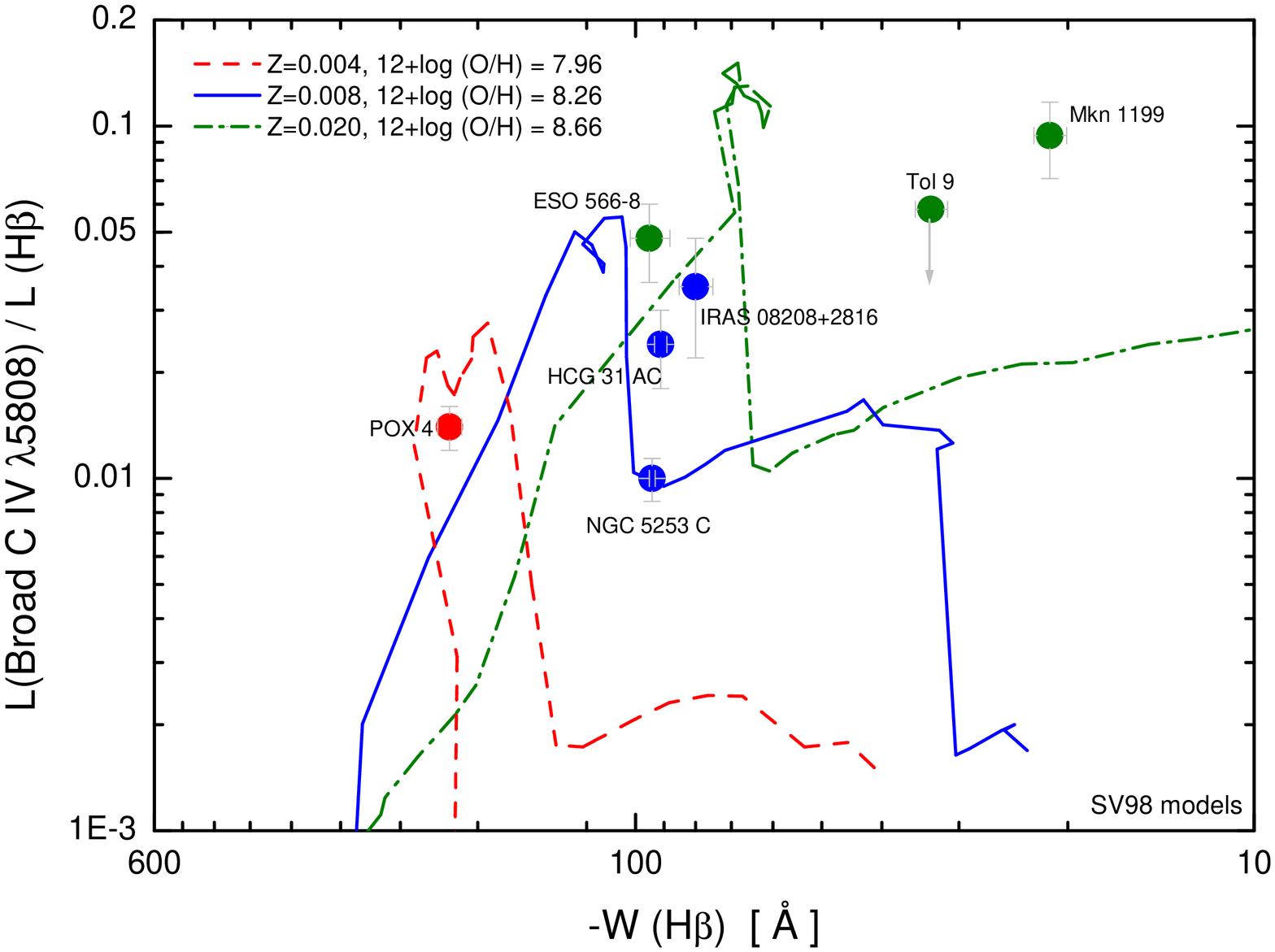} &
\includegraphics[angle=270,width=0.47\linewidth]{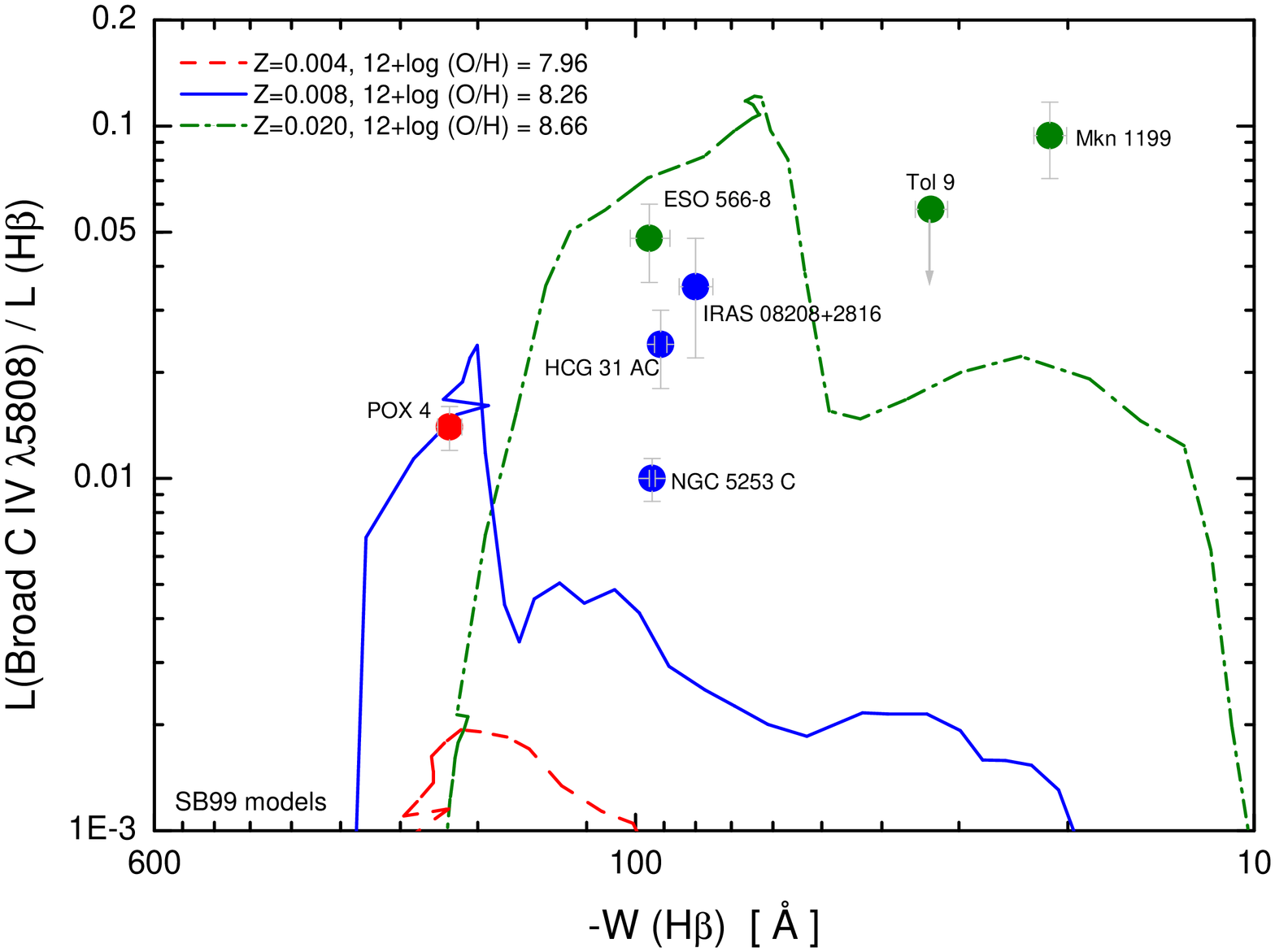} \\
\includegraphics[angle=270,width=0.47\linewidth]{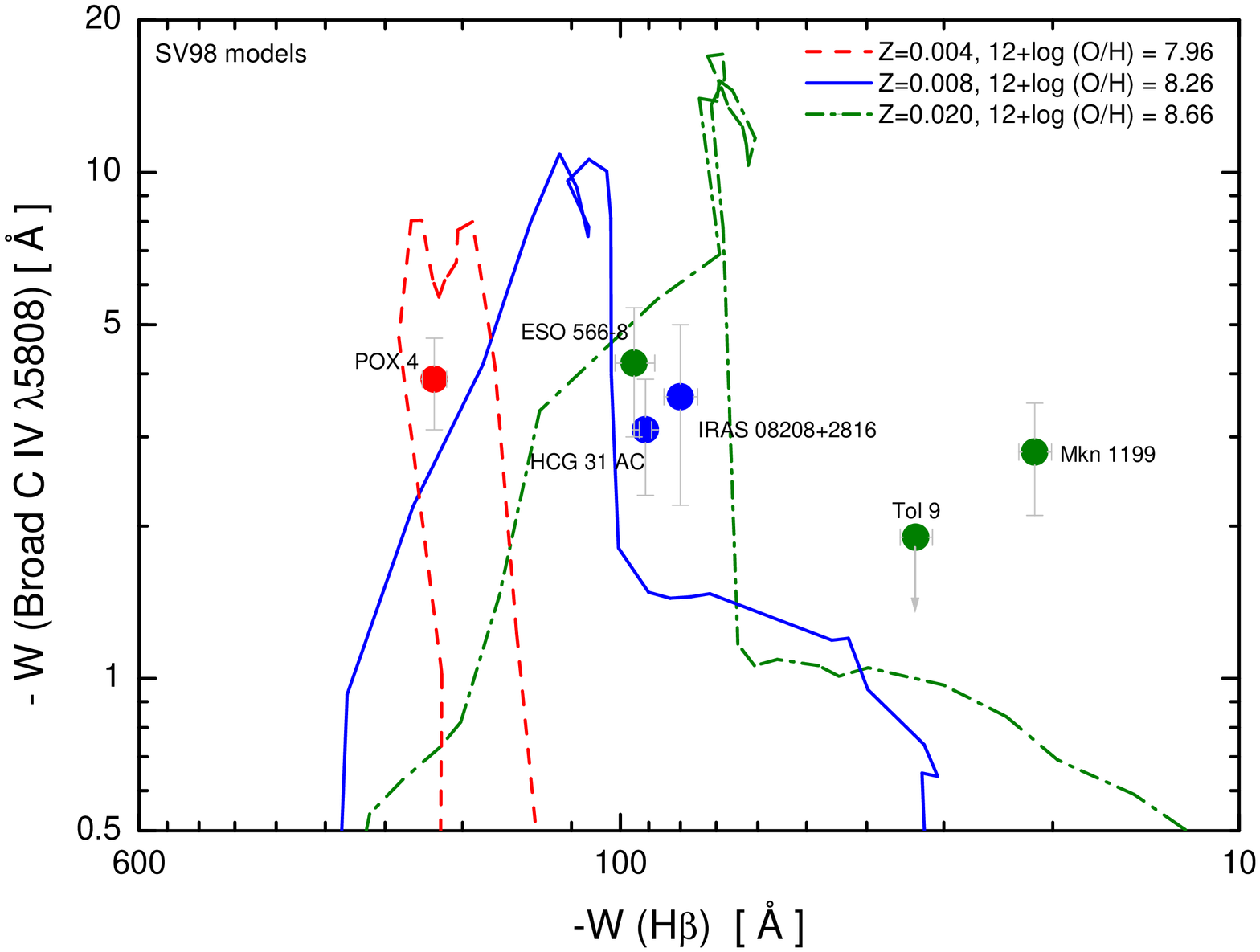} &
\includegraphics[angle=270,width=0.47\linewidth]{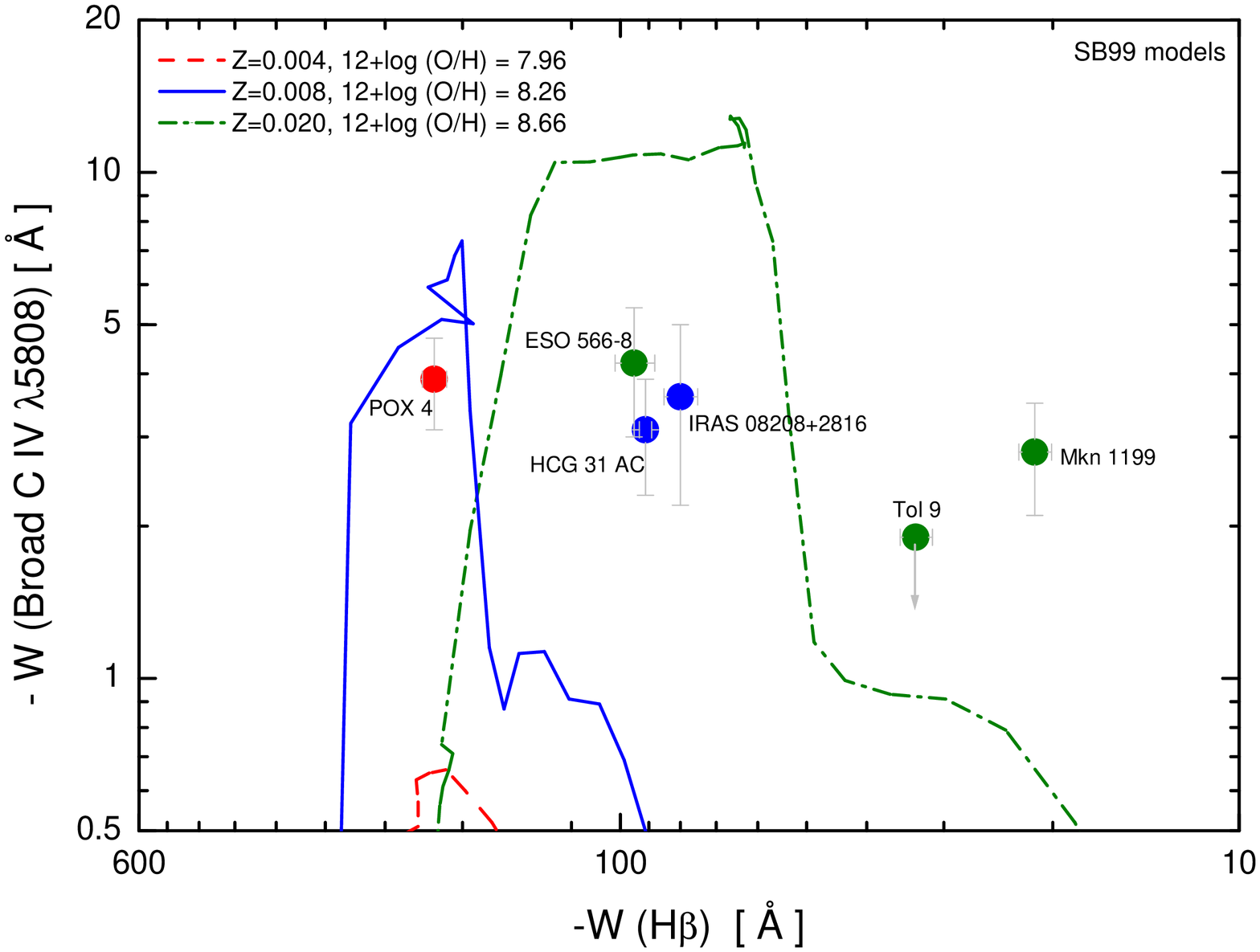} \\
\end{tabular} 
\protect\caption[ ]{\footnotesize{Comparison of the observed intensities (upper row) and equivalent width (lower row) of the broad \ion{He}{ii}  
$\lambda$4686 emission line with the predictions given by SV98 (left) and SB99 (right) models. Symbols are the same as in Fig.~\ref{heiimodel}. }}
\label{wrmodels2}
\end{figure*}

\begin{figure*}[t!]
\centering
\begin{tabular}{c@{\hspace{4pt}}c}
\includegraphics[angle=270,width=0.47\linewidth]{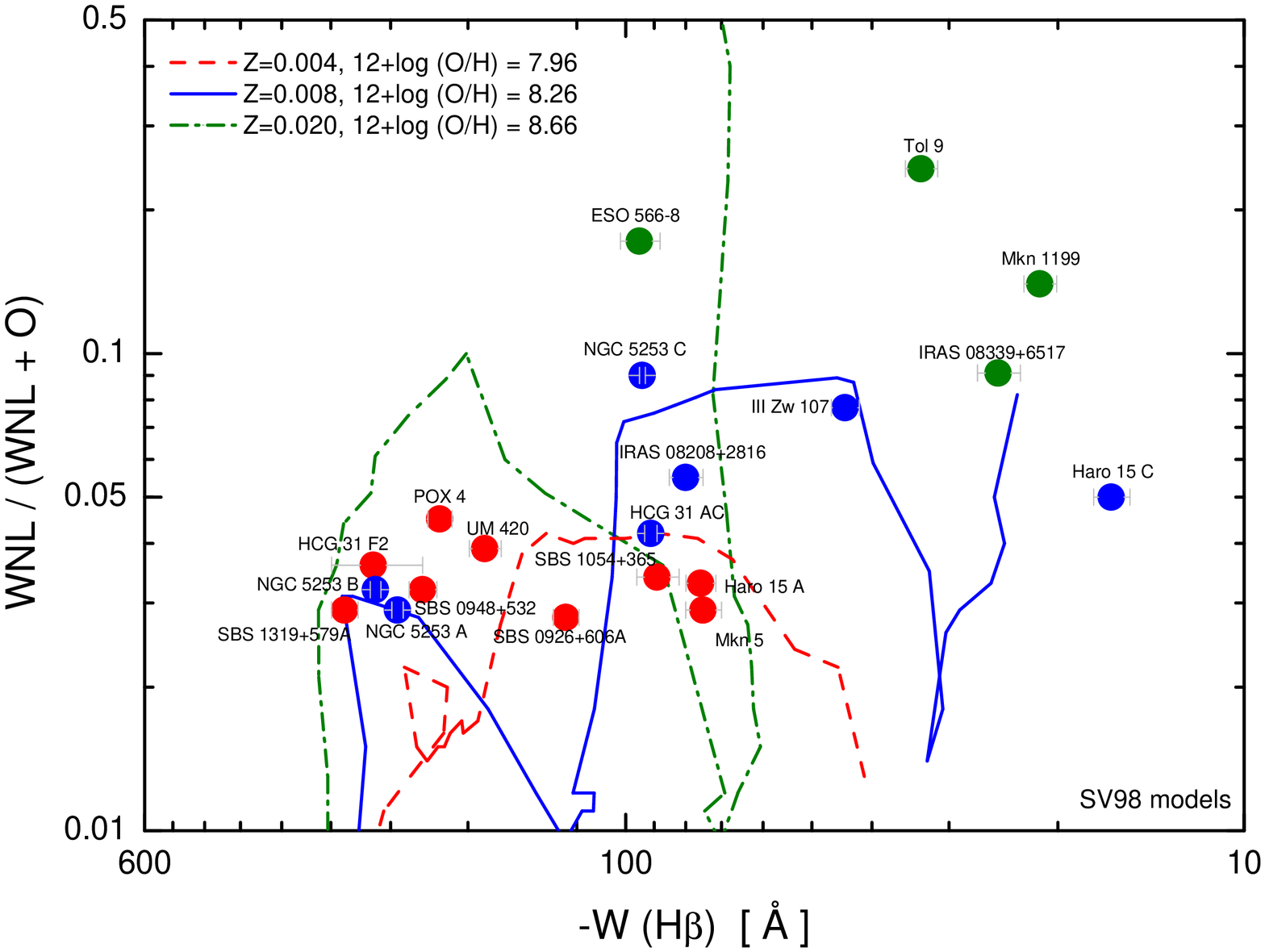} &
\includegraphics[angle=270,width=0.47\linewidth]{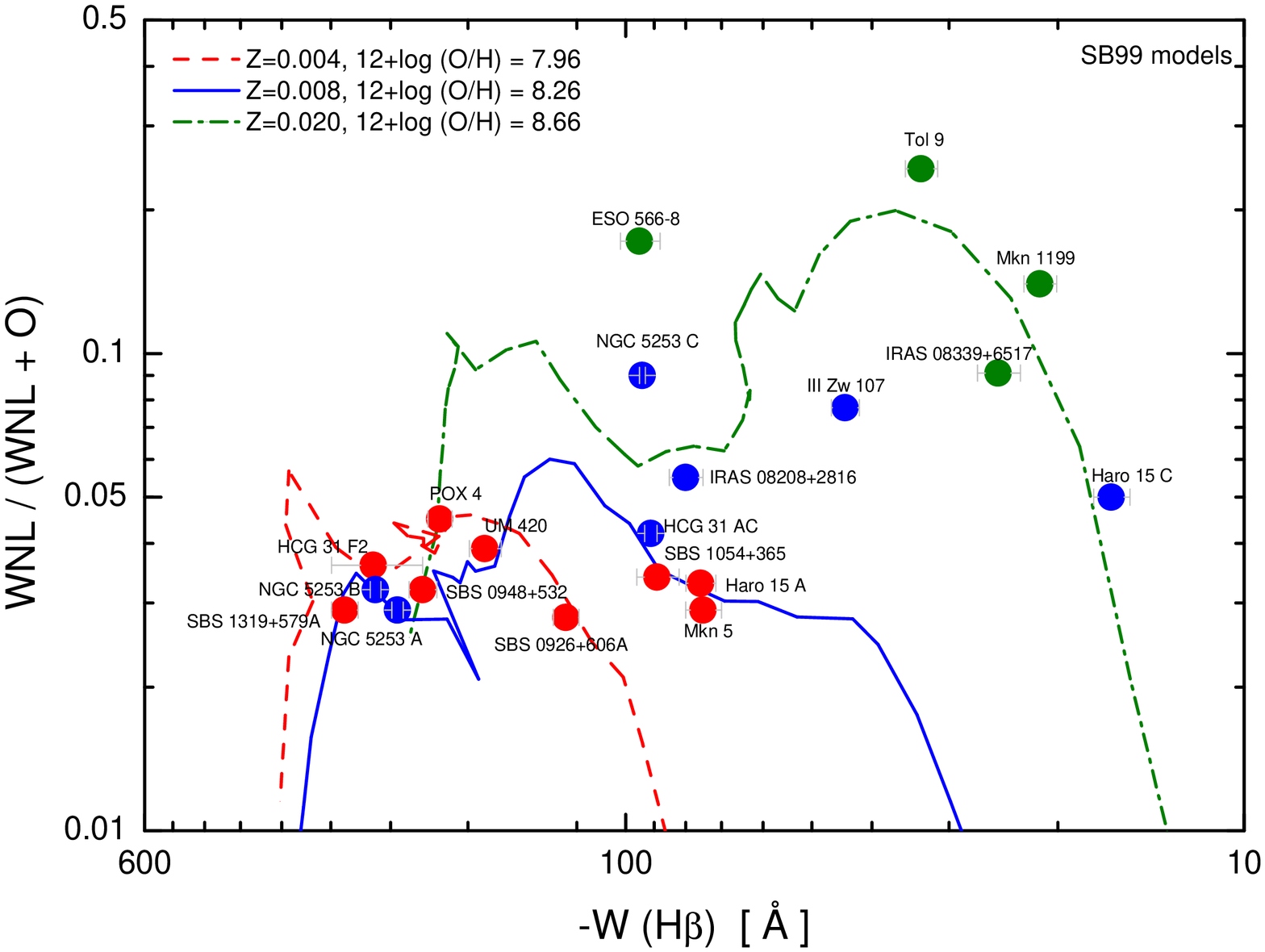} \\
\includegraphics[angle=270,width=0.47\linewidth]{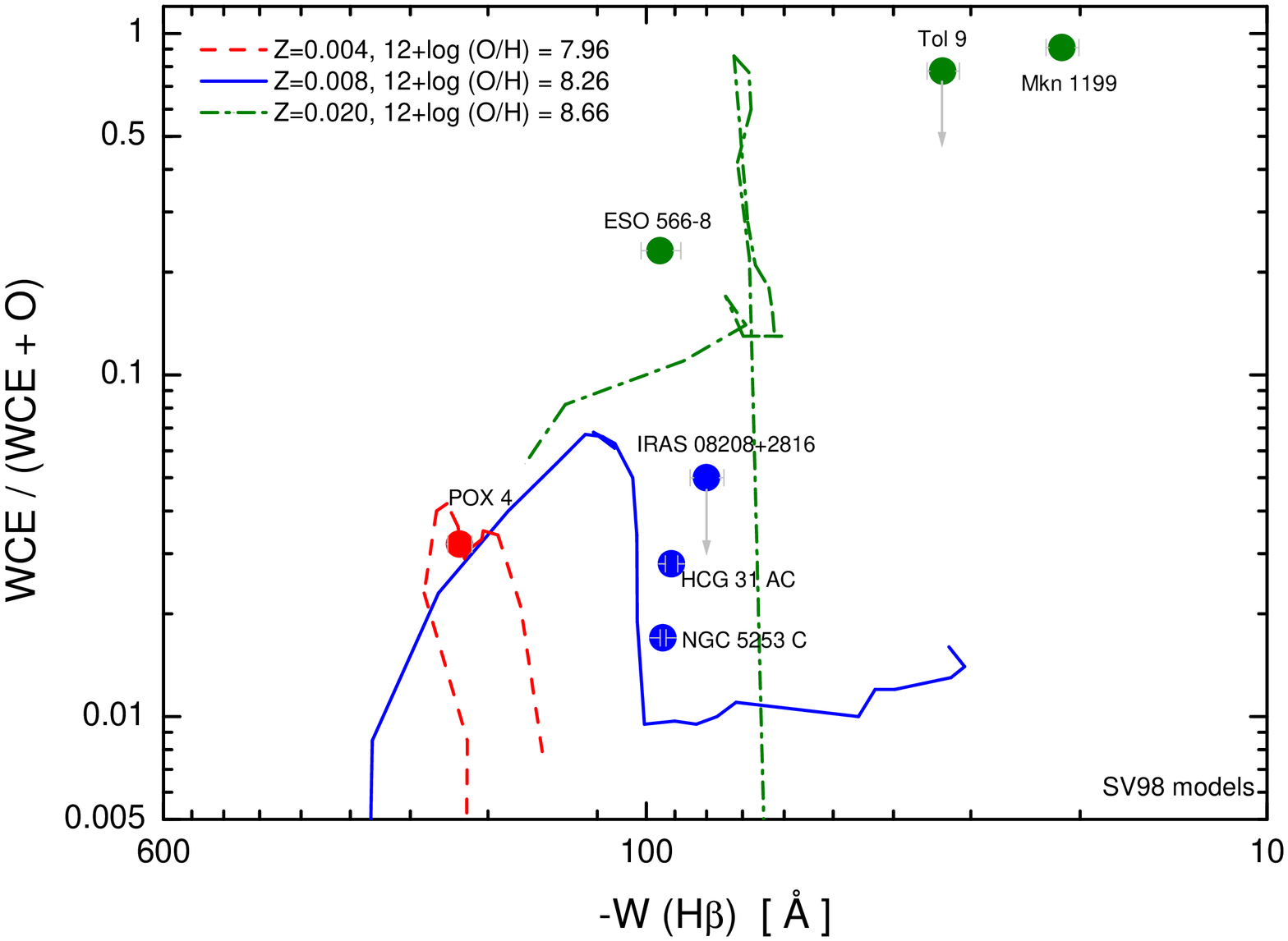} &
\includegraphics[angle=270,width=0.47\linewidth]{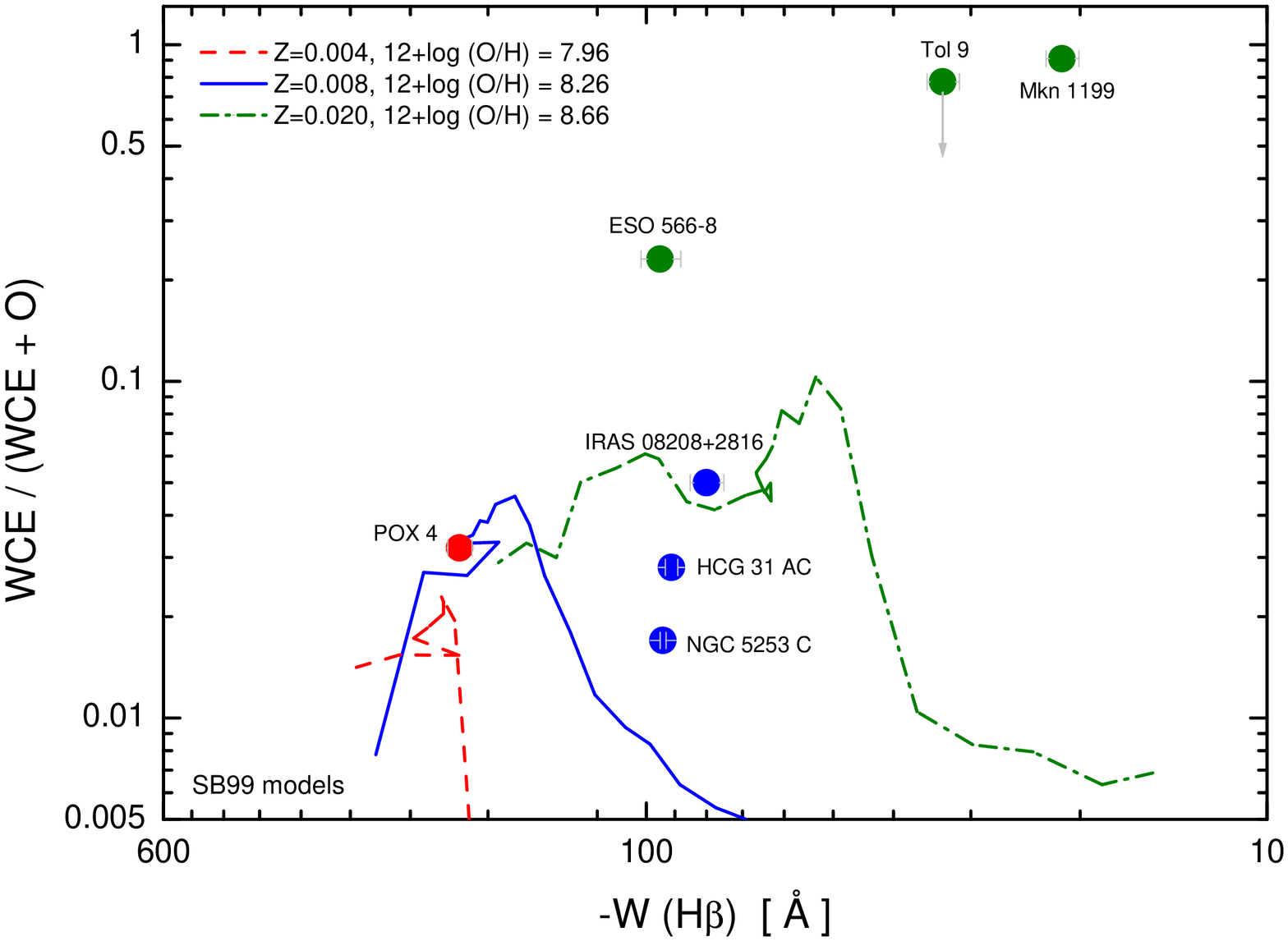} \\
\includegraphics[angle=270,width=0.47\linewidth]{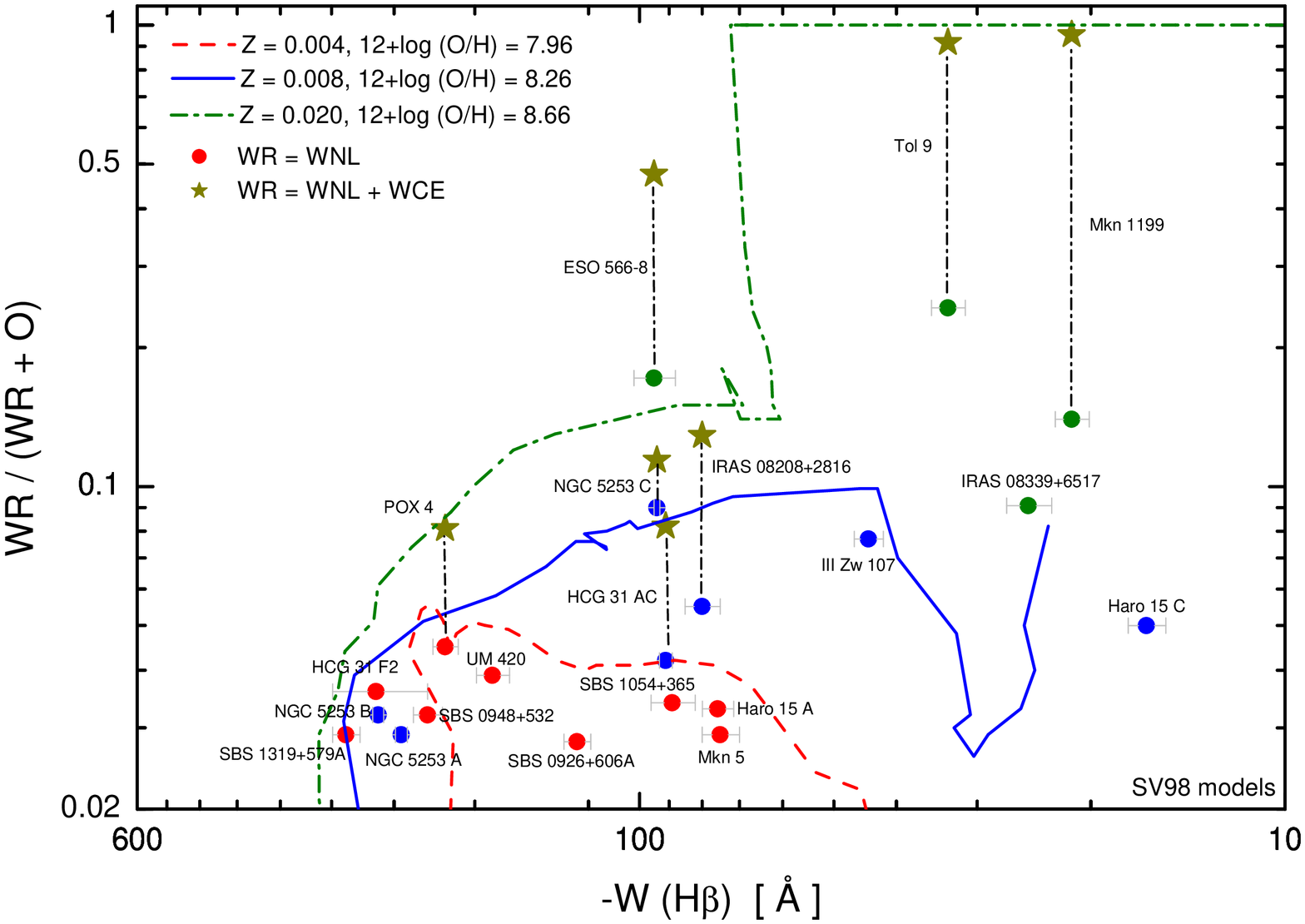} &
\includegraphics[angle=270,width=0.47\linewidth]{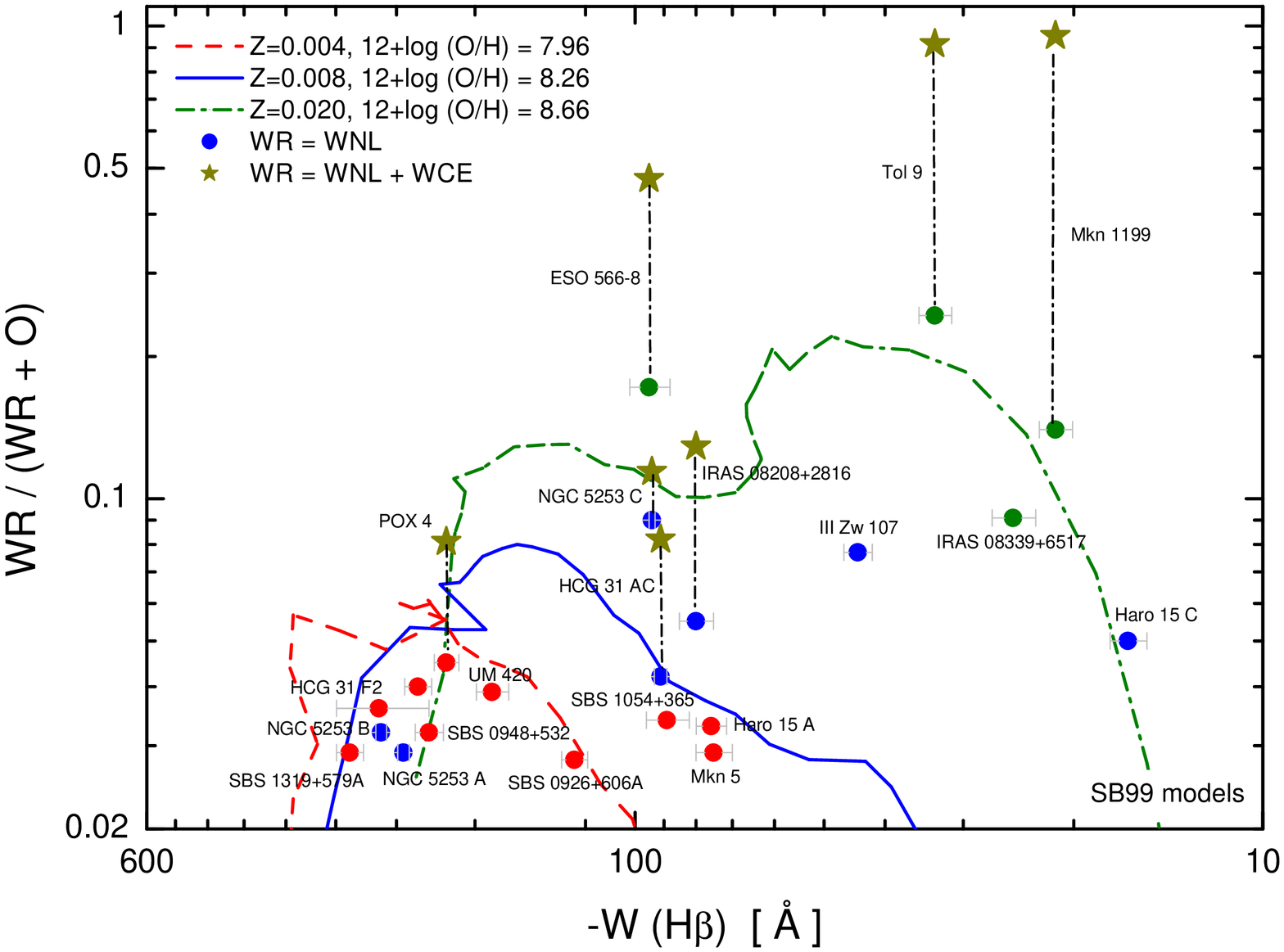} \\
\end{tabular} 
\protect\caption[ ]{\footnotesize{Comparison of the WNL/(WNL+O) (upper row) WCE/(WCE+O) (central row) and WR/(WR+O) (lower row) ratios with the  
predictions given by SV98 (left) and SB99 (right) models. Symbols are the same as in Fig.~\ref{heiimodel}.}}
\label{wrmodels3}
\end{figure*}

\begin{figure*}[t!]
\includegraphics[angle=270,width=0.47\linewidth]{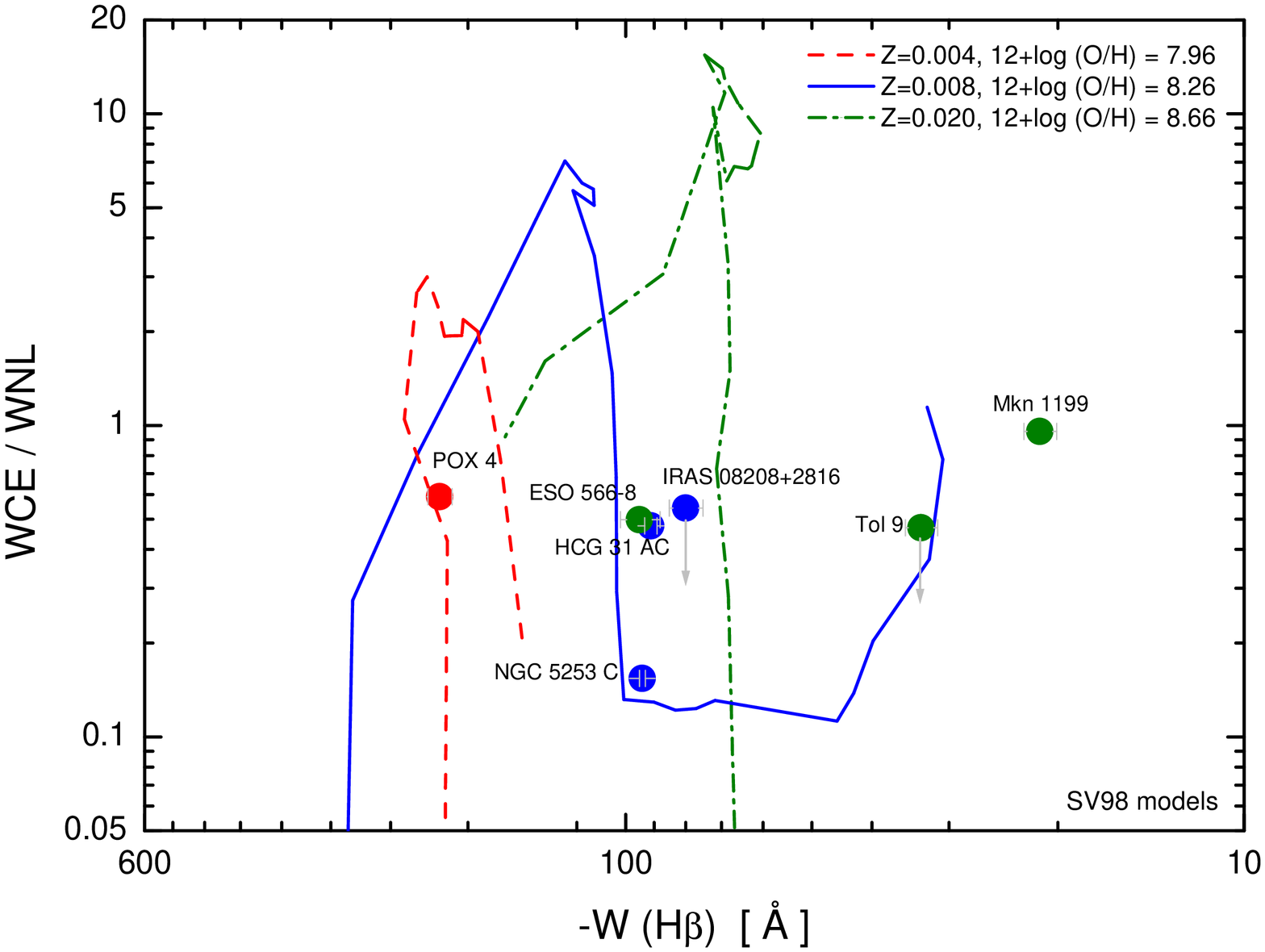}
\includegraphics[angle=270,width=0.47\linewidth]{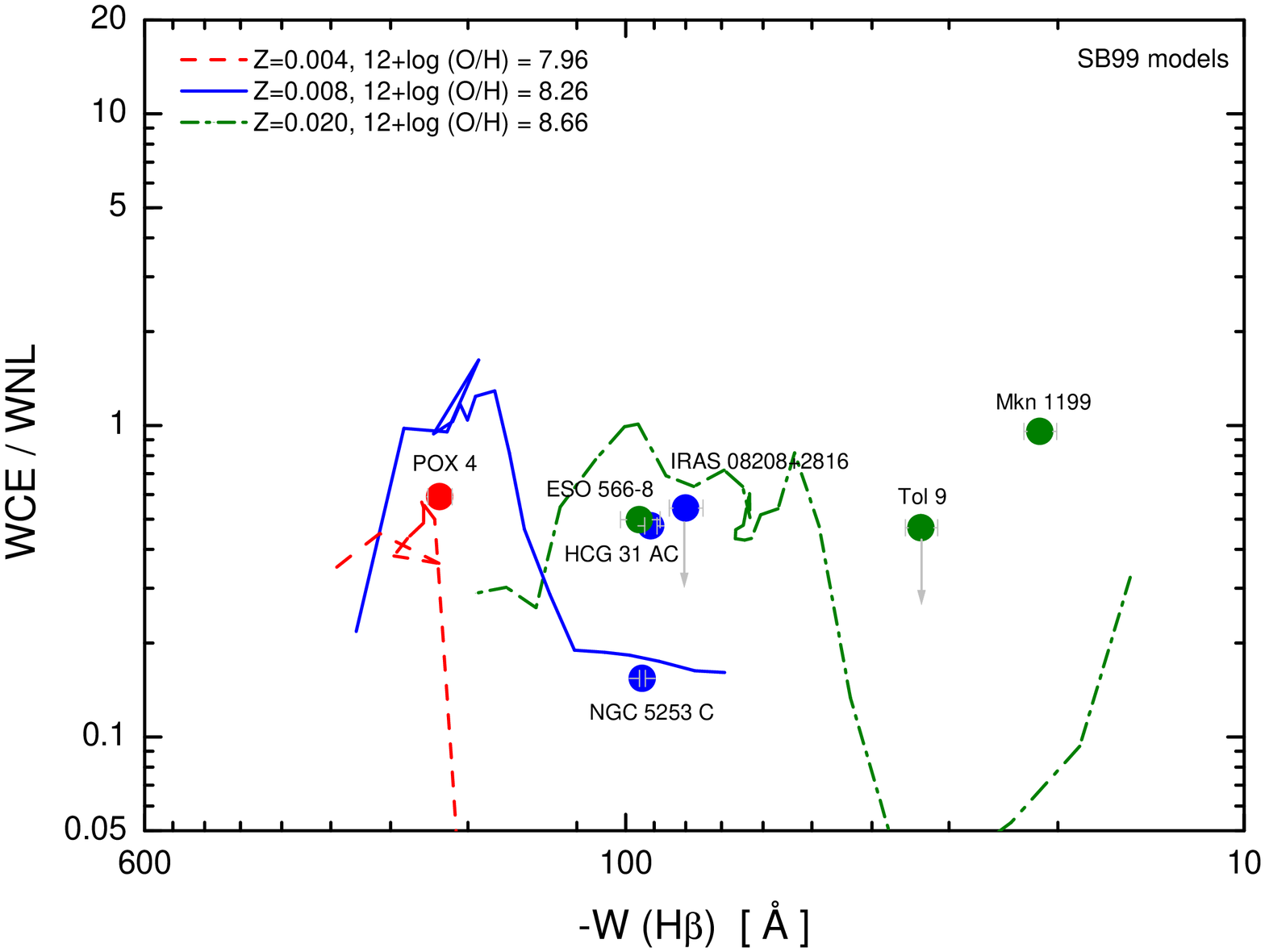}
\protect\caption[ ]{\footnotesize{Comparison of the derived WCE/WNL ratios with the predictions given by the SV98 (left) and SB99 (right) models. Symbols  
are the same as in Fig.~\ref{heiimodel}. }}
\label{wrmodels4}
\end{figure*}

Figures~\ref{heiimodel}--\ref{wrmodels4} compare the predictions of the SV98 and SB99 theoretical models with our observational data. 
As the luminosity of the WR features and their relative strength compared to other spectral properties depends on age, metallicity, star-formation history,
and the IMF, the comparison of the observational data with the models should be carefully done.   
In all these figures the green dashed-dotted line represents the model predictions for solar metallicities ($Z$=0.020), the continuous blue line is the  
prediction of the $Z$=0.008 model and the red dashed line indicates the predictions of the $Z$=0.004 models. The color of the observational points  
is chosen according to the model that better matches their metallicity. 
We always represent in the $x$-axis the \Hb\ equivalent width, \WHb, which is a good indicator for the age of the most recent massive star-formation  
event. Indeed, the hydrogen-ionizing flux of a star cluster gradually decreases as the most massive stars disappear with time, and hence the absolute  
value of the width of \Hb\ decreases with time (see Papers~I and II). BKD08 questioned the validity of this parameter for a large sample of WR  
galaxies spanning a wide range in metallicities and sizes, moreover if the spectrum covers a big region in the galaxy including young and old stellar  
populations. This is not typically our case, because all the spectra analyzed here were mainly obtained in brightest \Ha-selected regions within the  
galaxies (see Paper~II) in which the underlying old stellar population is not dominating the optical light. 
Hence, while the spectroscopic SDSS data used by BKD08 randomly selected a very heterogeneous sample of objects with a fixed fiber size (3$\arcsec$),  
we first carefully analyzed the position of the brightest \HII regions within our sample of WR galaxies (Paper~I) and then extracted the  
spectroscopic data using efficient apertures (Paper~II).     
Furthermore, both the analysis of the \Ha\ imagery and the spectra indicate that all objects have experienced an almost instantaneous star-formation  
event that happened between 3 and 6 \Myr\ ago.
Although some few cases of \BCDG s seem to indicate continuous star formation for some few Myr (i.e., Mkn~209, see P\'erez-Montero \& D\'{\i}az, 2007),
the majority of the
analyses of similar low-metallicity galaxies 
[i.e., Guseva et al. 2000; Cair\'os et al.
2001, 2007; Gil de Paz \& Madore 2005] support the instantaneous star-formation scenario. 
Hence we consider the comparison of the observational data with a model of a single star-formation event
appropriate for our analysis, and that \WHb\ is a reliable indicator of the age of the starbursts.

Figure~\ref{heiimodel} compares the predictions of the nebular \ion{He}{ii} $\lambda$4686 following the SV98 models with our data. The diagrams plot  
the intensity (left panel) and the equivalent width (right panel) of this nebular line as a function of the \Hb\ equivalent width.  
Notice that several high-metallicity objects --green dots--  with  \WHb$\geq-60$ \AA\ are displaced somewhat to the right of the model predictions in  
both diagrams. We think that this effect is a consequence of a probable underestimation of the \WHb\ in these objects, which have more complex  
star-formation histories and a considerable contribution of the old underlying population than the low- and intermediate-metallicity objects. We  
should keep this in mind when comparing observed and predicted properties for high-metallicity objects in the next figures. 
In particular, we should expect a much more complex star-formation history in a spiral galaxy than in a \BCDG . In our study, only Mkn~1199,  
Mkn~1087, ESO~566-8, ESO~566-7, IRAS~08339+6517 and Tol~9, all with 12+log(O/H)$>$8.4, show clear complex star-formation histories, although today
all of them are hosting a strong starburst. These objects are plotted with green dots in Figs.~\ref{heiimodel}--\ref{wrmodels4}. Perhaps their  
\WHb\ have been slightly underestimated, but the comparison of their observed emission line properties with the predictions given by the  
photoionization models provided by \citet{SSL01} (see Fig.~10 in Paper~IV) indicates that the underestimation of the \WHb\ value is not very  
important (10-20 \AA\ at almost). Hence, although the \WHb\ derived in high-metallicities objects may be somewhat underestimated, we think that it  
is not the case for low- and intermediate-metallicity objects because (i) the contribution of the underlying stellar population, although it does  
exist, is not dominating in the strong star-formation bursts 
(see Sect.~4 in Paper~I and its Figs.~37-39) and (ii) the ages of the star-forming bursts obtained from $W$(\Hb) agree with those determined using  
independent methods (see Sect~4 in Paper~II and its Fig.~39).

From Fig.~\ref{heiimodel} (left) is also evident that galaxies with higher metallicities show lower \ion{He}{ii} $\lambda$4686 intensities.  
As we already discussed, this line is not or only very weakly detected in our highest metallicities objects. Both the \ion{He}{ii} $\lambda$4686 intensity  
and its equivalent width (Fig.~\ref{heiimodel}, right) decrease with increasing  \WHb.
The model predictions by SV98 generally agree well with our data, but perhaps they overpredict the intensity of the nebular \ion{He}{ii} line at high  
metallicities, as \citet{GIT00} previously noticed. These authors also pointed out that objects with detected and nondetected WR features mingle  
indistinctly in these plots, and hence not only WR stars but also O stars may be contributing to the ionization of the He$^+$. Precisely this  
was one of the conclusions reached by BKD08, as we already discussed in Sect.~4.

Figure~\ref{wrmodels1} compares the observed intensities and equivalent width of the broad \ion{He}{ii} $\lambda$4686 emission line with the  
predictions given by SV98 and SB99 models. Again the theoretical predictions agree well with the observational data, but 
we observe that the solar-metallicities model clearly overpredict the broad \ion{He}{ii} $\lambda$4686 intensities and equivalent widths by a factor  
4--8. Although part of this disagreement maybe a consequence of the more complex star-formation histories of the high-metallicity galaxies (and
hence an increase of the \WHb\ because of the higher importance of older stellar populations), we consider that some of the disagreement between  
predictions and observational data is real. \citet{PSGD02} also reported that theoretical models overestimate the WR fluxes in solar and super-solar  
metallicity objects. 

Actually, we hold that the agreement we see at low metallicities may be just incidental. 
The SV98 and SB99 models consider the solar values of the intensities of the broad WR lines independently of the metallicity, but as we explained before  
(Sect.~5) they are drastically reduced at low metallicities. Therefore, the predictions given by these models \emph{should not} agree with the  
observational data at low metallicities.
Hence, models using the same parameters as SV98 and SB99 do agree but including metallicity-dependent WR line luminosities would \emph{underpredict} the  
observed broad \ion{He}{ii} $\lambda$4686 intensity (and its equivalent width).
For example, considering the formulae presented before (Eq.~\ref{formula1}), the model with \abox=7.96 (red dashed line in the figures) would predict  
broad \ion{He}{ii} intensities that are 65\% of the intensities shown in Fig.~\ref{wrmodels1}, which makes their equivalent widths also much lower than  
those seen in the figure. 

The same situation is found when comparing the observed broad \ion{C}{iv} $\lambda$5808 emission line with the model predictions  
(Fig.~\ref{wrmodels2}). For example, for the central burst in POX~4, the predicted $I$(\ion{C}{iv} $\lambda$5808)/$I$(\Hb) is  
$\sim$0.015 following the SV98 models, quite in agreement with the actual value observed in this object (0.014). Considering the effect of lower WR  
luminosities at lower metallicities (Eq.~\ref{formula2}), the value predicted by the models should be $\sim$0.01, which is around 70\% of the observed  
value. 
Hence, the SV98 and SB99 theoretical models are indeed underpredicting the intensities of the broad WR lines at low metallicities. This result agrees
with that reported by BKD08 and previously noticed by other authors (e.g. Legrand et al. 1997; Guseva et al. 2000).
The best explanation of this discrepancy seems to be that the binary channel for producing WR stars is important at low metallicities, so that single  
star models are not efficient enough to reproduce the observed values.
The inclusion of binaries in the models led to a prolongated WR phase \citep{VanBeverVanbeveren03} and to an increase in the WR/(WR+O) ratio  
\citep{Eldridge08}. The binary channel ingredient should be taken into account in new models of evolving starbursts showing WR features, as it  
has not been included before, although there is some work in process \citep{EldridgeStanway09}. These new models should also consider the metallicity
dependence of the WR luminosities.

Besides this problem, from Fig.~\ref{wrmodels2} is also clear that the predictions given by the SB99 models are worse than those predicted by the SV98  
models. The SB99 predictions quite disagree for low- and intermediate-metallicity objects. Even without considering the diminution of the WR  
luminosities, the SB99 models predict $I$(\ion{C}{iv} $\lambda$5808)/$I$(\Hb)$\sim$0.2 for POX~4, which is only the 14\% of the observed value. Hence we  
suggest that we detect more WCE stars than those predicted by the models, which difference is even more pronounced at lower metallicities.   

Let us now compare the observed WR/(WR+O) ratios with the theoretical predictions (Fig.~\ref{wrmodels3}). 
The numbers of WNL and WCE stars predicted by the models do not depend on the luminosities of the broad emission features assumed for a single WR  
star (only the predicted WR luminosities depend on them) 
and hence the comparison with our observational data should be still valid. The SB99 models better reproduce the WNL/(WNL+O) ratio than the SV98 models  
(Fig.~\ref{wrmodels3}, top row): the WNL/(WNL+O) ratio predicted by the SB99 models agrees well with the observed values for all metallicities, while  
the SV98 models 
underestimate the \WHb\ in which the maximum of the WR/(WR+O) ratio is expected, which is very evident at low metallicities (differences $>$100 \AA). If  
our observational points were affected by underlying absorption, this effect would be even higher. 
However, both kinds of models strongly disagree with the observations at high metallicities when compared with the observed WCE/(WCE+O) ratio  
(Fig.~\ref{wrmodels3}, central row). The disagreement is particularly important (more than one order of magnitude) for Mkn~1199.

The lower row in Fig.~\ref{wrmodels3} shows the comparison of the derived WR/(WR+O) ratio with the theoretical predictions given by SV98 and SB99. In  
both cases we considered the ratios involving WNL and WR=WNL+WCE stars. The number of WCE stars estimated for Tol~9 and IRAS~08208+2816 is  
tentative, so their derived WR/(WR+O) ratio is an upper limit to the real one. As we commented before, there is a disagreement between the SV98  
predictions and the observations at low metallicities, as these tend to show higher \WHb\ values for the derived WR/(WR+O) ratio. If WCE are  
typically present in those bursts, the models would also underpredict the WR/(WR+O) ratio. Considering again the low-metallicity galaxy POX~4, the  
SV98 predictions for its WR/(WR+O) ratio are $\sim$4 times smaller than the observed value considering the detection of both WNL and WCE stars in  
this object. This difference is also observed in objects with intermediate metallicites, although SV98 models work better in the solar-metallicity  
range. 

On the other hand, the SB99 models seem to overpredict the WR/(WR+O) ratio at low luminosities. As we saw that the observed WNL/(WNL+O) ratio agrees well  
with the predictions in SB99 models, we may interpret this result in two ways: (i) models are overpredicting the number of WCE stars or (ii) we  
are not detecting WCE stars in these low-metallicity objects. In the only low-metallicity galaxy where we detect both WNL and WCE stars the agreement  
is not bad: the SB99 models account for around 80\% of the WR population detected in POX~4. As we see, in all regions in which both subtypes of WR stars  
are detected the models underestimate the total number of WR stars. Hence the problem seems to be a consequence of the predictions of the number of  
WCE stars, but no the number of WNL stars. As we will see below, the derived WNL/WCE ratio computed for these objects is higher than that predicted by  
more recent models \citep{MeynetMaeder05,EldridgeVink06} and also higher than that empirically found counting individual WR stars in nearby galaxies. 

\section{The WCE/WNL ratio}

Table~\ref{wrpop} compiles the WCE/WNL ratio computed in the galaxies for which both WNL and WCE stars have been detected. Figure~\ref{wrmodels4}  
compares the WCE/WNL ratios with the results given by the SV98 and SB99 models. The SV98 models clearly predict an increase of the maximum of the WCE/WNL  
ratio with increasing metallicity, but that is not seen in the SB99 models. Actually, the predictions given by these two kinds of models clearly disagree.  
The SV98 models seem to better reproduce the observed ratios at low and intermediate metallicities, but because of all the caveats about the predictions  
of the numbers of WNL and WCE stars we consider that the SV98 and SB99 models are not valid.

\begin{figure}[t!]
\includegraphics[angle=270,width=\linewidth]{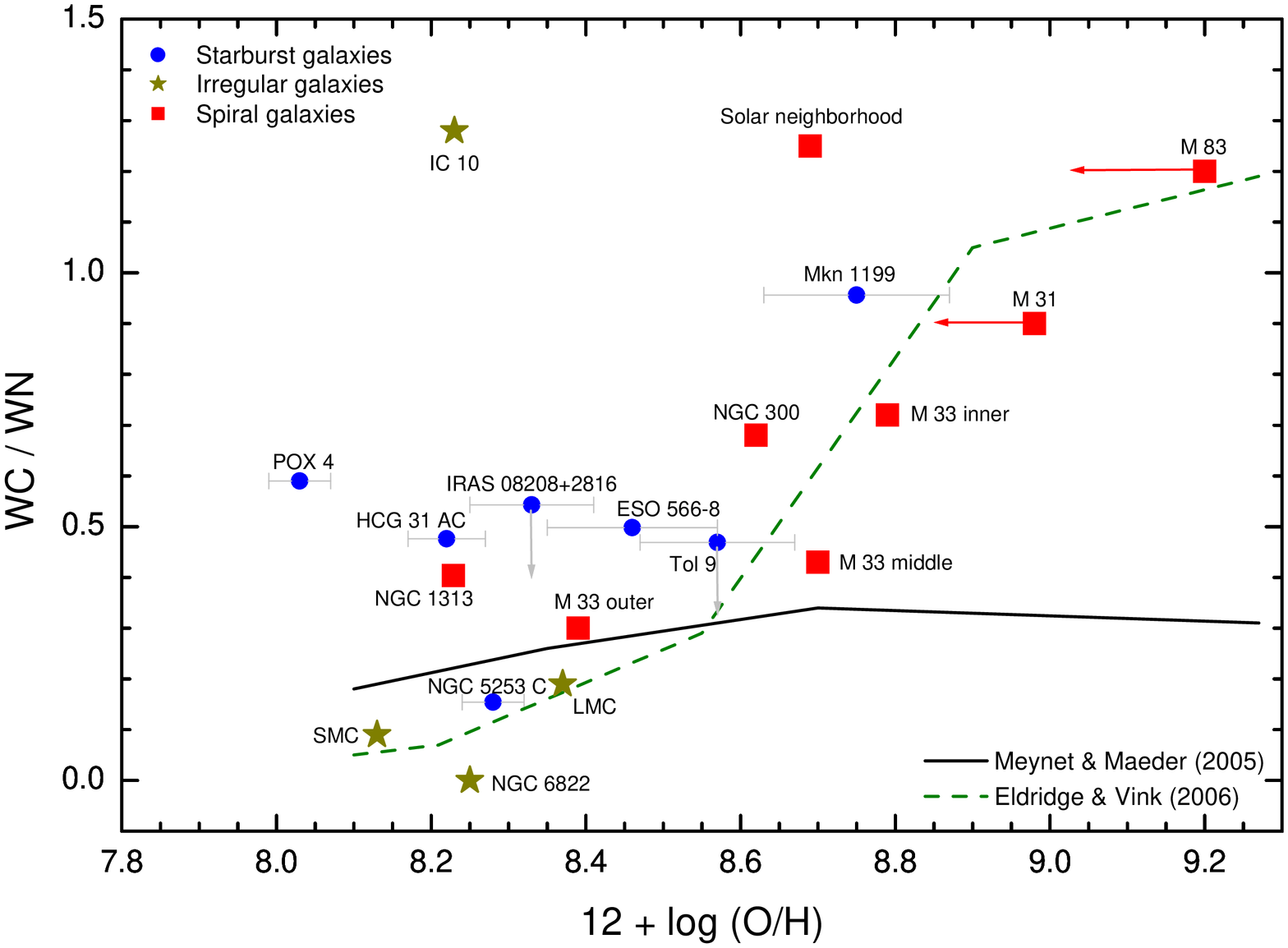}
\protect\caption[ ]{\footnotesize{Comparison between the observed WCE/WNL ratio and the metallicity for the starburst galaxies analyzed in this work  
(blue circles). We include the WC/WN ratios derived for nearby spiral (red squares) and irregular (yellow stars) galaxies (Massey \& Johnson, 1998;  
Crowther et al. 2003, 2007; Schild et al. 2003; Hadfield et al. 2005; Hadfield \& Crowther 2007). Several regions within M~33 are shown because of  
the strong metallicity gradient in this galaxy. The evolutionary model predictions by Meynet \& Maeder (2005, black continuous line) including  
rotation and Eldridge \& Vink (2006, green dashed line) including a metallicity-dependence of the WR winds are also shown.}}
\label{wcwnabox}
\end{figure}

A much more interesting analysis is the metallicity-dependence of the WCE/WNL ratio. Figure~\ref{wcwnabox} shows the WCE/WNL ratio as a function of  
the oxygen abundance for our starbursting galaxies (blue circles) and its comparison with the values obtained for nearby spiral (red squares) and  
irregular (yellow stars) galaxies for which the WR content has been studied in detail \citep{Crowther07}. Except for the particular case of the  
starbursting Local Group galaxy IC~10 \citep{MasseyHolmes02,Crowther+03}, for which the true WR populations are still somewhat uncertain, the WCE/WNL  
ratio clearly increases with the metallicity. That trend also agrees with the SV98 and SB99 models predictions (not shown in Fig.~\ref{wcwnabox}), for  
example, the maximum of the WR/(WR+O) ratio in SV98 models increases from 0.025 to 1 between metallicities of 7.36 and 8.96 --in units of \abox--,  
respectively.  
Following  \citet{Crowther07}, we plot in Fig.~\ref{wcwnabox} the predictions given by the most recent WR models, \citet{MeynetMaeder05}, which  
allow for rotational mixing but without a WR metallicity dependence, and \citet{EldridgeVink06}, which neglect rotational mixing but include a  
metallicity-scaling for WR stars. At high metallicities,  the \citet{EldridgeVink06} models agree well with previous observations and also with our value  
for Mkn~1199, a spiral galaxy experiencing a strong star-formation event because of the interaction with a nearby companion (see Papers~I and II). 

However, the situation is quite different in the intermediate- and low-metallicity regime. Previous data of spiral and irregular galaxies agree well  
with both kind of models, but except for region C within NGC~5253, they clearly underestimate the WCE/WNL ratio in our sample of starburst galaxies.  
That is the same result that \citet{GIT00} found in their analysis. They assumed that the WC/WN ratio empirically derived by counting individual WR  
stars in nearby galaxies corresponds to continuous star formation, that should be $\sim$0 for galaxies with \abox$<$8.1. But  
low-metallicity galaxies are experiencing an instantaneous starburst, and hence their derived WCE/WNL may be as high as $\sim$3 for $Z$=0.004  
--\abox$\sim$7.96-- according to SV98 models. This observation also agrees very well with the high ratio derived for IC~10, WCE/WNL$\sim$1.25. 
We therefore conclude that the differences in the WCE/WNL ratio found between starbursting low-metallicity dwarf galaxies and the typical nearby  
star-forming galaxies are a consequence of their very different star-formation histories, and that new models should definitively include this  
component.  

\begin{figure}[t!]
\includegraphics[angle=270,width=\linewidth]{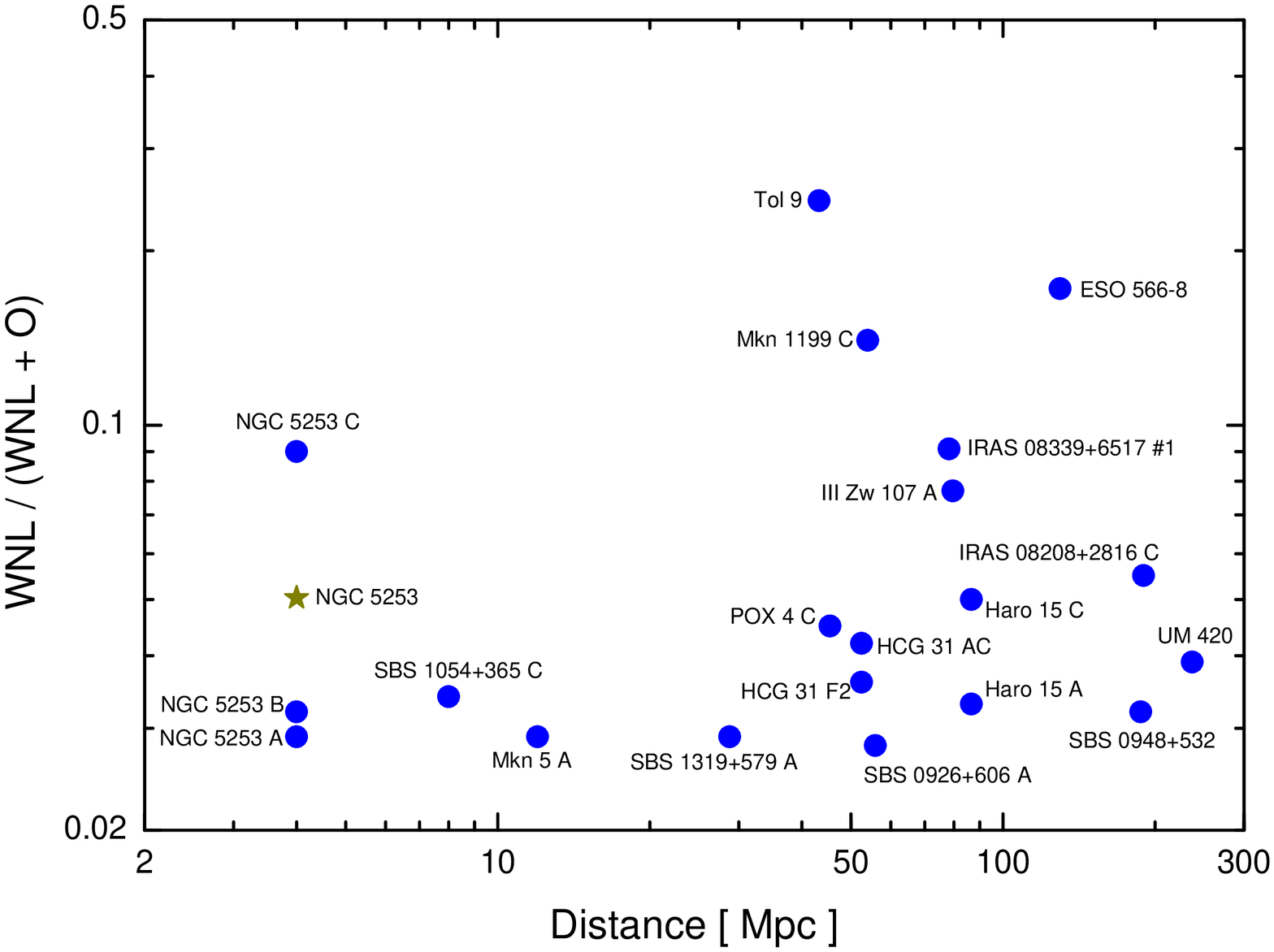}
\protect\caption[ ]{\footnotesize{WNL/(WNL+O) ratio vs. the distance to the galaxies. Notice that we also indicate the region within the galaxy in  
which the WR features are detected. The average value of the WNL/(WNL+O) ratio for the \BCDG\ NGC~5253 is plotted with a yellow star.}}
\label{wrdist}
\end{figure}

However, we must keep in mind the effect of the slit position and the size of the aperture on the derived WR properties, which is particularly  
evident in nearby objects such as IC~10 or NGC~5253. Indeed, we see that different regions within these galaxies have very different WR populations.  
To investigate this issue, we plot in Fig.~\ref{wrdist} the metallicity-dependence WNL/(WNL+O) ratio as a function of the distance to the galaxies  
(see Table~1 in Paper~I). The distance range of our sample galaxy is between 4~Mpc (NGC~5253) and 237~Mpc (UM~420), with the majority of the objects  
within a distance of 100~Mpc. We do not see any clear correlation in this diagram, so we can discard any distance effect in our conclusions. We  
remind that we did not consider all flux of the galaxy with a fixed aperture, but analyzed the spectra of the brightest \HII regions found in  
each galaxy using our \Ha\ maps (Paper~I) and choosing the size of the aperture according to these regions (Paper~II). Hence the effects of dilution of  
the WR features (including areas in which the WR stars are not presented) were minimized in our analysis.

\section{The WR dependence on the IMF}

The detection of WR stars of both WN and WC subtypes and the derived WC/WN ratio provides strong constraints for stellar evolution models through the  
determination of the upper cut-off limit of the IMF. As WC stars are difficult to observe in low-metallicity environments, these studies are  
usually performed in metal-rich objects. 
Actually, the slope of the IMF and its dependence on metallicity is still a controversial issue. 
Some studies support a Salpeter-like IMF at solar metallicities \citep{BresolinKennicutt02,PSGD02,FCCG04,Bre05}, but
other observations have questioned this assumption at different metallicities. \citet*{SCP99} and \citet*{HGJL99} suggested a flatter IMF at lower  
metallicities. \citet{FCCG04} underlined a dependence of the IMF slope on galaxy metallicity, in which low metallicity galaxies show a Salpeter-like  
IMF while high-metallicities galaxies either show a steeper IMF or experience an extended burst. A somewhat similar result was recently found by  
\citet{Zhang07}, who found that 
the IMF slope increases with increasing metallicity, with the slope index ranging from $\alpha\sim$1.0 for $Z$=0.001 to $\alpha\sim$3.3 for  
$Z$=0.020.

\begin{figure}[t!]
\includegraphics[angle=270,width=\linewidth]{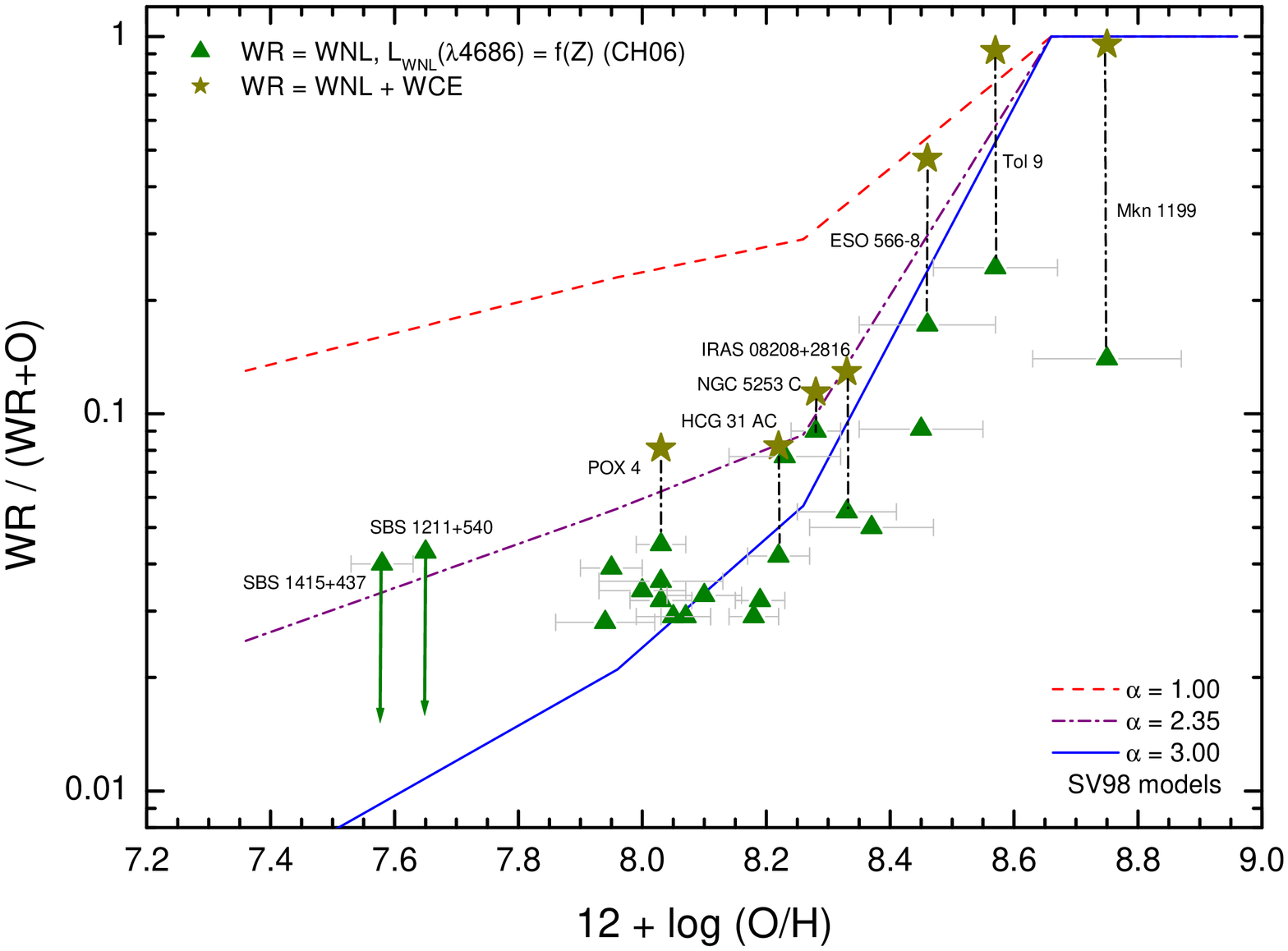}
\protect\caption[ ]{\footnotesize{WR/(WR+O) vs. metallicity and the predictions of SV98 models assuming three different IMFs, $\alpha$=1.00 (flatter,  
red dashed line), 2.35 (Salpeter, dashed-dotted purple line) and 3.00 (steeper, blue continuous line). As in Fig.~\ref{wrabox}, green  
triangles indicate WR=WNL considering a metallicity-dependent broad \ion{He}{ii} $\lambda$4686, and dark yellow stars plot WR=WNL+WCE assuming the  
metallicity-dependence of both WNL and WCE luminosities. Dashed lines connect the WR=WNL and the WR=WNL+WCE values for the galaxies for which we have WCE data,  
these objects have been labelled. We include the upper limit to the WR/(WR+O) ratio estimated for the very low-metallicity galaxies SBS~1211+540 and  
SBS~1415+437.}}
\label{wrimf}
\end{figure}

BKD08 pointed out that taking into account the considerable number of parameters that should be considered when modeling starbursts with significant WR  
populations (age of the burst, star-formation history, metallicity, dependence of the WR features on stellar rotation and wind loss, inclusion of the  
binary channel...) it is not a good idea to include the slope of the IMF as an additional parameter. Just as an exercise, we compared the  
predictions given by SV98 models for an instantaneous burst with the same properties we explained before but changing the slope of the IMF,  
$\alpha$=1.00, 2.35 (Salpeter) and 3.00. Figure~\ref{wrimf} plots the observed WR/(WR+O) ratio versus the oxygen abundance and the maximum of the  
WR/(WR+O) predicted by these three models. The observational data plotted in this figure are the same as in Fig.~\ref{wrabox};  the WNL and WCE  
stars numbers were computed assuming a metallicity-dependent luminosity of the WR features, as we explained before. As we see, models with flatter  
IMFs tend to predict higher WR values and at all metallicities, so that a flat IMF implies a relatively large population of massive stars, leading to  
a larger WR population.

From Fig.~\ref{wrimf} is evident that our data do not support the hypothesis that low-metallicity objects have flatter IMFs. Actually it is quite  
the opposite, the model with $\alpha$=3.00 seems to agree better with the data. The assumption of metallicity-independent WR luminosities  
would move our data points to lower WR/(WR+O) values, and hence favore an even steeper slope for the IMF at low metallicities. However, we  
notice that the majority of the data points assumes WR=WNL, but we have seen that the WCE stars should have a non-negligible contribution to the  
total WR star population even at low metallicities. Actually, the model with a Salpeter-like IMF agrees very well with the observed WR/(WR+O) value  
of the objects for which we have detected both WNL and WCE stars. We then conclude that the assumption of a Salpeter-like IMF is valid and that the  
metallicity-dependence on the WR luminosities is probably playing a key factor in the discrepancies between data and models found by   
\citet{Zhang07}.

A final remark about IMFs in star-forming galaxies. Recent studies \citep{WeidnerKroupa05,WeidnerKroupa06} have revealed that the total stellar  
population of the new formed stars of all young star clusters within the same galaxy follows a distribution function steeper than the canonical IMF  
in the high-mass regime. The so-called \emph{integrated galactic initial mass function} (IGIMF) deviates increasingly from the underlying canonical  
IMF with decreasing star-formation and total galaxy mass. Blue Compact Dwarf galaxies, such as the typical low-metallicity objects in which WR stars are detected, should   
have a very different IGIMF than spiral galaxies like the Milky Way. This parameter should be therefore also considered when modelling the evolution  
of the stellar populations in starbursts and star-forming galaxies.


\section{Conclusions}

We presented a detailed analysis of the broad stellar features originated by winds of Wolf-Rayet stars in a sample of starbursting galaxies  
previously classified as WR galaxies. The photometric properties of these galaxies were compiled in Paper~I, while their spectroscopic analysis was  
shown in Paper~II. The metallicity of these galaxies lies between 7.58 and 8.75 --in units of \abox--. Our goals are to locate the WR-rich stellar  
clusters, derive the number of O, WN and WC stars and compare the results with previous observations and the predictions of evolutionary synthesis  
models. 
Our main results are the following:

\begin{enumerate}
\item The blue \WRBUMP, mainly composed by the broad, stellar \ion{He}{ii} $\lambda$4686 emission line and attributed to WN stars, is unambiguously  
detected in 12 regions and it is probably observed in eight other regions. 
Aperture effects and/or the position of the slit within the starburst are clearly playing a fundamental role in the detection of the WR features. We  
consider that multi-aperture spectroscopy or narrow-band imagery using filters sensitive to WR features and the adjacent continuum later followed by  
slit spectroscopy may be more appropriate for this kind of analysis. 
The advent of the optical 3D spectroscopy is already showing the localization of WR features with star-forming regions 
(i.e. Kehrig et al. 2008; Garc\'{\i}a-Lorenzo et al. 2008; James et al. 2009,2010; L\'opez-S\'anchez et al. 2010). 
\item The broad \ion{C}{iv} $\lambda$5808 emission line (red \WRBUMP), attributed to WC stars, is clearly detected in four regions, and it is probably  
observed in two other objects. 
\item We have performed a detailed fitting of the spectra considering the broad stellar and narrow nebular emission lines in both the blue and the  
red \WRBUMP s. In the blue \WRBUMP\ we usually found a contribution of the nebular and broad \ion{He}{ii} $\lambda$4686 components. 
\item The nebular \ion{He}{ii} $\lambda$4686 is detected in 17 regions and it is probably observed in six other knots. We confirm the trend that  
galaxies with lower metallicities have a higher $I$(nebular \ion{He}{ii} $\lambda$4686)/$I$(\Hb) ratio. Some of our lowest-metallicity regions show  
the nebular but not the broad \ion{He}{ii} emission line.     
The SV98 models 
seem to underpredict the intensity of this emission line in low metallicity objects.  Both findings agree with the conclusions reached by BKD08, who  
suggested that O stars and not only WR stars contribute to the ionization of the He$^+$ at low metallicities.
\item We used the intensity ratios of the broad \ion{He}{ii} $\lambda$4686 and \ion{C}{iv} $\lambda$5808 lines to derive the number of O, WN and WC  
stars within each region. We have considered two methods: (i) assuming constant luminosities for the WR features, as it is commonly done and it   
used in theoretical models, and (ii) assuming metallicity-dependent luminosities for the WR features, as both recent observations and new WR stars  
models indicate (CH06). This second method gives higher number of WR stars at lower metallicities. 
We derive an empirical estimation of the WNL/(WNL+O) ratio using the intensity of the broad \ion{He}{ii} $\lambda$4686 line assuming the  
metallicity-dependent of the WR luminosities.
\item As expected following theoretical models, the total number of WR stars increases with increasing metallicity. However, we observe a possible  
flattening of the WR/(WR+O) ratio for objects with \abox$<$8.2, because all the low-metallicity galaxies show a constant value of $\sim$0.03 -- 0.04,  
but more data are needed to confirm this trend. We also conclude that the contribution of the WCE stars is not negligible at low metallicities. 
\item The comparison of the observational data with the theoretical predictions given by the most accurate models available to date, SV98 and SB99,  
is difficult because of all the parameters involved, namely, age, metallicity, star-formation history, assumption of the IMF and WR stars properties  
(variation of the WR luminosities with the metallicity, effect of star rotation, contribution of the WR binnary channel). New models including all  
these factors are absolutely needed to perform an appropriate comparison with the observational data. The available SV98 and SB99 models seem to work  
better at higher metallicities, but more complex star-formation histories are needed to explain the observed WR luminosities and ratios in this  
regime. At low metallicities the SV98 and SB99 models fail to reproduce the observed WR intensities because these models do not consider the decreasing  
of the WR luminosities with decreasing metallicity. 
Our data agree with the SV98 models considering a Salpeter-like IMF for all metallicites. 
\item Finally, the WCE/WNL ratios observed in our starbursting galaxies are quite different than those empirically found in nearby star-forming and  
irregular galaxies. We consider that this effect is a consequence of the very different star-formation histories that these objects are  
experiencing. 
\end{enumerate}

\begin{acknowledgements}

We are very grateful to the referee, Daniel Kunth, who helped us to improve the quality of this manuscript.
\'A.R. L-S. \emph{deeply} thanks to Universidad de La Laguna (Tenerife, Spain) for forcing
him to translate his PhD thesis from English to Spanish; he had to translate it from Spanish to English to complete this publication. 
\'A.R. L-S. also thanks to all the people at the CSIRO/Australia Telescope National Facility, especially to B\"arbel Koribalski, 
for their support and friendship while translating his PhD. 
The authors are very grateful to A\&A language editor, A. Peter, for his kind revision of the manuscript. 
This work has been partially funded by the Spanish Ministerio de Ciencia y Tecnolog\'{\i}a (MCyT) under project AYA2004-07466. 
This research has made use of the NASA/IPAC Extragalactic Database (NED) which is 
operated by the Jet Propulsion Laboratory, California Institute of Technology, 
under contract with the National Aeronautics and Space Administration. 
This research has made extensive use of the SAO/NASA Astrophysics Data System Bibliographic Services (ADS).

\end{acknowledgements}

\listofobjects


\end{document}